\documentclass[final,5p,times,twocolumn,authoryear]{elsarticle}

\usepackage{amsmath,amssymb,amsfonts,amsthm,mathtools,bm}
\usepackage[utf8]{inputenc}
\usepackage[english]{babel}
\usepackage{microtype}
\usepackage[dvipsnames]{xcolor}
\usepackage{graphicx}
\usepackage{float}
\usepackage{dblfloatfix}       
\usepackage{subcaption}
\usepackage{placeins}         
\usepackage{booktabs}
\usepackage{enumitem}
\usepackage[ruled,vlined,linesnumbered]{algorithm2e}
\usepackage{csquotes}
\usepackage[hidelinks]{hyperref}
\usepackage{xcolor}

\newtheorem{remark}{Remark}

\newtheorem{lemma}{Lemma}

\allowdisplaybreaks
\def\E{\mathbb{E}}
\def\l{\langle}

\journal{Neural Networks}

\begin{document}

\begin{frontmatter}

\title{\boldmath Do Hopfield Networks Dream of Stored Patterns?\\
A Statistical--Mechanical Theory of Dreaming in Multidirectional Associative Memories}

\author[a,b,c]{Adriano Barra}
\author[d]{Fabrizio Durante}
\author[a,c]{Andrea Ladiana\corref{cor1}}
\ead{andrea.ladiana@uniroma1.it}
\author[c,d,e]{Michela Marra Solazzo}

\affiliation[a]{organization={Dipartimento di Scienze di Base Applicate all'Ingegneria, Sapienza Universit\`a di Roma},
            city={Rome},
            country={Italy}}
\affiliation[b]{organization={Istituto Nazionale di Fisica Nucleare, Sezione di Roma1},
            country={Italy}}
\affiliation[c]{organization={Istituto Nazionale d'Alta Matematica, GNFM},
            city={Roma},
            country={Italy}}
\affiliation[d]{organization={Dipartimento di Matematica e Fisica, Universit\`a del Salento},
            city={Lecce},
            country={Italy}}
\affiliation[e]{organization={Istituto Nazionale di Fisica Nucleare, Sezione di Lecce},
            country={Italy}}
\cortext[cor1]{Corresponding author}

\begin{abstract}
We introduce the Dreaming $L$-directional Associative Memory (DLAM), a multi-layer Hebbian architecture in which off-line dreaming and supervised heteroassociative coupling coexist within a single energy function, placing our approach within the framework of energy-based models (EBMs). The replica-symmetric free energy, derived via the Guerra interpolation scheme, yields self-consistency equations governing the order parameters across the control-parameter space. The effective local field decomposes into signal, intra-layer dreaming noise, and inter-layer noise. Dreaming improves retrieval by differentially attenuating high-eigenvalue interference modes of the empirical correlation matrix, suppressing inter-pattern crosstalk while preserving the signal.
Dreaming and inter-layer coupling prove synergistic, opening retrieval regions unreachable by either mechanism alone, as confirmed by Monte Carlo simulations for $L=3$. Their interplay is most pronounced on pattern disentanglement: given a mixture state as input, the network splits the constituent patterns one-per-layer, recovering each modality-specific pattern from a common cue that simultaneously blends noisy evidence from all sensory channels.
Phase diagrams are planar projections of the hyperspace $(\alpha,\beta,\rho,t)$—where $\alpha$ is the storage load, $\beta$ the fast-noise inverse temperature, $\rho$ the dataset entropy, and $t$ the sleeping time. In the $(\rho,t)$-plane, the diagrams reveal a data–computation trade-off: off-line consolidation substitutes for additional training data, extending to heteroassociative architectures a phenomenon previously established for autoassociative networks.
Enriching the standard Hopfield model with heteroassociativity and dreaming gives rise to EBMs capable of complex tasks beyond classical pattern recognition, contributing to a modern theory of neural information processing.
\end{abstract}


\end{frontmatter}

\section{Introduction}\label{sec:intro}

Among the most provocative ideas to emerge from the intersection of neuroscience and statistical mechanics is the hypothesis that sleep serves a computational function.  \citet{CrickMitchison1983} proposed that rapid-eye-movement sleep operates as a form of reverse learning, selectively weakening parasitic synaptic configurations accumulated during waking experience that would otherwise corrupt stored memories.  This early proposal is now situated within a broad experimental literature documenting the active role of sleep in memory consolidation, replay, and synaptic reorganisation \citep{DiekelmannBorn2010,RaschBorn2013,KlinzingNiethardBorn2019}.  Translated into the framework of Hopfield networks \citep{Hopfield1982}, this amounts to an off-line adjustment of the coupling matrix that suppresses inter-pattern crosstalk and sharpens the basins of attraction around genuine stored patterns.  The idea is noteworthy because the role it assigns to sleep is at its core \emph{algorithmic}: the sleeping network performs a spectral filtering of its own synaptic matrix, a computation that the waking network, constrained to process incoming data, cannot easily carry out on-line.  This interpretation is broadly consonant with synaptic-homeostasis accounts of sleep \citep{TononiCirelli2006}, although the present kernel is a mathematical idealisation rather than a biological claim.  The analogy between this biological picture and the mathematical operation of regularising a noisy empirical covariance matrix provides both a conceptual bridge and a technical starting point for the present work.

The analytical confirmation of this picture has unfolded over several decades.  The storage capacity of the classical Hopfield network, limited to $\alpha_c \simeq 0.138$ by inter-pattern interference \citep{AmitGutfreundSompolinsky1985,HopfieldFeinsteinPalmer1983}, can be raised substantially when the coupling matrix is subjected to a dreaming-type correction, whether through Hebbian unlearning, pseudo-inverse regularisation, or the spectral kernel introduced in \citet{FachechiETAL2019dreaming} and \citet{AlemannoBarraETAL2023dreaming}.  In the supervised learning-from-examples framework, where the network observes noisy copies of each archetype rather than the archetypes themselves, dreaming carries a dual benefit: it suppresses inter-pattern crosstalk \emph{and} compensates for finite-sample noise, thereby acting as both a capacity booster and a data-saving mechanism \citep{AlemannoETAL2023small,FachechiETAL2025regularization,JSTAT2024barra,SerricchioETAL2025daydreaming}.  These results, however, have been obtained exclusively in single-layer architectures (auto-associative memories).  When the task is intrinsically multimodal, that is, when a memory comprises coordinated information arriving through several sensory channels, a single layer cannot represent the relational structure between modalities, and dreaming within that layer alone cannot exploit cross-modal regularities.  Recent work has shown that an assembly of Hopfield networks can display emergent capabilities that are absent at the level of a single network \citep{agliari2025networks}: in particular, a layered architecture can spontaneously perform \emph{pattern disentanglement}, recovering the individual components of a composite signal, in analogy with identifying the notes composing a chord, a task that is by construction impossible for any single-layer model.  This ``more is different'' principle, in the spirit of P.~W.~Anderson, motivates the multi-layer architecture developed in the present work.

The natural formalism for multimodal memory is the $L$--directional associative memory (LAM), illustrated in Fig.~\ref{fig:dlam_architecture} (left panel), in which $L$ layers of neurons, each encoding a distinct modality, interact through Hebbian heteroassociative couplings that bind co-occurring patterns across layers \citep{KobayashiTAM,agliari2025generalized, alessandrelli2025supervised, alessandrelli2025beyond}.  The architecture has a principled biological motivation: the mammalian cortex stores multimodal representations in anatomically distributed areas and retrieves them through associative completion \citep{McClellandMcNaughtonOReilly1995,RissmanWagner2012}: a visual cue can evoke an auditory memory, and a partial stimulus can elicit its full multisensory context \citep{BarkerWarburton2020}.  Pattern completion across modalities and pattern separation within a modality are widely regarded as complementary operations subserved by hippocampal--cortical circuits \citep{YassaStark2011,Rolls2013}; the LAM provides a tractable mathematical abstraction of this organisation.  From the statistical-mechanics viewpoint, inter-layer coupling extends the retrieval region relative to independent layers, but the capacity remains limited by inter-pattern interference, and the problem becomes especially acute when the training data is noisy or scarce.

These two lines of research, single-layer dreaming and multi-layer heteroassociation, have developed largely in parallel, yet the regime where their combination matters most is precisely the one that neither addresses on its own: when the training data is noisy or scarce, when the cues are ambiguous, and when successful recall requires both intra-modal noise suppression and cross-modal support.  A layer whose local data is too poor for standalone retrieval might be rescued by information relayed from better-informed partner layers, but only if its own intra-layer landscape is sufficiently clean for the heteroassociative signal to take effect.  The interplay between the two mechanisms is non-trivial and cannot be anticipated by studying them in isolation.

In this paper we construct the \emph{Dreaming $L$--directional Associative Memory} (DLAM), a single Hamiltonian that integrates intra-layer spectral dreaming with inter-layer Hebbian coupling, and we develop a closed replica-symmetric description of its thermodynamics.  The analytical framework rests on a Hubbard--Stratonovich linearisation of both quadratic sectors, followed by a Guerra interpolation that yields the quenched free energy under the replica-symmetric (RS) ansatz.  The central analytical result is a closed system of self-consistency equations governing the fixed points of the order parameters (listed in Table~\ref{tab:orderparams}) whose effective local field decomposes into a signal term, enhanced by dreaming; a dreaming-noise term, suppressed by dreaming; and an inter-layer noise term.  This decomposition provides a clear field-theoretic mechanism for the dreaming benefit and yields the phase diagram as a function of storage load, dataset entropy, temperature, and inter-layer coupling, revealing re-entrant structure and synergistic dreaming--cooperation regions not reached by either mechanism alone in the explored regime.  Monte Carlo simulations on three-layer networks confirm the analytical predictions across the explored parameter space.  The finding with the most immediate significance concerns the task of \emph{disentanglement}: presenting all layers simultaneously with a common cue that blends the noisy modal evidence for a single target archetype across all sensory channels, and requiring every layer to converge to its own modality-specific version of that archetype.  Without dreaming the cross-modal mixture cue excites inter-pattern correlations that trap each layer in a spurious state; with moderate dreaming the spectral filtering reorganises the landscape and joint disentanglement success rises substantially.  In the $(\rho,t)$-plane, the numerical phase diagrams further establish a data--computation trade-off: off-line consolidation substitutes for additional training examples, extending to the multi-layer setting results recently obtained in single-layer networks \citep{AlemannoETAL2023small,FachechiETAL2025regularization}.

The remainder of the paper is organised as follows.  Section~\ref{sec:model} defines the DLAM architecture, introduces the dreaming kernel and the inter-layer coupling, and formulates the disentanglement task that serves as the central test of the theory.  Section~\ref{sec:analytical} develops the replica-symmetric free energy, states the self-consistency equations, and analyses the phase structure.  Section~\ref{sec:numerical} presents Monte Carlo results, beginning with disentanglement and data saving, and proceeding through the full phase structure, finite-size scaling, design rules, dynamics, and robustness to heterogeneous data quality.  Section~\ref{sec:discussion} discusses connections to related work, implications, and limitations.  Detailed derivations are collected in a companion Appendix.

\section{Model}\label{sec:model}

\subsection{Architecture}\label{subsec:architecture}

The DLAM comprises $L$ interacting layers, each populated by $N$ binary neurons $s_{i,l}\in\{-1,+1\}$, $i=1,\dots,N$, $l=1,\dots,L$.  Each layer represents a distinct sensory modality or data channel; the multi-layer organisation is what allows the network to bind information across modalities, in contrast with single-layer architectures that treat each modality independently.  The network stores $K$ archetypes, each of which is \emph{multimodal} in the sense that it consists of an $L$-tuple of independently drawn Rademacher vectors, one per layer: $\bm{\xi}_l^\mu\in\{-1,+1\}^N$ for $\mu=1,\dots,K$ and $l=1,\dots,L$. Within each layer, the component $\bm{\xi}_l^\mu$ is a standard Hopfield pattern; the qualifier \emph{multimodal} simply records that a single index $\mu$ simultaneously addresses one Rademacher pattern in every layer. All archetypes are quenched and hidden from the network.

\begin{figure*}[htbp]
    \centering
    \includegraphics[width=0.95\linewidth]{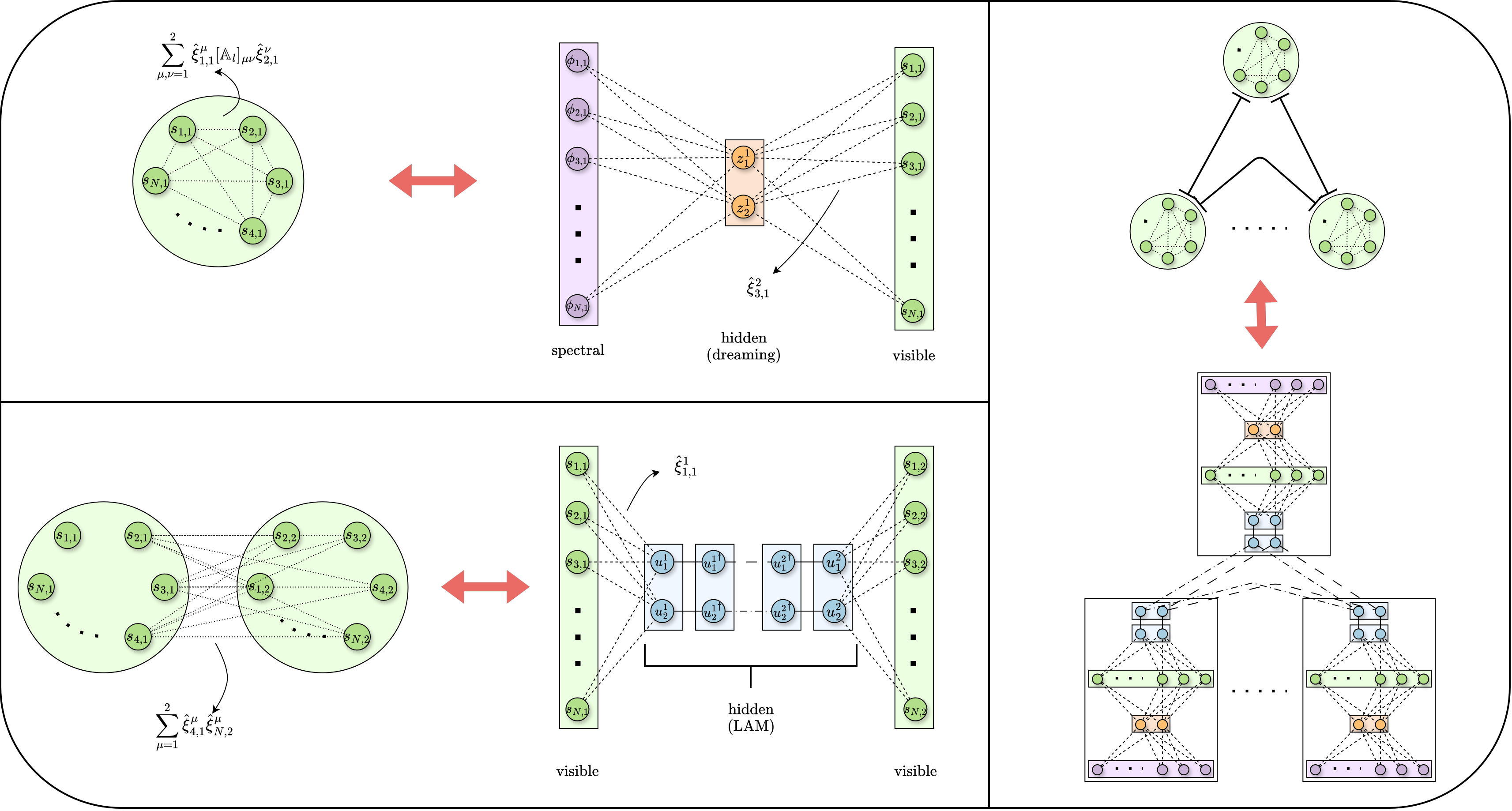}
    \caption{\textbf{Architectural equivalence of the Dreaming L-directional Associative Memory (DLAM).} \textbf{Top Left:} Integral representation of a simple single-layer dreaming model. Through Hubbard-Stratonovich linearisation \citep{barra2012equivalence}, the intra-layer dreaming interactions among visible neurons ($s_{i,l}$, green nodes) are decoupled into a Restricted Boltzmann Machine (RBM)-like structure mediated by site-indexed ($\phi_{i,l}$, purple nodes) and pattern-indexed ($z_\mu^l$, orange nodes) Gaussian fields. \textbf{Bottom Left:} Integral representation of a LAM model with $L=2$ layers, where inter-layer heteroassociative bindings are mediated by complex conjugate auxiliary fields ($u_{\mu}^{l}, u_{\mu}^{l\dagger}$, blue nodes). \textbf{Right:} A stylized illustration of the complete, fully coupled DLAM architecture integrating both mechanisms across $L$ interacting layers. This unified bipartite representation forms the analytical foundation for the Guerra interpolation and the replica-symmetric theory.}
    \label{fig:dlam_architecture}
\end{figure*}

The network learns from data, not from the archetypes directly: it never experiences the true vectors $\bm{\xi}_l^\mu$. For each archetype $\mu$ and layer $l$, it observes $M$ independent noisy copies generated by flipping each bit with probability $(1-r_l)/2$:
\begin{equation}
  \eta_{i,l}^{\mu,k} = \chi_{i,l}^{\mu,k}\,\xi_{i,l}^{\mu},
  \qquad
  \E[\chi_{i,l}^{\mu,k}] = r_l\in(0,1],
\end{equation}
where $r_l$ controls data fidelity.  The empirical archetype $\hat\xi_{i,l}^\mu := M^{-1}\sum_{k=1}^M \eta_{i,l}^{\mu,k}$ is the network's best estimate of the true pattern, and its fluctuations at finite sample size are governed by the dataset entropy \citep{AlemannoBarraETAL2023dreaming}
\begin{equation}
  \rho_l := \frac{1-r_l^2}{M\,r_l^2}.
  \label{eq:rho_def}
\end{equation}
The control and order parameters used throughout the analysis are collected in Table~\ref{tab:orderparams}.
When $\rho_l=0$ the empirical archetypes coincide with the true ones; as $\rho_l$ grows, the dataset becomes noisier, either because the observations are corrupted ($r_l$ small) or because fewer examples are available ($M$ small).  The parameter $\rho_l$ thus conflates two distinct sources of degradation into a single effective noise measure, and the data--computation trade-off explored later amounts to the observation that dreaming can compensate for large $\rho_l$.  We also define $\Gamma_l := r_l^2(1+\rho_l)$ and $\Gamma_{lm}:=\sqrt{\Gamma_l\Gamma_m}$, which serve as normalisation constants throughout.

Retrieval quality is measured by the Mattis magnetisation $m_l^\mu := N^{-1}\sum_i \xi_{i,l}^\mu s_{i,l}$, which quantifies the alignment of the network state with the true archetype.  We also define $\hat m_l^\mu := N^{-1}\sum_i \hat\xi_{i,l}^\mu s_{i,l}$ (alignment with the empirical archetype) and the rescaled variant $m_{\eta,l} := \hat m_l^1/[r_l(1+\rho_l)]$, which factors out the trivial dependence on dataset quality. \footnote{The same macroscopic order parameters are used throughout for both the standard retrieval problem and the disentanglement task.  What changes between the two settings is the cueing/initialisation protocol, not the order-parameter structure of the theory.}

\subsection{Dreaming kernel}\label{subsec:dreaming}

The dreaming mechanism operates within each layer through the empirical correlation matrix $(\Sigma_l)_{\mu\nu} := (N\Gamma_l)^{-1}\sum_i \hat\xi_{i,l}^\mu\hat\xi_{i,l}^\nu$.  As the sleep time $t_l$ grows, the effective coupling matrix
\begin{equation}
  \mathbb{A}_l(t_l) := (1+t_l)\bigl(\mathbb{I}+t_l\,\Sigma_l\bigr)^{-1}
  \label{eq:Al_def}
\end{equation}
progressively suppresses the eigenmodes of $\Sigma_l$ responsible for inter-pattern crosstalk \citep{AgliariBarra2019rigorous,FachechiETAL2019dreaming}.
  The spectral interpretation is immediate: in the eigenbasis of $\Sigma_l$, an eigenvalue $\lambda$ of $\Sigma_l$ is rescaled to $(1+t_l)/(1+t_l\lambda)$, which is a strictly decreasing function of $\lambda$.  Consequently, eigenmodes with larger $\lambda$ are suppressed more strongly than those with smaller $\lambda$.  As $t_l\to\infty$, eigenvalue $\lambda$ saturates at $1/\lambda$, so the weight assigned to each mode scales inversely with its original magnitude.  Crucially, the modes that dominate the inter-pattern crosstalk (the eigenvectors of $\Sigma_l$ carrying the largest eigenvalues) are differentially attenuated relative to the signal direction, which contributes across all eigenmodes but is concentrated in the condensed spin-space eigenvector.  The net effect is a flattening of the effective coupling spectrum that suppresses spurious interference without destroying the signal: the kernel acts as a \emph{spectral regulariser} that reduces the variance of the effective couplings rather than zeroing individual eigenmodes.

The intra-layer Hamiltonian constructed with this kernel reads
\begin{equation}
  \mathcal{H}_{\mathrm{Dream}}(\bm s) = -\sum_{l=1}^{L}\frac{1}{N\Gamma_l}
  \sum_{i,j}\sum_{\mu,\nu}
  \hat\xi_{i,l}^\mu\bigl[\mathbb{A}_l(t_l)\bigr]_{\mu\nu}\hat\xi_{j,l}^\nu\,s_{i,l}\,s_{j,l}.
  \label{eq:Hdream}
\end{equation}

{\color{black}
The regularising action of the dreaming kernel admits a direct geometric interpretation in the macroscopic space of empirical overlaps. Let $\hat{\bm{m}}_l(\bm{s}) \in \mathbb{R}^K$ denote the vector with components $\hat{m}_l^\mu = N^{-1}\sum_i \hat\xi_{i,l}^\mu s_{i,l}$. For $t_l=0$, the intra-layer energy reduces to the negative squared Euclidean norm, $\mathcal{H} \propto - \hat{\bm{m}}_l^\top \hat{\bm{m}}_l$. This isotropic metric preserves the correlated noise embedded in the empirical data. 

Conversely, the kernel $\mathbb{A}_l(t_l) = (1+t_l)(\mathbb{I} + t_l \Sigma_l)^{-1}$ introduces a non-trivial metric tensor. Since the empirical covariance $\Sigma_l$ is positive semi-definite, the operator $\mathbb{I} + t_l \Sigma_l$ is strictly positive definite for any $t_l > 0$, ensuring that its inverse is a well-defined precision matrix. Consequently, the intra-layer Hamiltonian $\mathcal{H}_{\mathrm{Dream}} \propto - \hat{\bm{m}}_l^\top \mathbb{A}_l(t_l) \hat{\bm{m}}_l$ computes a Tikhonov-regularised Mahalanobis squared norm of the state vector. In the asymptotic limit $t_l \to \infty$, the tensor $\mathbb{A}_l(t_l)$ behaves as $\Sigma_l^{-1}$ on the non-null eigenspace of the patterns. The dreaming mechanism thus performs a spatial whitening of the overlap space, replacing the Euclidean geometry with a Mahalanobis metric that natively decorrelates the macroscopic state variables.

}

\subsection{Inter-layer coupling}\label{subsec:LAM}

The dreaming kernel operates within each layer independently; the second ingredient of the DLAM is a cross-modal binding that couples layers through their shared archetype structure.  When archetype~$\mu$ is stored simultaneously in layers~$l$ and~$m$, the heteroassociative term rewards spin configurations in which both layers align with their respective empirical estimates of~$\mu$.  This is the mathematical counterpart of the biological observation that multimodal memories are stored in distributed cortical areas and retrieved through associative completion.  Formally, the inter-layer coupling reads
\begin{equation}
  \mathcal{H}_{\mathrm{LAM}}(\bm s) = -\sum_{l<m}\frac{g_{lm}}{N\Gamma_{lm}}
  \sum_{\mu=1}^K\sum_{i,j}\hat\xi_{i,l}^\mu\,\hat\xi_{j,m}^\mu\,s_{i,l}\,s_{j,m}.
  \label{eq:Hlam}
\end{equation}

\subsection{Full Hamiltonian}\label{subsec:hamiltonian}

The complete DLAM energy function is the sum
\begin{equation}
  \mathcal{H}_{\mathrm{DLAM}}(\bm s)
  := \,\mathcal{H}_{\mathrm{Dream}}(\bm s) + \,\mathcal{H}_{\mathrm{LAM}}(\bm s),
  \label{eq:Hdlam}
\end{equation}
The retrieval behaviour is governed by four classes of control parameters: the storage load $\alpha:=K/N$, the inverse temperature $\beta$, the dataset entropies $\{\rho_l\}$, and the sleep times $\{t_l\}$.\footnote{\textbf{Symmetric coupling parametrisation.}\label{rem:g_param} For $L$-layer networks with symmetric architecture it is natural to adopt the following parametrisation of the coupling tensors.  One assigns a common \emph{auto-coupling} strength $g_{\mathrm{auto}} := g_{ll}$ to all diagonal entries ($l{=}m$) and a common \emph{hetero-coupling} strength $g_{\mathrm{hetero}} := g_{lm}$ ($l{\neq}m$) to all off-diagonal entries of the coupling matrix $\{g_{lm}\}$.  Concretely, this amounts to replacing $1 \mapsto g_{\mathrm{auto}}$ in $\mathcal{H}_{\mathrm{Dream}}$ and $g_{lm} \mapsto g_{\mathrm{hetero}}$ uniformly in $\mathcal{H}_{\mathrm{LAM}}$.  All numerical experiments in Section~\ref{sec:numerical} use this parametrisation, and all caption entries labelled $g_{\mathrm{auto}}$ and $g_{\mathrm{hetero}}$ refer to these two scalars.  The \emph{auto/hetero mixing parameter} $a \in [0,1]$ is defined via $g_{\mathrm{auto}} = a\,g_0$ and $g_{\mathrm{hetero}} = (1{-}a)\,g_0/(L{-}1)$, where $g_0 = g_{\mathrm{auto}} + (L{-}1)\,g_{\mathrm{hetero}}$ is the total coupling budget per layer.  At $a{=}1$ the architecture is purely autoassociative; at $a{=}0$ it is purely heteroassociative.  This parametrisation is a modelling choice: the general theory in Sections~\ref{sec:analytical}--\ref{sec:numerical} holds for arbitrary $\{g_{lm}\}$.}

The Hamiltonian~\eqref{eq:Hdlam} contains several known models as limiting cases: at $L{=}1$ it reduces to a single-layer dreaming network; at $t_l{=}0$ for all $l$ it yields a multidirectional associative memory built from noisy empirical patterns; with the inter-layer couplings removed, it factorises into $L$ independent dreaming problems \citep{AgliariBarra2019rigorous,FachechiETAL2019dreaming}.
 The key architectural point is that, unlike a two-stage pipeline in which dreaming and heteroassociation are applied sequentially, both mechanisms live inside the same energy landscape and interact through the shared neural dynamics.

\subsection{Cost function vs. Loss function: a Machine Learning perspective}\label{subsec:cost_loss}

It is instructive to recast the DLAM Hamiltonian, traditionally viewed as an energy function in statistical mechanics, into a Loss function typical of the Machine Learning framework. This establishes a rigorous bridge between the physical dynamics of retrieval and the minimization of a Mean Squared Error (MSE) objective.

Let us rewrite the empirical Mattis magnetization as $\hat{m}_l^\mu(\bm{s}) := \frac{1}{N}\sum_i \hat\xi_{i,l}^\mu s_{i,l}$. The total energy $\mathcal{H}_{\mathrm{DLAM}}$ can be expressed entirely in terms of these macroscopic overlaps:
\begin{equation}
  \mathcal{H}_{\mathrm{DLAM}} = -N \sum_{l=1}^{L} \frac{1}{\Gamma_l} \sum_{\mu,\nu} \hat{m}_l^\mu [\mathbb{A}_l(t_l)]_{\mu\nu} \hat{m}_l^\nu - N \sum_{l<m} \frac{g_{lm}}{\Gamma_{lm}} \sum_{\mu=1}^{K} \hat{m}_l^\mu \hat{m}_m^\mu.
\end{equation}

To frame the dreaming sector as a reconstruction task, we introduce the rotated empirical archetypes $\tilde{\bm{\xi}}_l^\mu := \sum_\nu \bigl[\mathbb{A}_l(t_l)^{1/2}\bigr]_{\mu\nu} \hat{\bm{\xi}}_l^\nu$, with corresponding rotated overlaps $\tilde{m}_l^\mu(\bm{s}) = \frac{1}{N}\sum_i \tilde{\xi}_{i,l}^\mu s_{i,l}$. This rotation exactly absorbs the dreaming kernel, allowing us to define a natural $L_2$ Loss function measuring the distance between the network state and the target (or its reflection):
\begin{equation}
  \tilde{\mathcal{L}}_{l,\pm}^\mu(\bm{s}) := \frac{1}{2N} \left\| \tilde{\bm{\xi}}_l^\mu \pm \bm{s}_l \right\|^2 = \frac{1}{2N} \left\| \tilde{\bm{\xi}}_l^\mu \right\|^2 + \frac{1}{2} \pm \tilde{m}_l^\mu(\bm{s}).
\end{equation}

The transformation $\tilde{\bm{\xi}}_l^\mu = \mathbb{A}_l(t_l)^{1/2} \hat{\bm{\xi}}_l^\mu$ is rigorously well-defined: because the empirical correlation matrix $\Sigma_l$ is a Gram matrix, it is symmetric positive semi-definite. Consequently, the operator $\mathbb{I} + t_l \Sigma_l$ has eigenvalues bounded below by $1$, ensuring that the dreaming kernel $\mathbb{A}_l(t_l)$ is strictly positive definite for any $t_l \ge 0$. Thus, its principal square root exists, is unique, and maintains the geometry of the pattern space.

Simple algebra reveals that the intra-layer dreaming energy is exactly proportional to the product of these symmetric losses:
\begin{equation}
  -\sum_{\mu,\nu} \hat{m}_l^\mu [\mathbb{A}_l(t_l)]_{\mu\nu} \hat{m}_l^\nu = \sum_{\mu=1}^K \Bigl( \tilde{\mathcal{L}}_{l,+}^\mu(\bm{s})\,\tilde{\mathcal{L}}_{l,-}^\mu(\bm{s}) - C_l^\mu(t_l)^2 \Bigr),
\end{equation}
where $C_l^\mu(t_l) = \frac{1}{2N}\|\tilde{\bm{\xi}}_l^\mu\|^2 + \frac{1}{2}$ is a state-independent structural constant.

Simultaneously, the hetero-associative LAM sector acts as a multi-view \emph{alignment loss}. By rewriting the coupling term as 
\begin{equation}\label{eq:LAM-Loss-Decomp}
    \hat{m}_l^\mu \hat{m}_m^\mu = \frac{1}{2}(\hat{m}_l^\mu)^2 + \frac{1}{2}(\hat{m}_m^\mu)^2 - \frac{1}{2}(\hat{m}_l^\mu - \hat{m}_m^\mu)^2,
\end{equation}
the LAM energy explicitly minimizes the cross-modal discrepancy:
\begin{equation}
  \mathcal{L}_{\mathrm{cross}}^{\mu, l, m}(\bm{s}) := \frac{1}{2} \left( \hat{m}_l^\mu(\bm{s}) - \hat{m}_m^\mu(\bm{s}) \right)^2.
\end{equation}

The decomposition of the heteroassociative LAM energy establishes a formal link with modern self-supervised learning. The identity in Eq.~\eqref{eq:LAM-Loss-Decomp}
illustrates that inter-layer dynamics extend beyond simple cross-modal alignment (i.e., the minimization of the discrepancy $\frac{1}{2}(\hat{m}_l^\mu - \hat{m}_m^\mu)^2$). Specifically, the positive auto-overlap terms $\frac{1}{2}(\hat{m}_l^\mu)^2$ function as variance-maximizing regularizers. In the context of self-supervised frameworks 
these terms mitigate representation collapse, preventing the network from converging to uninformative trivial solutions in the pursuit of cross-layer synchronization \citep{bardes2022vicreg}.

Thus, the DLAM thermal relaxation inherently performs a dual optimization: the dreaming kernel minimizes the intra-layer MSE (effectively whitening the empirical noise), while the LAM couplings minimize the inter-layer contrastive loss, forcing distinct sensory channels to converge onto a unified, coherent multisensory representation.

\subsection{Disentanglement task}\label{subsec:disentanglement_task}

The preceding definitions equip the DLAM with the machinery for both intra-layer noise suppression and cross-modal binding.  The most stringent test of whether these mechanisms work together is not standard pattern retrieval, which probes only the depth of a single attractor, but \emph{disentanglement}: the simultaneous recovery of each layer's modality-specific representation from a single common cue that blends noisy evidence from all sensory channels at once.

Concretely, a target archetype $\mu^\star$ is drawn at random.  For each layer $l\in\{1,2,3\}$\footnote{Throughout the numerical experiments we instantiate the model with $L=3$ layers, but the disentanglement framework generalises naturally to an arbitrary number of layers $L$.}, a single noisy training example $\eta_{i,l}^{\mu^\star, k_l}$ is drawn independently  (example index $k_l\in\{1,\ldots,M\}$ drawn uniformly at random) from the layer-$l$ dataset for archetype $\mu^\star$ (using the notation of Section~\ref{subsec:architecture}).  A weighted superposition of these three modal examples is formed and used as the initial state of \emph{all} layers simultaneously:
\begin{equation}
  s_{i,l}^{\mathrm{init}} = \mathrm{sign}\bigl(\alpha_1\,\eta_{i,1}^{\mu^\star,k_1} + \alpha_2\,\eta_{i,2}^{\mu^\star,k_2} + \alpha_3\,\eta_{i,3}^{\mu^\star,k_3}\bigr),
  \qquad l = 1,2,3,
  \label{eq:simplex_cue}
\end{equation}
with $(\alpha_1,\alpha_2,\alpha_3)\geq 0$ and $\alpha_1+\alpha_2+\alpha_3=1$.  Ties (i.e.\ $\alpha_1\,\eta_{i,1}^{\mu^\star,k_1} + \alpha_2\,\eta_{i,2}^{\mu^\star,k_2} + \alpha_3\,\eta_{i,3}^{\mu^\star,k_3} = 0$) are broken by independently assigning $+1$ or $-1$ with equal probability $\tfrac12$; this event has positive probability on simplex edges and must be handled consistently to keep the initial state in the binary space $\{-1,+1\}^N$.  The construction implicitly assumes a \emph{shared neuron-index space}: the $i$-th component of each modal sample $\eta_{i,l}^{\mu^\star}$ refers to the same index position across layers, so that the coordinatewise sum is geometrically well-defined.  This common-basis convention is stated here explicitly: it is not a structural requirement of the DLAM (the layers may in general have independent index sets), but it is the natural choice for the present benchmark.  The weight vector $(\alpha_1,\alpha_2,\alpha_3)$ lives on the 2-simplex: its vertices, e.g.\ $(1,0,0)$, correspond to pure single-view cues (cue derived entirely from layer~1's training data), the edges to two-view mixtures, and the barycentre $(\tfrac{1}{3},\tfrac{1}{3},\tfrac{1}{3})$ to equal weighting of all three modal views.

A trial is successful only if the tail-averaged magnetisation $m_l = N^{-1}\langle\bm{s}_l\cdot\bm\xi_l^{\mu^\star}\rangle$ exceeds a threshold $m_{\mathrm{th}}$ on \emph{all} layers simultaneously.  This is a genuine disentanglement requirement: all three layers start from the exact same mixed vector, one that belongs to none of the individual layers' pattern spaces, and each must recover its own modality-specific target archetype.

The difficulty varies systematically with $(\alpha_1,\alpha_2,\alpha_3)$.  Near the barycentre every layer receives a diluted but nonzero signal from its own modal view, so that inter-layer cooperation can bootstrap all three layers simultaneously.  At a simplex vertex, by contrast, only one layer receives useful signal while the other two start from essentially zero overlap with their targets; joint retrieval must then rely entirely on the heteroassociative coupling to rescue the uninformed layers, a far harder task when the intra-layer basins are contaminated by noise.  The central prediction of the DLAM theory is that, by suppressing the inter-pattern correlations encoded in each $\Sigma_l$, the dreaming kernel $\mathbb{A}_l(t_l)$ cleans the basins sufficiently for the heteroassociative signal to propagate and rescue even uninformed layers, expanding the region of joint success from the barycentre toward the vertices.  This prediction is confirmed quantitatively in Section~\ref{sec:numerical}.

\begin{figure*}[t]
\centering

\begin{subfigure}[t]{0.31\textwidth}
  \vspace{0pt}
  \centering

  \includegraphics[trim={0 0 2.9cm 0}, clip, height=4.8cm, keepaspectratio]{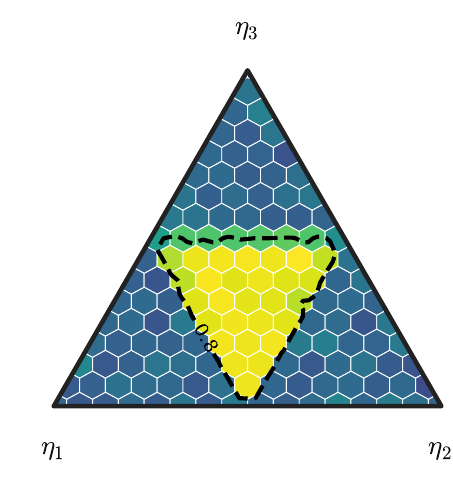}
  \label{fig:disentanglement:a}
\end{subfigure}\hfill
\begin{subfigure}[t]{0.31\textwidth}
  \vspace{0pt}
  \centering
  \includegraphics[height=4.8cm, keepaspectratio]{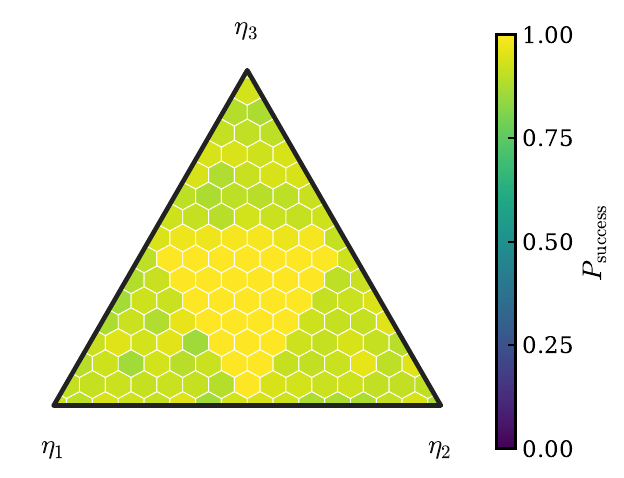}
  \label{fig:disentanglement:b}
\end{subfigure}\hfill
\begin{subfigure}[t]{0.31\textwidth}
  \vspace{0pt}
  \centering
  \includegraphics[width=\linewidth]{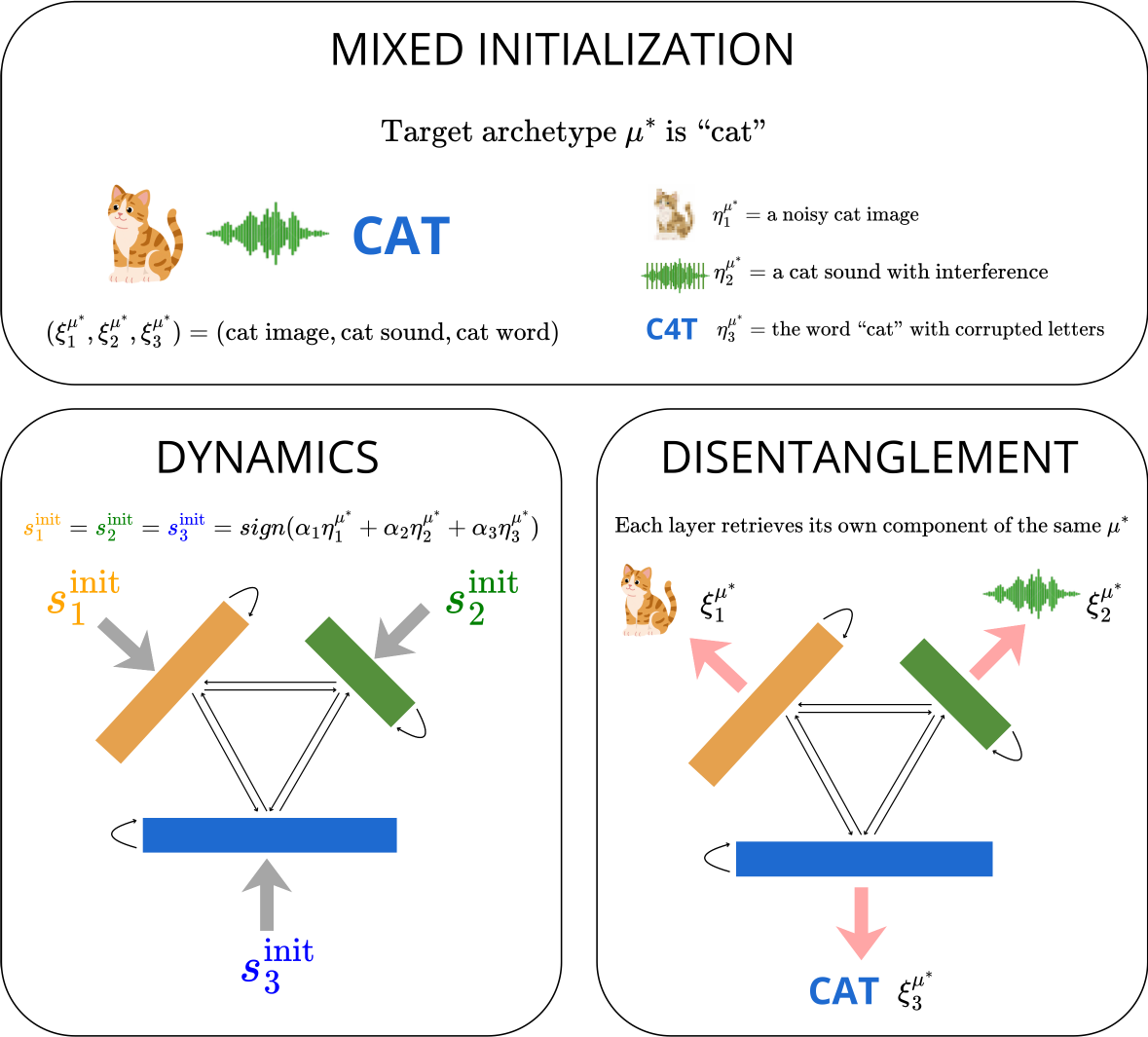}
  \label{fig:disentanglement:c}
\end{subfigure}
\caption{\textbf{Disentanglement under dreaming.}
\emph{Left and centre}: joint success rate on the cue-weight simplex~\eqref{eq:simplex_cue}.
\textbf{Left:}~LAM only ($t{=}0$): retrieval succeeds only near the barycentre, where every layer receives some signal from its own modal view and inter-layer cooperation can bootstrap joint recovery; near the vertices, where a single view dominates and two layers start uninformed, joint convergence fails.
\textbf{Center:}~Dreaming+LAM ($t{=}64$): the high-success region expands from the barycentre toward the vertices, covering nearly the entire simplex; even vertex cues (single-view evidence) now yield reliable joint retrieval.  Dashed black contour at $P{=}0.8$.  Colour encodes $P_{\mathrm{success}}$ on the \textit{viridis} scale (dark = failure, bright = success).
\textbf{Right:}~Conceptual illustration of the disentanglement task: a common cross-modal cue blending noisy samples from all three sensory layers is presented simultaneously to every layer as the initial state; each layer must then converge independently to its own modality-specific archetype.  Without dreaming, inter-pattern correlations trap each layer in a spurious mixed attractor; with dreaming, the spectral kernel $\mathbb{A}_l(t)$ suppresses off-diagonal correlations and carves out well-separated per-modality basins, enabling every layer to jointly recover its correct target from the shared cue.
Simplex panels~(a,b): $N{=}400$, $K{=}72$ ($\alpha{=}0.18$), $M{=}30$, $r{=}0.75$, $\beta{=}10$, $g_{\mathrm{auto}}{=}0.82$, $g_{\mathrm{hetero}}{=}0.41$, cue noise $\varepsilon{=}0.10$, $m_{\mathrm{th}}{=}0.80$; $200$ Glauber steps, $40$ independent replicates.}
\label{fig:disentanglement}
\end{figure*}

\section{Statistical--mechanical analysis}\label{sec:analytical}
This section develops the replica-symmetric theory for the DLAM.  Rather than reproducing every algebraic step, we present the derivation strategy, state the main results, and focus on the physical interpretation of the self-consistency equations and the effective-field decomposition.  Full details (Hubbard--Stratonovich decoupling, Guerra interpolation, one-body calculations, and explicit LAM scalar forms for $L{=}3$) are in the companion Appendix.

\subsection{Hubbard--Stratonovich decoupling and partition function}
\label{subsec:HS}

The goal is to compute the intensive quenched pressure
\begin{equation}
  \mathcal{A}(\beta) := \lim_{N\to\infty}\frac{1}{N}\,
  \E\bigl[\log\mathcal{Z}_{N,L}(\beta\mid\mathcal{D})\bigr],
  \label{eq:quenched_pressure}
\end{equation}
where $\mathcal{Z}_{N,L}$ is the partition function and $\E$ denotes expectation over the quenched dataset $\mathcal{D}$.  Both the dreaming and the LAM sectors of~\eqref{eq:Hdlam} are quadratic in the spins; both are linearised by independent Hubbard--Stratonovich (HS) transformations.

\smallskip\noindent\textbf{Rescaled patterns and overlaps.}
Set
\begin{align}
  J_{i,l}^\mu &:= \frac{\hat\xi_{i,l}^\mu}{\sqrt{\Gamma_l}},
  &
  p_l^\mu(\bm s) &:= \frac{1}{\sqrt{N}}\sum_{i=1}^{N}J_{i,l}^\mu\,s_{i,l},
  \label{eq:J_p_def}
\end{align}
so that $\hat\xi_{i,l}^\mu\hat\xi_{j,m}^\mu/\Gamma_{lm}=J_{i,l}^\mu J_{j,m}^\mu$ and the LAM energy takes the form $-\beta\sum_{l<m}g_{lm}\sum_\mu p_l^\mu p_m^\mu$.  Under the replica-symmetric condensation ansatz, mode $\mu=1$ is macroscopic (signal) and modes $\mu\ge 2$ are noise; the signal--noise split reads
\begin{equation}
  \sum_{\mu=1}^{K}p_l^\mu p_m^\mu
  = p_l^1 p_m^1 + \sum_{\mu\ge 2}p_l^\mu p_m^\mu.
  \label{eq:condensation}
\end{equation}

{\color{black}
\noindent The condensed empirical Mattis magnetisation is
\begin{equation}
  \widehat M_l(\bm s)
  := \widehat m_l^1(\bm s)
  := \frac{1}{N}\sum_{i=1}^{N}\hat\xi_{i,l}^1\,s_{i,l}.
  \label{eq:Mhat_def}
\end{equation}
}

\smallskip\noindent\textbf{Dreaming sector.}
The intra-layer dreaming term is linearised by $N$ site-indexed Gaussian fields $\phi_{i,l}\sim\mathcal{N}(0,1)$, collected in the product measure $D\phi_l:=\prod_{i}(2\pi)^{-1/2}e^{-\phi_{i,l}^2/2}d\phi_{i,l}$, and packaged into the complexified multi-spin
\begin{equation}
  k_{i,l} := s_{i,l} + i\sqrt{\frac{t_l}{\beta(1+t_l)}}\,\phi_{i,l}.
  \label{eq:k_def}
\end{equation}

{\color{black}
\noindent In this HS representation, the purely spin overlap
$\widehat M_l(\bm s)$ is lifted to the complexified condensed dreaming overlap
\begin{equation}
  \eta_l(\bm s,\bm\phi)
  :=
  \frac{1}{NM}\sum_{i=1}^{N}\sum_{k=1}^{M}
  \eta_{i,l}^{1,k}\,k_{i,l}
  =
  \frac{1}{N}\sum_{i=1}^{N}
  \hat\xi_{i,l}^{1}\,k_{i,l}.
  \label{eq:eta_complex_def}
\end{equation}
Here $\eta_{i,l}^{1,k}$ denotes the $k$-th noisy example of the condensed archetype in layer $l$.
}

The weight $\sqrt{t_l/[\beta(1+t_l)]}$ in the imaginary part is chosen so that Gaussian integration over $\bm\phi$ reproduces the dressed dreaming kernel $\mathbb{A}_l(t_l)=(1+t_l)(\mathbb{I}+t_l\Sigma_l)^{-1}$ rather than the bare identity (verified by completing the square; Appendix~A of the companion).  The non-condensed dreaming modes are further linearised by pattern-indexed fields $z_\mu^l$ drawn from the measure
\begin{equation}
  Dz_l :=
  \prod_{\mu\ge 2}\frac{dz_\mu^l}{\sqrt{2\pi}}\,
  \exp\Bigl(-\frac{(z_\mu^l)^2}{2(1+t_l)}\Bigr),
  \label{eq:Dz_def}
\end{equation}
so that $z_\mu^l\sim\mathcal{N}(0,1+t_l)$.

\smallskip\noindent\textbf{Inter-layer sector.}
For the non-condensed LAM modes, complex auxiliary fields $u_\mu^l,u_\mu^{l\dagger}\in\Sigma$ ($l=1,\dots,L$, $\mu\ge 2$) are introduced with a non-diagonal Gaussian measure $Duu^\dagger$ whose cross-covariance encodes the off-diagonal connectivity matrix $\mathbf{G}_0\in\mathbb{R}^{L\times L}$ defined by
\begin{equation}
  (\mathbf{G}_0)_{lm} :=
  \begin{cases}
    g_{lm}, & l\neq m,\\
    0,       & l = m.
  \end{cases}
  \label{eq:G0_def}
\end{equation}
Integrating out $u,u^\dagger$ against $Duu^\dagger$ regenerates $\exp\bigl(\frac\beta 2\sum_\mu\bm p^{\mu\top}\mathbf{G}_0\bm p^\mu\bigr)$, recovering the non-condensed inter-layer coupling exactly (Lemma~1 of the companion appendix).

\smallskip\noindent\textbf{Exact decoupled partition function.}
Assembling the condensation split~\eqref{eq:condensation} with the three HS linearisations yields the exact representation
\begin{equation}
\begin{aligned}
  \mathcal{Z}(\beta\mid\mathcal{D})
  &= \sum_{\bm s}\,e^{\mathcal{S}_{\mathrm{LAM}}(\bm s)}
    \prod_{l=1}^{L}\int D\phi_l\;
    e^{\beta\mathcal{S}_{\mathrm{Dream}}(\bm s,\bm\phi)}\\
  &\quad\times
    \mathcal{N}_{\mathrm{Dream}}(\bm s,\bm\phi)\,
    \mathcal{N}_{\mathrm{LAM}}(\bm s),
\end{aligned}
  \label{eq:Z_decoupled}
\end{equation}
with signal factors
\begin{align}
  \mathcal{S}_{\mathrm{LAM}}(\bm s)
  &:= \beta\sum_{l<m}g_{lm}\,p_l^1(\bm s)\,p_m^1(\bm s),
  \label{eq:S_LAM_main}\\[2pt]
  \beta\mathcal{S}_{\mathrm{Dream}}(\bm s,\bm\phi)
  &:= \sum_{l=1}^{L}
     \frac{\beta N(1+t_l)}{2\Gamma_l}\,\eta_l(\bm s)^2,
  \label{eq:S_Dream_main}
\end{align}
and noise factors
\begin{align}
  \mathcal{N}_{\mathrm{Dream}}(\bm s,\bm\phi)
  &:= \prod_{l=1}^{L}\int Dz_l \;\times
     \nonumber\\
  &\quad\times\exp\Bigl(
     \sqrt{\tfrac{\beta}{\Gamma_l N}}
     \sum_{\mu\ge 2}\sum_{i=1}^{N}
     \hat\xi_{i,l}^\mu z_\mu^l k_{i,l}\Bigr),
  \label{eq:N_Dream_main}\\[4pt]
  \mathcal{N}_{\mathrm{LAM}}(\bm s)
  &:= \int Duu^\dagger \; \times
     \nonumber\\
  &\quad\times\exp\Bigl(
     \sqrt{\tfrac{\beta}{N}}
     \sum_{\mu\ge 2}\sum_{l=1}^{L}\sum_{i=1}^{N}
     J_{i,l}^\mu s_{i,l} u_\mu^l\Bigr).
  \label{eq:N_LAM_main}
\end{align}
The factorisation~\eqref{eq:Z_decoupled} is exact; no approximation has been made.  Its three hidden-variable sectors correspond to the three families of auxiliary nodes in Fig.~\ref{fig:dlam_architecture}: the $\phi_{i,l}$ (purple) mediate intra-layer dreaming corrections; the $z_\mu^l$ (orange) absorb non-condensed dreaming fluctuations; and the complex pair $u_\mu^l,u_\mu^{l\dagger}$ (blue) implement the inter-layer LAM connectivity.

The quenched average is handled by a Guerra-type interpolation: an auxiliary parameter $x\in[0,1]$ continuously deforms the interacting system ($x{=}1$) into a factorised one-body problem ($x{=}0$), and the fundamental theorem of calculus yields $\mathcal{A}(1) = \mathcal{A}(0) + \int_0^1 dx\,(d\mathcal{A}/dx)$.  The streaming term $d\mathcal{A}/dx$ is evaluated via Gaussian integration by parts (the Stein--Nishimori identity) and, under the RS self-averaging hypothesis, reduces to a deterministic expression involving the order parameters listed in Table~\ref{tab:orderparams}.

\begin{table}[t]
\centering
\caption{Control parameters and order parameters of the DLAM.  Layer indices $l,m\in\{1,\dots,L\}$ with $l<m$ for coupling parameters.}
\label{tab:orderparams}
{\footnotesize
\begin{tabular}{@{}ll@{}}
\toprule
Symbol & Description \\
\midrule
\multicolumn{2}{@{}l}{\textit{Control parameters}} \\
\midrule
$\alpha := K/N$ & storage load \\
$\beta$ & inverse temperature \\
$\rho_l$ & dataset entropy, eq.~\eqref{eq:rho_def} \\
$t_l$ & sleep time (dreaming budget, per layer) \\
$g_{lm}$ & inter-layer coupling strength \\
\midrule
\multicolumn{2}{@{}l}{\textit{Order parameters (RS ansatz)}} \\
\midrule
$\hat M_l$ & empirical Mattis magnetisation \\
$\bar\eta_l$ & auxiliary signal variable (dreaming sector) \\
$\bar q_l$ & Edwards--Anderson overlap (off-diagonal, dreaming) \\
$\bar Q_l$ & diagonal $k$-variable overlap (dreaming) \\
$\bar p_l,\bar P_l$ & conjugate dreaming-noise overlaps \\
$H_l$ & spin-glass overlap (LAM sector) \\
$\bar\Pi_l$ & LAM noise order parameter \\
\bottomrule
\end{tabular}}
\end{table}

\subsection{Replica-symmetric (RS) free energy}

Assembling the one-body contribution at $x{=}0$ and the streaming contributions from the three sectors yields the RS pressure:
\begin{align}
  \mathcal{A}^{\mathrm{RS}}_{\mathrm{DLAM}} &= 
  -\sum_{l<m}\frac{\beta\,g_{lm}}{\Gamma_{lm}}\hat M_l\hat M_m 
  -\frac{\alpha}{2}\log\det\mathbf{C}(H) + \nonumber\\
  &\quad + \frac{\alpha}{2}\mathbb{D}_{\mathrm{LAM}}(H) - \frac{\alpha\beta}{4}\sum_{l=1}^{L}\bar\Pi_l(1{-}H_l) + \nonumber\\
  &\quad + \sum_{l=1}^{L} \Bigg[ \log 2 + \E\bigl[\log\cosh\Psi_l\bigr]
     - \tfrac{1}{2}\log\bar D_l
     + \tfrac{\alpha}{2}\log\tfrac{1{+}t_l}{S_l} + \nonumber\\
  &\quad + \frac{\alpha\beta\bar q_l(1{+}t_l)}{2S_l}
     - \frac{\alpha\bar p_l\,t_l}{2\bar D_l(1{+}t_l)}
     + \frac{\beta(\bar D_l{-}1)(1{+}t_l)}{2\bar D_l\,t_l}  +\nonumber\\
  &\quad - \frac{\beta(1{+}t_l)\bar\eta_l^2(t_l{+}\bar D_l)}{2\Gamma_l\,\bar D_l}
     + \frac{\alpha\beta}{2}(\bar p_l\bar q_l - \bar P_l\bar Q_l) \Bigg].
  \label{eq:A_total}
\end{align}
Here $\mathbf{C}(H)$ is the $2L{\times}2L$ LAM covariance matrix, $\mathbb{D}_{\mathrm{LAM}}(H)$ its quadratic correction, and $\bar D_l := 1 + \alpha\frac{t_l}{1+t_l}(\bar P_l - \bar p_l)$, $S_l := 1 - \beta(1{+}t_l)(\bar Q_l - \bar q_l)$.
It is worth noting that for an arbitrary number of layers $L$ the terms governing the inter-layer noise sector do not admit a general closed-form algebraic representation. 
We provide in the companion Appendix the formulas for the specialization $(L=3)$.

\subsection{Self-consistency equations and the effective field}

Extremising~\eqref{eq:A_total} with respect to each order parameter yields the closed RS self-consistency system.  Its structure is most perspicuous when written in terms of the effective local field
\begin{equation}
  \Psi_l = \underbrace{I_l\,\iota\vphantom{\hat X}}_{\text{LAM noise}}
  + \underbrace{\sigma_l\,\varphi\vphantom{\hat X}}_{\text{dreaming noise}}
  + \underbrace{\hat X_l\,(u_l + v_l)}_{\text{signal}},
  \label{eq:Psi}
\end{equation}
where $\iota,\varphi\sim\mathcal{N}(0,1)$ are independent standard Gaussians, $\hat X_l\stackrel{d}{=}\hat\xi_{i,l}^1$ with $\E[\hat X_l^2]=\Gamma_l$, and
\begin{equation}
\begin{aligned}
  I_l = \sqrt{\tfrac{\alpha\beta}{2}\bar\Pi_l},&\quad
  \sigma_l = \frac{\sqrt{\alpha\beta\,\bar p_l}}{\bar D_l},\quad\\
  u_l = \sum_{m\neq l}\frac{\beta g_{lm}}{\Gamma_{lm}}\hat M_m,&\quad
  v_l = \frac{\beta(1{+}t_l)}{\Gamma_l}\frac{\bar\eta_l}{\bar D_l}.
  \label{eq:field_coeffs}
\end{aligned}
\end{equation}

This decomposition is the central object of the theory.  The signal reaching neuron $i$ on layer $l$ has two sources: $v_l$, the autoassociative signal enhanced by dreaming, and $u_l$, the heteroassociative signal relayed from neighbouring layers.  These compete against two independent noise channels: $\sigma_l$, the residual interference from the $K{-}1$ non-recalled patterns within the same layer (dreaming noise), and $I_l$, the noise generated by cross-talk in the inter-layer sector (LAM noise).  Pattern recall succeeds when the signal dominates both noise terms.

The self-consistency equations, obtained by setting the derivatives of $\mathcal{A}^{\mathrm{RS}}$ to zero, are:

\paragraph{Signal sector.}
\begin{equation}
  \hat M_l = \E\bigl[\hat X_l\tanh\Psi_l\bigr],
  \qquad
  \bar\eta_l = \frac{\hat M_l}{t_l+\bar D_l}.
  \label{eq:sc_signal}
\end{equation}
Here $\hat M_l$ is the \emph{empirical} Mattis magnetisation $N^{-1}\sum_i\hat\xi_{i,l}^1 s_{i,l}$, i.e.\ the overlap with the stored empirical archetype rather than the hidden true archetype $\bm\xi_l^1$.  The two are related by $m_l^1 = \hat M_l\,r_l/\Gamma_l + \mathcal{O}(N^{-1/2}) = \hat M_l/[r_l(1+\rho_l)] + \mathcal{O}(N^{-1/2})$ in probability (see Section~\ref{subsec:architecture} and the rescaled overlap $m_{\eta,l}$, which satisfies $m_{\eta,l}\approx m_l^1$ in the thermodynamic limit); the true-archetype overlap $m_l^\mu$, which is the quantity plotted in simulation figures, is recovered from $\hat M_l$ via this rescaling.

\paragraph{Dreaming overlap sector.}
\begin{align}
  \bar q_l &= \E\bigl[\tanh^2\Psi_l\bigr],
  \label{eq:sc_q}\\
  \bar Q_l &= \bar q_l + \frac{1-\bar q_l}{\bar D_l^2} - \frac{t_l}{\beta(1{+}t_l)\bar D_l},
  \label{eq:sc_Q}\\
  \bar p_l &= \beta\frac{(1{+}t_l)^2}{S_l^2}\bar q_l,\qquad
  \bar P_l = \frac{1{+}t_l}{S_l} + \beta\frac{(1{+}t_l)^2}{S_l^2}\bar q_l.
  \label{eq:sc_pP}
\end{align}

\paragraph{LAM sector.}
\begin{equation}
  H_l = \E\bigl[\tanh^2\Psi_l\bigr],\qquad
  \bar\Pi_l = \frac{2}{\beta}\Bigl[\partial_{H_l}\log\det\mathbf{C}(H) - \partial_{H_l}\mathbb{D}_{\mathrm{LAM}}(H)\Bigr].
  \label{eq:sc_LAM}
\end{equation}

\begin{remark}
The overlap $\bar q_l$ (dreaming sector) and the spin-glass Edwards--Anderson parameter $H_l$ (LAM sector) satisfy the same functional equation $\E[\tanh^2\Psi_l]$, but are closed by different companion equations; they are not redundant.  For $L{=}3$ the LAM derivatives in~\eqref{eq:sc_LAM} admit fully explicit scalar form; the expressions are given in the Appendix.
\end{remark}

\subsection{Phase boundaries and the data-saving mechanism}\label{subsec:phase_boundaries}

The system~\eqref{eq:sc_signal}--\eqref{eq:sc_LAM} admits three classes of fixed points:
\begin{itemize}[leftmargin=1.5em,nosep]
\item \emph{Retrieval}: $\hat M_l > 0$, $\bar q_l > 0$, the network has recalled a stored pattern (i.e., the signal prevails).
\item \emph{Spin glass}: $\hat M_l = 0$, $\bar q_l > 0$, pattern information is lost and the system retains frozen disorder.
\item \emph{Paramagnetic}: $\hat M_l = 0$, $\bar q_l = 0$, where the noise prevails.
\end{itemize}
The phase boundaries, namely the critical load $\alpha_c$, the critical temperature $T_c$, and the critical dataset entropy $\rho_c$, are the loci where the retrieval fixed point ceases to exist or becomes thermodynamically subdominant, yielding a phase diagram whose boundaries delimit the retrieval, paramagnetic, and spin-glass states.

\begin{figure*}[htbp]
\centering
\begin{subfigure}[t]{0.48\textwidth}
  \centering
  \includegraphics[width=\linewidth]{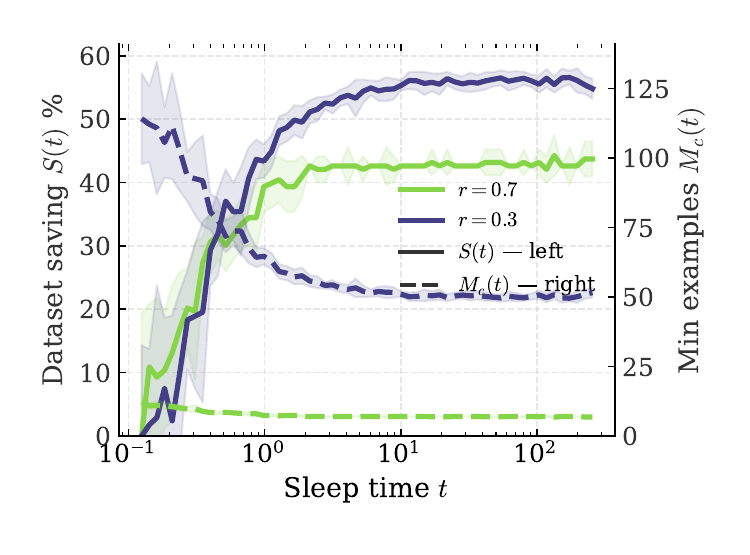}
  \caption{Unstructured random data}
  \label{fig:datasaving:a}
\end{subfigure}\hfill
\begin{subfigure}[t]{0.48\textwidth}
  \centering
  \includegraphics[width=\linewidth]{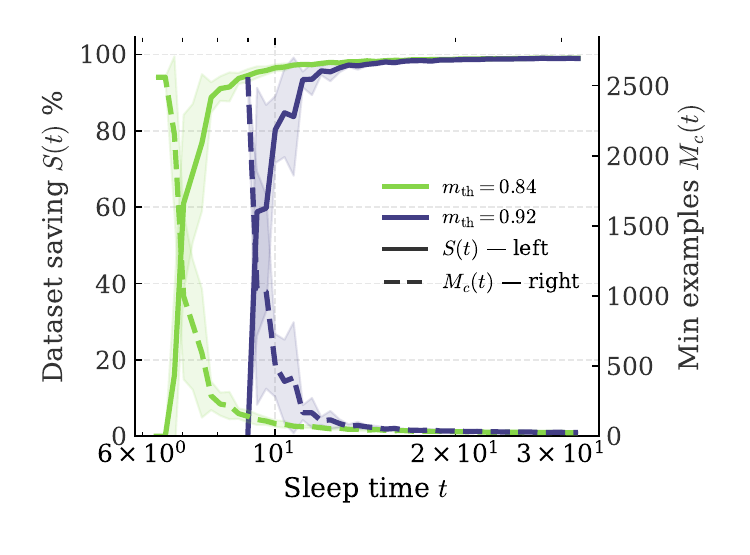}
  \caption{MNIST structured data}
  \label{fig:datasaving:b}
\end{subfigure}
\caption{\textbf{Data saving under dreaming.}
In both panels the bright (yellow-green) curve corresponds to the high-quality data condition and the dark (blue-purple) curve to the low-quality condition.
\textbf{Left:}~Critical number of training examples $M_c(t)$ (dashed, right axis) and dataset saving $S(t)=1-M_c(t)/M_c(0)$ (solid, left axis) versus sleep time, for synthetic data with $r{=}0.70$ (bright) and $r{=}0.30$ (dark).  Moderate dreaming consistently reduces $M_c$, demonstrating the data--computation trade-off in its most direct form: off-line consolidation substitutes for a significant fraction of the required training examples, with savings reaching $40$--$60\%$ at moderate sleep times.  The effect is even more pronounced at lower $r$, highlighting the increased effectiveness of consolidation in the presence of higher initial data corruption.
\textbf{Right:}~Same dual-axis display on structured synthetic data (MNIST-like patterns, $N{=}784$, three horizontal bands used as independent views) at two retrieval thresholds ($m_{\mathrm{th}}{=}0.84$, bright; $m_{\mathrm{th}}{=}0.92$, dark): a stricter success criterion requires more examples at any fixed $t$, but dreaming erases most of that gap, confirming that the trade-off is robust to the choice of evaluation metric and to structured (non-random) pattern statistics.  \emph{Dataset pipeline for panel~(b):} MNIST digits ($28{\times}28$ pixels) are binarised by thresholding at the per-image median intensity (pixels above median set to $+1$, below to $-1$); 
$K{=}20$ randomly chosen digit classes serve as archetypes (one archetype per class, selected as the image closest to the class centroid); train/test sets are split 8:2 within each class, and $M$ denotes the number of binarised training images used per class per view.
Panel~(a): $N{=}300$, $K{=}30$, $r{\in}\{0.70,\,0.30\}$, $g_{\mathrm{auto}}{=}1.0$, $g_{\mathrm{hetero}}{=}0.3$, $\beta{=}10$; $M_c(t)$ determined via binary search (see text) with $n{=}15$ replicates and success criterion $P(\langle m\rangle \ge 0.9) \ge 0.8$.}
\label{fig:datasaving}
\end{figure*}

The role of dreaming is visible directly in the field coefficients~\eqref{eq:field_coeffs}.  As the sleep time $t_l$ increases:
\begin{itemize}[leftmargin=1.5em,nosep]
\item the dreaming-noise coefficient $\sigma_l$ decreases, because the kernel $\mathbb{A}_l$ differentially attenuates the high-eigenvalue modes of $\Sigma_l$ (which carry most of the inter-pattern crosstalk), reducing the effective noise overlap $\bar p_l$;
\item the signal coefficient $v_l$ increases, because the condensed retrieval direction is preserved (its effective coupling saturates at a finite value) while $\bar D_l$ stabilises.
\end{itemize}
This heuristic argument, based on simultaneous noise suppression and signal amplification, predicts that the critical boundary should shift outward with increasing $t$.  In the single-layer limit this mechanism has been established rigorously \citep{FachechiETAL2019dreaming,AlemannoBarraETAL2023dreaming}; in the multi-layer model, the self-consistency equations~\eqref{eq:sc_signal}--\eqref{eq:sc_LAM}, which are exact under the RS ansatz, provide a closed characterisation of the equilibrium phase structure.  Numerical solutions via fixed-point iterations confirm the predicted outward shift of the retrieval boundary across all parameter planes. Phase diagrams in the $(\rho,t)$ and $(\alpha,t)$ planes, both from the self-consistency equations and from direct Monte Carlo simulations on finite-size networks, are presented in Section~\ref{sec:numerical} and confirm the phenomenology quantitatively.  In the $(\rho,t)$ plane, the shift has a direct operational interpretation: at fixed load, the maximum tolerable dataset entropy $\rho_c(t)$ increases with $t$, so that fewer training examples are needed: the data--computation trade-off recently explored in the single-layer setting by \citet{AlemannoETAL2023small} and \citet{FachechiETAL2025regularization}.

The phase diagrams in the $(\alpha,T)$-plane, obtained by iterating the self-consistency equations for varying inter-layer coupling $g$ (presented in Section~\ref{subsec:critical}), reveal a re-entrant retrieval boundary at intermediate temperatures, driven by the determinant singularity in the LAM noise equations, and a monotonic expansion of the retrieval region with $g$.

\subsection{Consistency reductions}

The RS closure recovers the expected limiting theories in the appropriate regimes.  Removing the heteroassociative sector yields $L$ independent single-layer supervised-dreaming problems; taking the limit $t_l\to 0$ for all $l$ recovers a noisy-data multidirectional associative memory (note that a literal substitution $t_l=0$ is not defined from the displayed formula because of the explicit $1/t_l$ term; the limit is understood in the sense $t_l\to 0^+$, which regularises the expression); removing the dreaming sector recovers the supervised LAM closure.  The explicit reduction dictionaries are reported in the Appendix.

\section{Numerical results}\label{sec:numerical}

The analytical theory in Section~\ref{sec:analytical} provides closed-form self-consistency equations describing the evolution of order parameters within the DLAM control parameter space in the thermodynamic limit. This framework suffices to derive the phase diagram of the network. Nevertheless, the Monte Carlo experiments that follow serve three purposes: first, to confirm the predicted phenomenology at finite $N$; second, to complement the analytical phase diagrams with finite-size scaling and quantitative tests where the RS solution serves as a benchmark; and finally, to probe dynamical phenomena—such as convergence acceleration and the temporal structure of disentanglement—that lie beyond the scope of the present theory.

All simulations use $L{=}3$ symmetric layers with $N$ neurons each.  Two initialisation protocols are employed, matched to the tasks of:
\begin{itemize}[leftmargin=1.5em,nosep]
\item \emph{Pattern retrieval.}  Layer~1 (the cued layer) is initialised at overlap $m_0 = 1 - 2\varepsilon$ with the target archetype; layers~2 and~3 start from random configurations ($m_0\simeq 0$).  This protocol isolates the dynamics of cross-modal completion and is used in the temperature-dependence experiments.
\item \emph{Pattern Disentanglement.}  All layers are initialised with the same weighted cross-modal cue~\eqref{eq:simplex_cue}, blending single noisy examples drawn from each layer's dataset for a shared target archetype.  The network must recover the correct modality-specific pattern on every layer simultaneously.  This protocol is used in the majority of the experiments.
\end{itemize}
Retrieval is declared successful when the tail-averaged Mattis magnetisation exceeds a threshold $m_{\mathrm{th}}$ (typically $0.8$).

\paragraph{Dataset-saving metric}
In the experiments of Section~\ref{subsec:disentanglement} we measure the critical training-set size $M_c(t)$, the minimum number of examples per archetype such that retrieval is reliable at sleep time $t$.  Operationally, $M_c(t)$ is found by binary search: starting from a bracket $[M_{\min},M_{\max}]$ that is expanded geometrically until the upper bound yields reliable retrieval, the search bisects the interval until the gap is smaller than one example.  At each candidate $M$, $n{=}15$ independent trials are run; retrieval is deemed reliable if at least $80\%$ of trials achieve tail-averaged magnetisation $\langle m\rangle \ge 0.9$.  The dataset saving is then $S(t) = 1 - M_c(t)/M_c(0)$.

\subsection{Disentanglement under dreaming}\label{subsec:disentanglement}

We begin with the experiment that most directly tests the central prediction of the DLAM theory, as storage grows: can the dreaming kernel destabilise the glassy attractor that traps ambiguous mixture cues and promote the emergence of sharp, individual retrieval basins?  The disentanglement task is the right stress test because it probes not just the depth of a single basin but the global organisation of the free-energy landscape under competing attractors.

To probe this prediction we deploy the simplex initialisation~\eqref{eq:simplex_cue} defined in Section~\ref{subsec:disentanglement_task}, sweeping all cue compositions $(\alpha_1,\alpha_2,\alpha_3)$ on a $30{\times}30$ simplex grid and recording the joint success rate averaged over 40 independent replicates per point.

\begin{figure*}[htbp]
\centering
\begin{subfigure}[t]{0.32\textwidth}
  \centering
  \includegraphics[width=\linewidth]{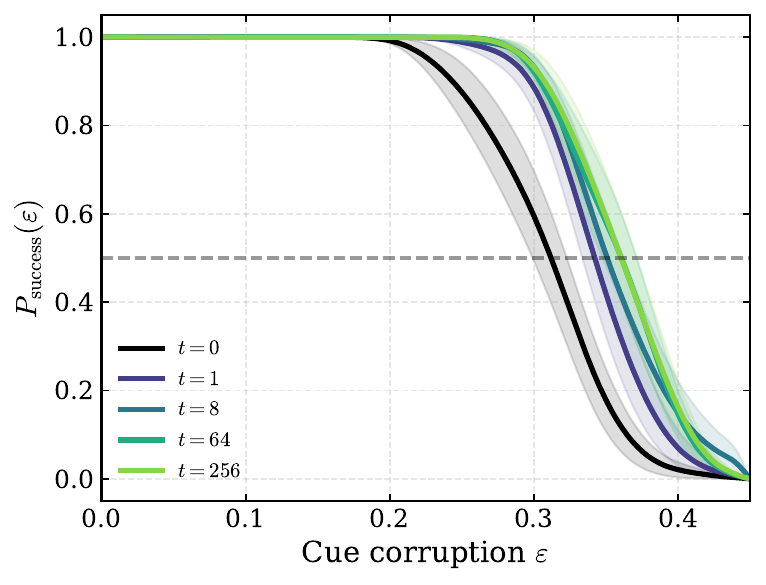}
  \label{fig:robustness:a}
\end{subfigure}\hfill
\begin{subfigure}[t]{0.32\textwidth}
  \centering
  \includegraphics[width=\linewidth]{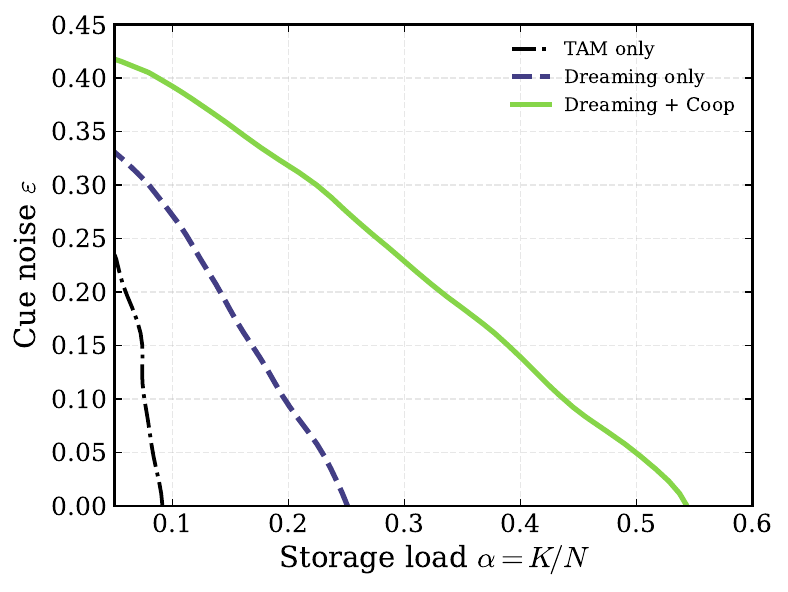}
  \label{fig:robustness:b}
\end{subfigure}\hfill
\begin{subfigure}[t]{0.32\textwidth}
  \centering
  \includegraphics[width=\linewidth]{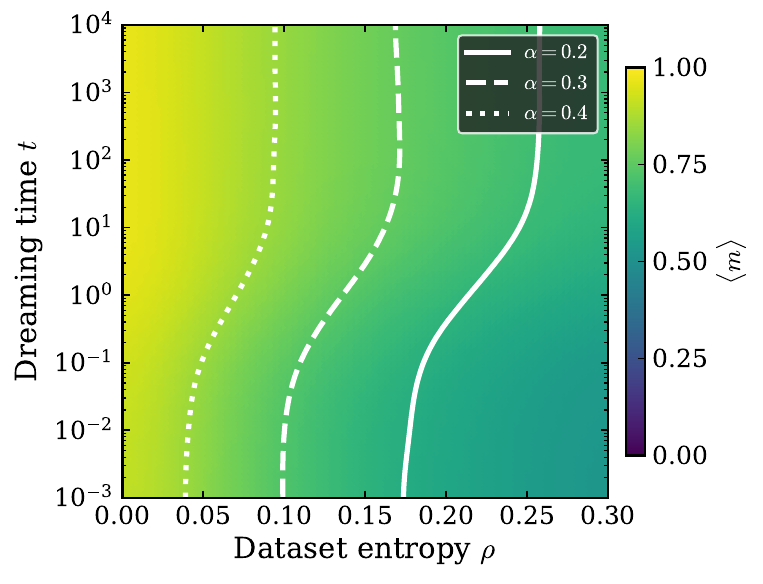}
  \label{fig:robustness:c}
\end{subfigure}
\caption{\textbf{Robustness and phase structure.}
\textbf{Left:}~Noise robustness: retrieval probability $P_{\mathrm{success}}(\varepsilon)$ versus cue corruption level for increasing sleep times $t$.  Curves are coloured with a viridis gradient (dark$\,=\,$short dreaming, bright$\,=\,$long dreaming); black marks the undreamed baseline $t{=}0$.  The critical corruption $\varepsilon_c$ at which retrieval drops to $50\%$ shifts markedly rightward with $t$, quantifying basin expansion.  Shaded bands: $\pm 1\sigma$ over independent realisations.
\textbf{Center:}~Two-dimensional phase boundary in the $(\alpha,\varepsilon)$-plane, comparing three regimes: LAM only (black dash-dotted), dreaming alone (dark viridis, dashed), and dreaming combined with inter-layer cooperation (bright viridis, solid).  The combined boundary encloses a substantially larger retrieval region, demonstrating genuine synergy.  The viridis heatmap shows $P_{\mathrm{success}}$ for the cooperative condition.
\textbf{Right:}~Phase diagram in the $(\rho,t)$-plane for three values of the storage load~$\alpha$.  The retrieval boundary (iso-contour at $\langle m\rangle{=}0.85$) separates the \emph{retrieval phase} (upper-left region: low dataset entropy~$\rho$ and large dreaming time~$t$, shown bright in the viridis heatmap) from the \emph{non-retrieval phase} (lower-right region, dark in the heatmap: high entropy or insufficient sleep).  Moving along the vertical axis at fixed~$\alpha$, increasing $t$ pushes the boundary rightward, meaning that a larger $\rho$ (poorer or scarcer data) is tolerated at fixed retrieval quality, equivalently, fewer training examples suffice for the same performance.  At higher load the boundary moves inward (lower $\rho_c$ at fixed $t$), but the relative gain from dreaming is preserved.  Background: Gaussian-smoothed $\langle m\rangle$ for $\alpha{=}0.4$.
Panel~(a): $N{=}500$, $M{=}30$, $r_1{=}0.50$, $r_{2,3}{=}0.80$, $\alpha{=}0.15$, $t{\in}\{0,1,8,64,256\}$, $\beta{=}10$, $50$ Glauber steps, $20$ replicates.
Panel~(c): $N{=}500$, $M{=}20$, $\alpha{=}0.40$ (background), $\beta{=}100$, $20$ steps, $10$ runs per cell on an $80{\times}80$ $(\rho,\log t)$ grid, retrieval threshold $\langle m\rangle{=}0.85$.}
\label{fig:robustness}
\end{figure*}

Fig.~\ref{fig:disentanglement}(a,b) display the central visual result of the paper.  Without dreaming, the joint success rate on the simplex is appreciable only near the barycentre, where every layer receives a diluted but nonzero signal from its own modal view and inter-layer cooperation can bootstrap all three layers.  Moving toward the vertices, where the cue is dominated by a single view and two layers start essentially uninformed, the uninformed layers cannot recover and joint retrieval fails.  With dreaming ($t{=}64$), the picture changes qualitatively: the high-success region expands from the barycentre toward the vertices, covering nearly the entire simplex.  The spectral filtering by $\mathbb{A}_l$ has cleaned the intra-layer basins so that the heteroassociative signal can rescue even uninformed layers; the transition is sharp in $t$, and above a threshold disentanglement becomes the generic outcome across the full simplex.

Alongside the benefits observed in the disentanglement task, the regularising action of dreaming provides a complementary and equally crucial advantage concerning data efficiency, an aspect that we quantify in terms of data saving (Fig.~\ref{fig:datasaving}). While the Hebbian dreaming mechanism was recently shown to reduce the volume of training examples needed for retrieval in single-layer networks \citep{AlemannoETAL2023small,FachechiETAL2025regularization}, our current simulations structurally extend and reaffirm this evidence within a considerably more complex multimodal architecture. As sleep time progresses, the critical number of training examples required to learn a pattern, $M_c(t)$, reliably decreases; the resulting relative dataset saving $S(t) = 1 - M_c(t)/M_c(0)$ scales rapidly up to a significant fraction of the required training data. This outlines a concrete, operational \emph{data--computation trade-off}: the network compensates for a sparse or highly noisy training set by expending computational effort during the off-line consolidation phase. The importance of this trade-off holds acute practical value since, in many real-world multimodal scenarios, collecting large batches of clean, aligned data across multiple sensors is far more cost-prohibitive and bottleneck-prone than allocating extra cycles for internal parameter tuning.

\subsection{Robustness and phase structure}\label{subsec:phase_diagrams}

The disentanglement results demonstrate a qualitative transformation of the free-energy landscape at a single operating point.  A natural follow-up question is whether this effect is robust across the parameter space explored, or whether it is a fragile anecdote tied to a particular choice of load and noise.  The answer, explored across three complementary projections of the phase diagram, is that the dreaming benefit is broad and systematic.

Fig.~\ref{fig:robustness}(a) quantifies the basin expansion.  The critical cue corruption $\varepsilon_c$, the noise fraction at which retrieval drops below $50\%$, shifts markedly rightward as $t$ increases: dreaming enlarges the retrieval basins in configuration space (equivalently, it broadens the set of corrupted or partial input cues from which the network still converges to the correct stored pattern), making the network robust to substantially noisier inputs.

The cooperative contribution is visible in Fig.~\ref{fig:robustness}(b).  In the $(\alpha,\varepsilon)$-plane three retrieval boundaries are compared: the baseline without dreaming or inter-layer coupling, dreaming alone, and the full dreaming+cooperation regime.  Inter-layer coupling shifts the phase boundary outward by a further margin, so that the combined regime encloses a retrieval region substantially larger than either mechanism alone.  Dreaming and inter-layer coupling are thus synergistic: they open operating regions not reached by either mechanism in isolation within the explored regime.

Fig.~\ref{fig:robustness}(c) maps the retrieval boundary in the $(\rho,t)$-plane for several values of the storage load~$\alpha$, the operational content of the data--computation trade-off.  As argued in Section~\ref{subsec:phase_boundaries}, the noise-suppression mechanism of the dreaming kernel increases $\rho_c$ with $t$: the boundary shifts rightward (to larger tolerable $\rho_c$), so that at fixed load the same retrieval quality is achieved with fewer training examples.  The picture is consistent with the analytical predictions and with the data-saving results of \citet{AlemannoETAL2023small,FachechiETAL2025regularization} established in the single-layer setting.  At higher $\alpha$ the boundary moves inward, because more patterns demand cleaner data, but the relative gain from dreaming is preserved, confirming that the spectral filtering mechanism operates independently of the storage load.

\subsection{Critical behaviour and finite-size scaling}\label{subsec:critical}

A natural question is whether the retrieval-to-noise transition visible in the phase diagrams is a genuine phase transition or a finite-size artefact.

\begin{figure*}[t]
\centering
\begin{minipage}[t]{0.64\textwidth}
  \centering
  \includegraphics[width=\linewidth]{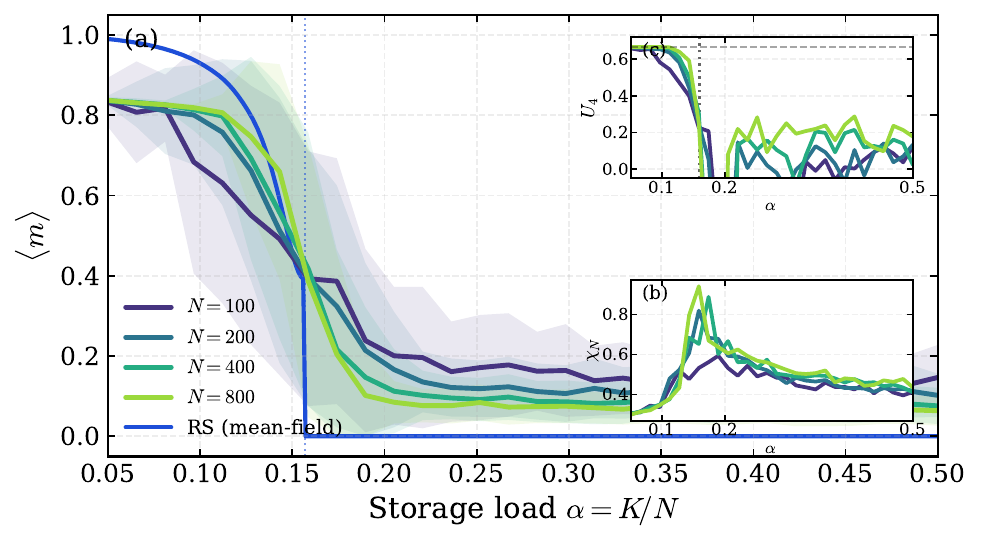}
\end{minipage}\hfill
\begin{minipage}[t]{0.33\textwidth}
  \centering
  \includegraphics[width=\linewidth]{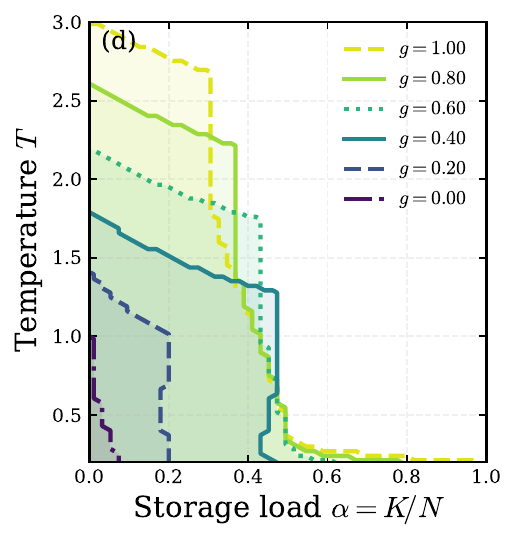}
\end{minipage}
\caption{\textbf{Critical behaviour and finite-size scaling.}
	\textbf{(a)}~Finite-size scaling of the retrieval transition: mean magnetisation $\langle m\rangle$ versus storage load $\alpha$ for increasing system sizes $N$. The blue solid line represents the theoretical prediction obtained from the numerical solution of the Replica Symmetric (RS) equations using the same physical parameters as the Monte Carlo simulations ($M{=}30$, $\beta{=}10$). The transition steepens progressively and provides the main visual summary of the critical crossover.
	\textbf{(b)}~Inset: susceptibility $\chi_N = N\,\mathrm{Var}(m)$, showing the growth of the critical peak with $N$.
	\textbf{(c)}~Inset: Binder cumulant $U_4$, whose crossing identifies a size-independent estimate of the critical load $\alpha_c$.
	\textbf{(d)}~Analytical retrieval regions in the $(\alpha,T)$-plane for six representative values of the inter-layer coupling, $g{=}0.0, 0.2, 0.4, 0.6, 0.8, 1.0$, obtained by numerical solution of the RS self-consistency equations~\eqref{eq:sc_signal}--\eqref{eq:sc_LAM}. The silhouettes are overlaid on the same axes, showing the expansion of the retrieval phase as coupling increases.
	Parameters for (a--c): $M{=}30$, $\beta{=}10$. Parameters for (d): $\rho{=}0.10$, retrieval regions identified via multi-start fixed-point iteration with threshold $\epsilon_M{=}10^{-3}$.}
\label{fig:critical}
\end{figure*}

The retrieval-to-noise transition visible in the preceding phase diagrams must be distinguished from a finite-size artefact before it can be interpreted as evidence of a genuine phase transition.  Standard finite-size scaling analysis provides the necessary test.

The magnetisation curves (Fig.~\ref{fig:critical}a) steepen progressively with $N$: as the system grows, the crossover from high to low magnetisation near the critical load sharpens markedly with system size, consistent with an increasingly abrupt thermodynamic transition.  The susceptibility $\chi_N = N\,\mathrm{Var}(m)$ (Fig.~\ref{fig:critical}b) develops a peak whose height grows with system size in a manner consistent with power-law divergence, $\chi_{\max}\sim N^{\gamma/\nu}$, the hallmark of a genuine critical point.  In the present setting, the Binder cumulant $U_4 = 1 - \langle m^4\rangle/(3\langle m^2\rangle^2)$ (Fig.~\ref{fig:critical}c) provides a convenient finite-size diagnostic of the retrieval transition: the curves for different $N$ display an approximately size-independent crossing near the critical load $\alpha_c$.  Taken together, these three diagnostics confirm that the boundary observed in the Monte Carlo experiments corresponds to a thermodynamic phase transition, not a gradual degradation caused by finite system size.

The analytical panel (Fig.~\ref{fig:critical}d), obtained by numerically iterating the RS self-consistency equations~\eqref{eq:sc_signal}--\eqref{eq:sc_LAM} to convergence, complements the finite-size analysis with a direct comparison among six representative couplings $g$.  Overlaying the six retrieval silhouettes on the same $(\alpha,T)$ axes makes the structural effect of cooperation immediate: as the inter-layer coupling $g$ increases, the region of stable retrieval expands outward substantially. 
Moreover, the qualitative agreement between these purely analytical phase maps and the threshold dynamics extrapolated from our finite-size Monte Carlo simulations supplies a strong, direct validation of the replica-symmetric framework.

\subsection{Design rules and trade-offs}\label{subsec:design_rules}

The preceding sections establish the qualitative and quantitative effects of dreaming on the retrieval region; we now turn to the complementary engineering question: \emph{given a finite sleep budget, how should it be allocated among the competing cognitive tasks?}  This perspective translates the physics of the preceding sections into actionable design principles for multi-layer architectures.

\begin{figure*}[t]
\centering
\begin{subfigure}[t]{0.32\textwidth}
  \centering
  \includegraphics[width=\linewidth]{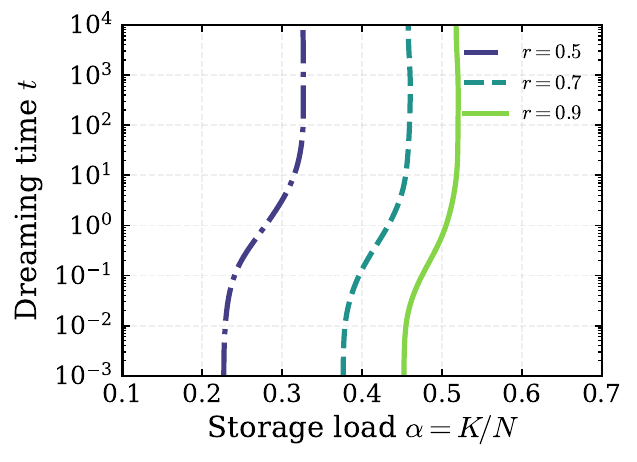}
  \label{fig:design:a}
\end{subfigure}\hfill
\begin{subfigure}[t]{0.32\textwidth}
  \centering
  \includegraphics[width=\linewidth]{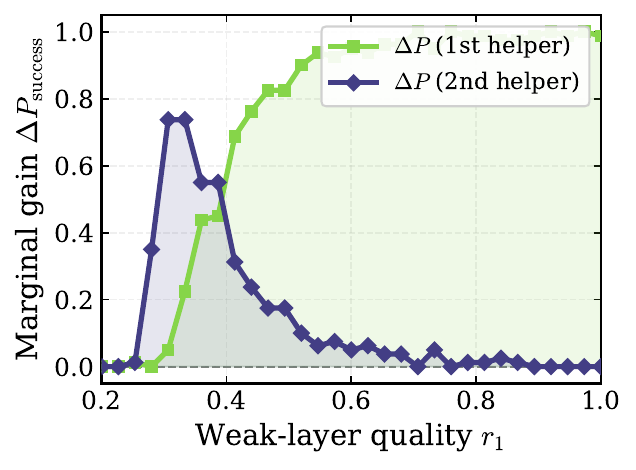}
  \label{fig:design:b}
\end{subfigure}\hfill
\begin{subfigure}[t]{0.32\textwidth}
  \centering
  \includegraphics[width=\linewidth]{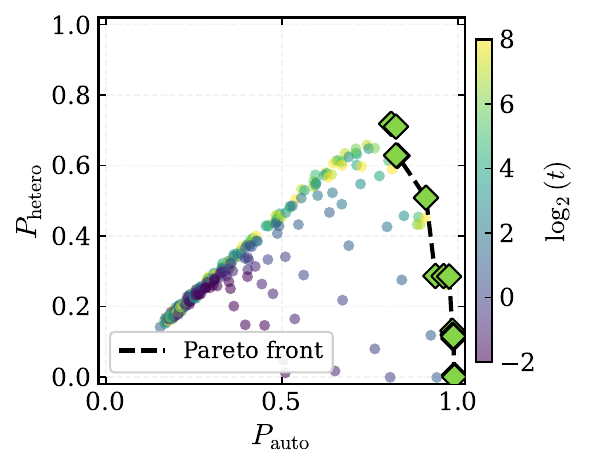}
  \label{fig:design:c}
\end{subfigure}
\caption{\textbf{Design rules for cooperative dreaming networks.}
\textbf{Left:}~Phase diagram in the $(\alpha,t)$-plane for three values of the data quality~$r$.  The retrieval boundary (iso-contour at $\langle m\rangle{=}0.85$) separates the \emph{retrieval phase} (upper-left region: low storage load~$\alpha$ and large dreaming time~$t$) from the \emph{non-retrieval phase} (lower-right region, where the load exceeds the network's effective capacity at the given sleep budget).  The boundary shifts rightward with increasing sleep time, meaning that higher storage loads are tolerated, moderate dreaming can nearly double the viable $\alpha$ at fixed~$r$.  At lower~$r$ (noisier or scarcer data) the boundary moves inward, but the relative gain from dreaming is preserved, confirming that spectral filtering remains effective regardless of raw data quality.  Background: $\langle m\rangle$ heatmap for $r{=}0.5$.
\textbf{Center:}~Marginal gain $\Delta P$ from additional helper layers as a function of the helper-layer data quality~$r_{2,3}$ (weak layer fixed at $r_1{=}0.50$).  Bright (yellow-green) curve: $\Delta P = P_{\mathrm{success}}(1\text{ helper}) - P_{\mathrm{success}}(0\text{ helpers})$; dark (blue-purple) curve: incremental gain of a second helper, $P_{\mathrm{success}}(2\text{ helpers}) - P_{\mathrm{success}}(1\text{ helper})$.  The first helper provides the dominant boost at high~$r_{2,3}$ (high-quality helpers), where cross-modal support is essential to rescue the weak layer; as helper quality decreases the marginal benefit diminishes.  The second helper peaks at intermediate~$r_{2,3}$ (the critical regime where the first helper partially rescues the weak layer but additional signal channels are still consequential) and remains positive throughout the explored range, confirming that richer cross-modal coupling never degrades performance.  Shaded bands indicate the signed contribution of each helper relative to the zero-helper baseline.
\textbf{Right:}~Pareto front in the stressed regime ($\alpha{=}0.20$, near capacity without dreaming).  Each scatter point represents one $(a,t)$ configuration, where $a$ is the Hamiltonian auto/hetero mixing parameter and $t$ the sleep time; colour encodes $\log_2 t$ on the viridis scale (dark $=$ little dreaming, bright $=$ long dreaming).  The horizontal axis $P_{\mathrm{auto}}$ measures retrieval success when the network is initialised from a partial within-layer cue (standard pattern reconstruction); the vertical axis $P_{\mathrm{hetero}}$ measures cross-modal completion success when the cue is provided in a different layer (heteroassociative propagation).  Without dreaming (dark cluster near the origin) both tasks fail in the high-load regime; increasing $t$ expands the jointly achievable performance region toward the upper-right.  Diamond markers trace the Pareto-optimal frontier: no configuration above this curve exists, so it defines the efficiency boundary for the simultaneous auto/hetero allocation problem.  Its convexity implies that a single intermediate sleep budget serves both tasks nearly as well as two specialised budgets.
Panel~(a): $N{=}500$, $M{=}20$, $r{=}0.5$ (background), $\beta{=}100$, $\alpha{\in}[0.1,0.7]$, $t{\in}[10^{-3},10^{4}]$ (log-sampled), $20$ steps, $10$ runs per cell.
Panel~(b): $N{=}500$, $\alpha{=}0.28$, $M{=}30$, $r_1{=}0.50$, $r_{2,3}{\in}[0.30,0.90]$, $t{=}4$, $g_{\mathrm{auto}}{=}0.82$, $g_{\mathrm{hetero}}{=}0.41$, $\beta{=}10$, $50$ steps, $30$ replicates.
Panel~(c): $N{=}500$, $\alpha{=}0.20$, $M{=}20$, $r{=}0.6$, $\beta{=}8$, $\varepsilon{=}0.15$, $t{=}2^k$ ($k{\in}\{{-2,\ldots,8}\}$), $200$ steps, $50$ replicates.}
\label{fig:design_rules}
\end{figure*}

Fig.~\ref{fig:design_rules}(a) maps the retrieval boundary in the $(\alpha,t)$-plane for several values of the data quality~$r$.  The phase boundary shifts rightward with increasing sleep time: at fixed~$r$, moderate dreaming nearly doubles the viable storage load.  Lower data quality pushes the boundary inward, but the relative gain from dreaming is preserved, confirming the generality of the spectral filtering mechanism.  The picture is consistent with the $(\rho,t)$ diagrams of Section~\ref{subsec:phase_diagrams} and provides a complementary view centred on storage capacity rather than dataset entropy.

Fig.~\ref{fig:design_rules}(b) addresses network heterogeneity from an architectural standpoint.  The marginal gain $\Delta P$ measures the increase in joint retrieval success when additional helper layers are introduced alongside the weak cued layer (fixed at $r_1{=}0.50$), as a function of the helper data quality $r_{2,3}$.  At high $r_{2,3}$ (high helper quality), the first helper provides a dramatic boost: cross-modal support from a clean partner injects the heteroassociative signal $u_l$ into the effective field~\eqref{eq:Psi} and can rescue a layer that would otherwise stall in the non-retrieval phase on its own.  As $r_{2,3}$ decreases toward the weak layer's quality, the helper becomes progressively less informative and the marginal benefit declines monotonically.  The second helper always contributes less than the first, consistent with diminishing returns, but its incremental gain peaks at intermediate $r_{2,3}$ values, precisely the critical regime where the first helper partially rescues the weak layer but the additional signal channel is still consequential.  Crucially, both marginal gains remain non-negative throughout the explored range (at fixed $r_1{=}0.50$, varying helper quality $r_{2,3}\in[0.30,0.90]$): additional helper layers never degrade performance within this regime, a practical consequence of the non-negative definiteness of the heteroassociative coupling term $u_l$ in the effective field~\eqref{eq:Psi}.  This observation leads to a practical design rule: in a constrained architecture, invest in one high-quality helper layer before adding further redundancy; the second helper is worth adding only when the first has been fully exploited.

Finally, the Pareto front (Fig.~\ref{fig:design_rules}(c)) crystallises the fundamental auto/hetero trade-off in the stressed regime ($\alpha{=}0.20$, near the capacity limit without dreaming).  Without dreaming (dark points clustered near the origin of the $P_{\mathrm{auto}}$--$P_{\mathrm{hetero}}$ plane) neither task is reliably achievable: the high storage load overwhelms both intra-layer pattern completion and cross-modal signal propagation.  Increasing $t$ (progressively brighter points) lifts the cloud toward the upper-right corner, jointly expanding the achievable performance space for both tasks.  The Pareto-optimal frontier (diamond markers) traces the efficiency boundary: no configuration can improve $P_{\mathrm{auto}}$ without degrading $P_{\mathrm{hetero}}$, or vice versa, above this curve.  The convexity of the frontier implies that intermediate allocations of the mixing parameter~$a$ and sleep budget~$t$ are Pareto-efficient: a single cooperative network can simultaneously achieve reliable auto-associative recall and reliable cross-modal completion without severe penalty on either objective.

\subsection{Dynamics and convergence}\label{subsec:dynamics}

The equilibrium analysis captures the asymptotic properties of the DLAM; a natural complementary question is whether dreaming also affects how rapidly the network reaches equilibrium and how the layers coordinate in time.  The answer is affirmative on both counts.  All simulations reported in this section employ the synchronous Glauber rule detailed in Algorithm~\ref{alg:glauber}; the full implementation is publicly available at \url{https://github.com/andrea-ladiana/dlam-mc-core}.

\begin{algorithm}[t]
\SetAlgoLined
\DontPrintSemicolon
\SetKwInOut{Input}{Input}
\SetKwInOut{Output}{Output}
\caption{Synchronous Glauber dynamics for the 3-layer DLAM}
\label{alg:glauber}
\Input{Empirical patterns $\{\hat{\bm\xi}_l^\mu\}$; dreaming kernels $\mathbb{A}_l(t_l)$ (eq.~\eqref{eq:Al_def}); couplings $\{g_{lm}\}$, normalisations $\{\Gamma_l,\Gamma_{lm}\}$; inverse temperature $\beta$; initial states $\bm{s}_l^{(0)}\in\{-1,+1\}^N$; number of steps $T$}
\Output{Magnetisation trajectories $\{m_l^{(t)}\}_{t=0}^{T}$, final states $\{\bm{s}_l^{(T)}\}$}
\BlankLine
\For{$t = 1, \ldots, T$}{
  \For{$l = 1, 2, 3$ \emph{\textup{(independently)}}}{
    \tcp{Pattern-space empirical overlaps ($K$-vector)}
    $\hat{m}_l^{\mu} \leftarrow \frac{1}{N}\sum_{j=1}^{N} \hat\xi_{j,l}^{\mu}\, s_{j,l}^{(t-1)}$, \quad $\mu=1,\ldots,K$ \;
    
    \tcp{Mix intra- and inter-layer signals}
    $v_l^{\mu} \leftarrow \frac{g_{ll}}{\Gamma_l}\bigl[\mathbb{A}_l(t_l)\,\hat{\bm{m}}_l\bigr]_{\mu} + \sum_{m \neq l} \frac{g_{lm}}{\Gamma_{lm}}\,\hat{m}_m^{\mu}$, \quad $\mu=1,\ldots,K$ \;
    
    \tcp{Back-project to site space: effective local field}
    $h_{i,l} \leftarrow \sum_{\mu=1}^{K} \hat\xi_{i,l}^{\mu}\, v_l^{\mu}$, \quad $i=1,\ldots,N$ \;
    
    \tcp{Synchronous Glauber flip}
    draw $u_i \overset{\mathrm{iid}}{\sim} \mathcal{U}[0,1]$ for $i=1,\ldots,N$ \;
    
    $s_{i,l}^{(t)} \leftarrow +1$ if $u_i < \tfrac{1}{2}\bigl(1+\tanh(\beta h_{i,l})\bigr)$, else $-1$ \;
  }
  $m_l^{(t)} \leftarrow \frac{1}{N}\sum_{i=1}^{N} \xi_{i,l}^{\mu^\star}\, s_{i,l}^{(t)}$, \quad $l=1,2,3$ \;
}
\end{algorithm}

\begin{figure*}[t]
\centering
\begin{subfigure}[t]{0.48\textwidth}
  \centering
  \includegraphics[width=\linewidth]{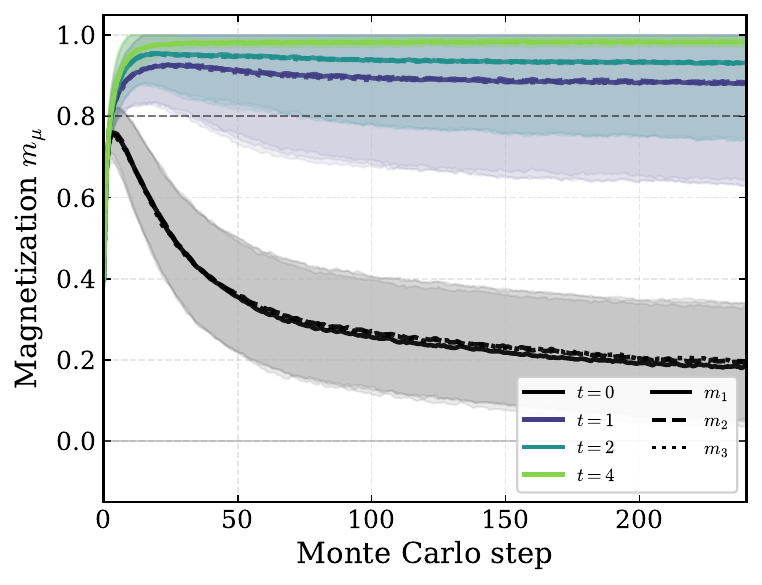}
  \label{fig:dynamics:a}
\end{subfigure}\hfill
\begin{subfigure}[t]{0.48\textwidth}
  \centering
  \includegraphics[width=\linewidth]{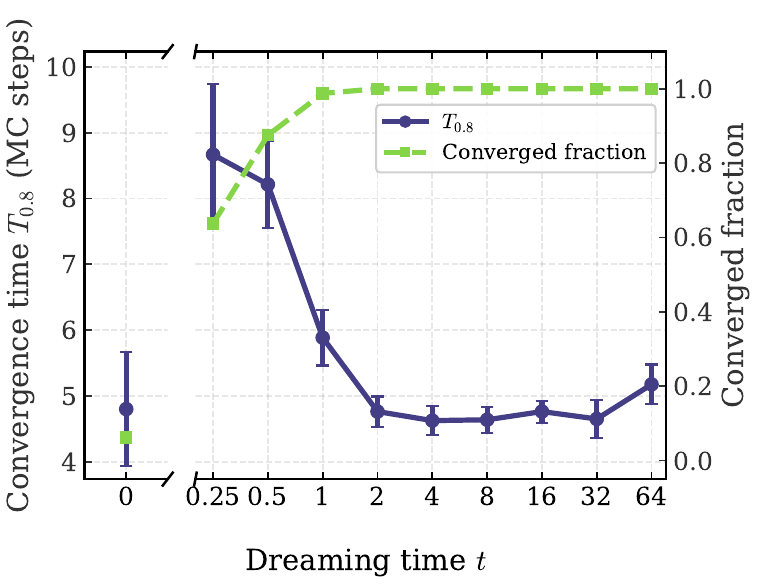}
  \label{fig:dynamics:b}
\end{subfigure}
\caption{\textbf{Dynamics and convergence acceleration.}
\textbf{Left:}~Magnetisation traces $m_1$, $m_2$, $m_3$ (solid, dashed, dotted) under the disentanglement protocol for selected dreaming times $t\in\{0,1,2,4\}$.  Without dreaming ($t{=}0$, black) the system stalls at a metastable plateau; increasing $t$ smooths the landscape and accelerates convergence.  Shaded bands: $\pm 1\sigma$.
\textbf{Right:}~Convergence acceleration: convergence time $T_{0.8}$ (steps to reach $m_1\ge 0.8$, left axis) and converged-trial fraction (right axis) versus dreaming time $t$.  Both metrics improve monotonically, demonstrating that dreaming accelerates retrieval.
Note the broken x-axis in the right panel, which combine the always-awake baseline at $t=0$ with a logarithmic scale for $t>0$.
Both panels: $N{=}500$, $M{=}30$, $r{=}0.70$, $80$ replicates.
Panel~(a): $\alpha{=}0.30$, $\beta{=}2$, $\varepsilon{=}0.10$, $g_{\mathrm{auto}}{=}0.82$, $g_{\mathrm{hetero}}{=}0.41$, $t_{\mathrm{dream}}{\in}\{0,1,2,4\}$, $240$ Glauber steps.
Panel~(b): $\alpha{=}0.20$, $\beta{=}5$, $\varepsilon{=}0.20$, $g_{\mathrm{auto}}{=}0.98$, $g_{\mathrm{hetero}}{=}0.12$, $t_{\mathrm{dream}}{\in}[0,64]$ (log-sampled), $40$ steps.}
\label{fig:dynamics}
\end{figure*}

The magnetisation trajectories (Fig.~\ref{fig:dynamics}a) provide direct evidence: without dreaming, layer~1 reaches a metastable plateau at intermediate magnetisation, a signature of the rugged free-energy landscape characteristic of the high-load regime.  With moderate dreaming ($t\ge 2$), these local minima are smoothed out, the plateau disappears, and all three layers converge rapidly to high magnetisation.  The inter-layer dynamics displays a clear causal structure: layers~2 and~3 lag behind layer~1 by 1--2 Monte Carlo sweeps, reflecting the time needed for the heteroassociative signal to propagate across layers.

The convergence time $T_{0.8}$, operationally defined as the number of Monte Carlo sweeps required for the cued layer's magnetisation to reach $m_1 \ge 0.8$ (Fig.~\ref{fig:dynamics}b), quantifies the acceleration: it decreases monotonically with dreaming time, while the fraction of trials that successfully converge increases in parallel.  Dreaming thus improves not only the asymptotic success rate but also the speed of retrieval, a distinction that matters in practice, since a network that converges slowly may be operationally useless even if it eventually reaches the correct attractor.

\subsection{Sleep-budget allocation under heterogeneous data quality}\label{subsec:heterogeneity}

A realistic multimodal system will have layers of unequal quality: one sensory channel may be clean and abundant while another is noisy or data-poor.  In that setting, the practically relevant question is how a finite dreaming budget should be distributed across weak and strong layers.

\begin{figure}[t]
\centering
\includegraphics[width=\linewidth]{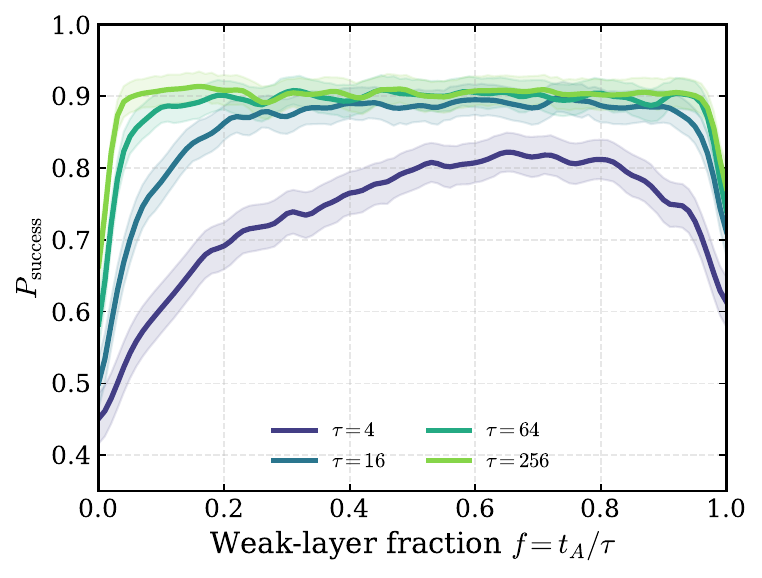}
\caption{\textbf{Cooperative sleep-budget allocation under heterogeneous data quality.}
A total budget $\tau$ is split between the weak layer (fraction $f$) and the helpers (fraction $(1{-}f)/2$ each).  A broad optimum at $f\in[0.2,0.7]$ shifts toward the helpers as $\tau$ grows, reflecting saturation of the weak layer's own dreaming.
Parameters: $N{=}500$, $\alpha{=}0.18$, $M{=}35$, $\beta{=}10$; weak layer entropy $\rho_A{=}0.40$, helper entropy $\rho_B{=}0.05$; $\tau{\in}\{4,16,64,256\}$, $g_{\mathrm{auto}}{=}0.82$, $g_{\mathrm{hetero}}{=}0.41$, cue noise $\varepsilon{=}0.15$, $m_{\mathrm{th}}{=}0.80$; $100$ Glauber steps, $200$ replicates.}
\label{fig:heterogeneity}
\end{figure}

Fig.~\ref{fig:heterogeneity} yields a practical design rule with non-trivial structure.  The experiment fixes the total sleep budget $\tau$ and sweeps the allocation fraction $f = t_A/\tau$ (fraction directed to the weak layer, so the two helpers each receive $(1-f)/2$), recording $P_{\mathrm{success}}$ at four representative values of $\tau\in\{4,16,64,256\}$.

Several features of the curves are noteworthy.  First, concentrating all dreaming on the weak layer ($f\to 1$) consistently underperforms: at $f{=}1$ the helpers receive no sleep at all, so their noise-dominant basins cannot relay a clean heteroassociative signal~$u_l$ to the weak layer, even if the latter's own intra-layer landscape has been perfectly cleaned.  Conversely, allocating all budget to the helpers ($f\to 0$) also underperforms: the weak layer is left undreamed, its non-condensed eigenmodes remain at their initial magnitude, and the inter-layer signal cannot overcome the strong intra-layer noise even with clean helpers.  The optimum $f^\star$ lies in the interior, robustly between $0.2$ and $0.7$ across all budgets.

Second, as $\tau$ grows (darker to brighter curves) the optimal fraction shifts leftward, from $f^\star\approx 0.5$--$0.7$ at small $\tau$ to $f^\star\approx 0.2$--$0.3$ at $\tau{=}256$.  The physical interpretation is saturation: once $t_A$ is large enough to suppress all non-condensed modes in the weak layer (approximately $t_A\sim 1/\alpha$), additional sleep on that layer yields no further noise reduction, so the marginal return on weak-layer investment drops to nearly zero.  At this point, budget is best redirected to the helpers, whose cleaner representations amplify the heteroassociative coefficient $u_l$ and provide a net lift to the weak layer through the coupling.  This shift is consistent with the effective-field decomposition~\eqref{eq:field_coeffs}: at saturation $\bar D_l\approx 1$ and $\sigma_l\approx 0$ for the weak layer, so further increases in $t_A$ have negligible effect on either the dreaming-noise coefficient or the signal coefficient $v_l$, and the dominant lever becomes the inter-layer term $u_l$, which benefits from helper dreaming.

Third, the success probability at the optimum ($P^\star\approx 0.8$--$0.95$ depending on $\tau$) substantially exceeds both extremes $f{=}0$ and $f{=}1$, confirming that a cooperative sleep allocation strategy consistently and significantly outperforms naive alternatives across all budget sizes.

\section{Discussion}\label{sec:discussion}

The preceding sections have established that the DLAM, a single cost function (Hamiltonian) combining off-line dreaming with heteroassociative coupling, produces a qualitative expansion of the retrieval region, a measurable data-saving effect, and a robust improvement on the demanding task of disentanglement.  We now place these findings in their broader context, interpret them from the three complementary viewpoints of statistical mechanics, machine learning, and biologically inspired computation, and state the limitations of the present analysis.

\subsection*{The mechanism: spectral filtering meets cross-modal support}

The effective-field decomposition~\eqref{eq:Psi} provides a compact explanation of why the two mechanisms are synergistic rather than merely additive.  Dreaming operates on the intra-layer sector: by differentially attenuating the high-eigenvalue modes of the empirical correlation matrix (which carry the inter-pattern interference), it reduces the dreaming-noise coefficient $\sigma_l$ while preserving the autoassociative signal $v_l$.  Inter-layer coupling, by contrast, introduces a second signal channel $u_l$ that is independent of the quality of layer~$l$'s own data and depends only on the retrieval state of the neighbouring layers.  The two improvements are complementary: one reduces noise, the other injects signal, and their joint effect exceeds the sum of their individual contributions, as confirmed by the synergistic retrieval regions documented in Fig.~\ref{fig:robustness}(b).

An important distinction separates the DLAM's dreaming mechanism from static decorrelation methods such as Kanter--Sompolinsky pseudo-inverse couplings \citep{kanter1987associative} or Gram--Schmidt orthogonalisation.  In the single-layer limit the two approaches are equivalent, but in the multidirectional setting the equivalence breaks down.  The cooperative effects arise because the dreaming kernel and the heteroassociative couplings share a common thermal dynamics: dreaming cleans the intra-layer representation, which in turn sharpens the inter-layer signal, which further stabilises the intra-layer state.  This feedback loop is absent when each layer is decorrelated independently before coupling, and it is responsible for the re-entrant phase structure visible in the corresponding $(\alpha,T)$ phase diagrams.

\subsection*{Data--computation trade-off and the role of disentanglement}

The data--computation trade-off revealed by the $(\rho,t)$ phase diagrams (Fig.~\ref{fig:robustness}c) has a direct practical reading from the machine-learning viewpoint: at fixed retrieval quality, off-line consolidation reduces the number of training examples required, so that computation substitutes for data acquisition.  This phenomenon was first identified in single-layer dreaming networks \citep{AlemannoETAL2023small,FachechiETAL2025regularization}; the present work shows that it persists, and is amplified, in the multi-layer architecture.

The effective-field decomposition~\eqref{eq:field_coeffs} also provides a principled criterion for \emph{when} dreaming is worth the computational investment: the benefit is largest when the dreaming-noise coefficient $\sigma_l$ is the dominant source of interference and the signal $v_l$ still has room to grow, that is, when the load is moderate and the data quality is low.  In the opposite regime (low load, high-quality data), the noise is already small and the marginal return on additional dreaming is negligible.  This observation rationalises the shape of the data-saving curves in Fig.~\ref{fig:datasaving} and provides architectural guidance for deciding whether off-line consolidation repays the computational cost.

Disentanglement provides the most challenging stress test for this trade-off.  The cross-modal mixture cue~\eqref{eq:simplex_cue} blends independent modal samples of a single target archetype from all three layers; at the barycentre every layer receives a diluted but nonzero starting signal, allowing bootstrap through inter-layer cooperation even without dreaming.  Toward the simplex vertices, however, one layer receives all the signal while the other two are essentially uninformed; joint convergence then requires the heteroassociative coupling to rescue the blind layers, which is impossible when the basins are contaminated by inter-pattern noise.  With dreaming, the spectral filter suppresses precisely these off-diagonal correlations, cleaning the basins so that even a single-view cue can propagate through the coupling and achieve joint retrieval.  The simplex visualisation (Fig.~\ref{fig:disentanglement}a,b) makes this transition vivid: the region of high joint success expands from the barycentre outward toward the vertices as dreaming is activated.  We note that disentanglement is not merely a technical benchmark: it captures the computational problem of multi-source memory integration, namely fusing partial, channel-specific evidence into separate, coherent per-modality representations, which is directly relevant to episodic memory consolidation, and the setting where the interaction between dreaming and heteroassociation is most consequential.

\subsection*{Biological resonance}

The biological reading of the DLAM is disciplined but substantive.  The architecture captures two features of cortical memory organisation, namely distributed multimodal storage and off-line consolidation \citep{FranklandBontempi2005,KlinzingNiethardBorn2019}, within a unified energy-based framework.  The heteroassociative couplings model the associative links between cortical areas that support cross-modal pattern completion, while the dreaming kernel models the synaptic rescaling that occurs during sleep.  The quantitative prediction that emerges, namely that sleep is most valuable when waking data is scarce, noisy, or ambiguous, resonates with experimental findings on the role of slow-wave sleep in memory consolidation under conditions of interference and weak encoding \citep{Denis2021Sleep}.  The disentanglement task of Section~\ref{subsec:disentanglement} provides a concrete instantiation: the cross-modal mixture cue~\eqref{eq:simplex_cue} can be read as the kind of multi-source, partially encoded memory trace that sleep is hypothesised to separate into clearly distinct representations \citep{CrickMitchison1983}, an interpretation supported by evidence of coordinated replay across hippocampal and cortical areas during sleep \citep{JiWilson2007}.

We emphasise that this biological framing is intended as a principled analogy, not as a claim of neural realism.  The DLAM assumes binary neurons, symmetric couplings and  substantial simplifications relative to cortical circuitry.  Nevertheless, the model captures the essential computational logic, namely noise suppression through off-line spectral filtering and cross-modal stabilisation through Hebbian binding, at a level of mathematical precision that permits quantitative predictions and phase-diagram analysis.  Furthermore, recent results suggesting that dreaming actively optimises memory even when imposing strict, biologically realistic bounds on true synaptic strengths \citep{marinari2026dreaming} lend analytical support to the plausibility of this paradigm under stringent physiological limits.  Whether the specific spectral mechanism of the DLAM has a counterpart in biological synaptic homeostasis is an empirical question that the present theory motivates but cannot answer.

\subsection*{Connections to modern associative memories}

The DLAM bears a structural resemblance to modern Hopfield networks \citep{ramsauer2020hopfield} and dense associative memories \citep{krotov2016dense}, in that all three families manipulate the energy landscape to sharpen retrieval basins.  The mechanism differs fundamentally: where modern Hopfield networks achieve sharper basins through higher-order (polynomial or continuous) energy functions, a pathway recently shown to profoundly expand associative capacities across both neurobiology and modern machine learning \citep{krotov2021large}, the DLAM achieves a comparable effect through off-line spectral filtering of a \emph{quadratic} energy, followed by on-line thermal relaxation.  The two approaches are not mutually exclusive: one could in principle apply dreaming-type regularisation to a dense associative memory, and a formal comparison of their capacity scaling and phase structure is an interesting direction for future work.

\subsection*{Limitations}

Several limitations should be stated explicitly.  The replica-symmetric ansatz is expected to break down near the spin-glass transition, where full replica-symmetry breaking (RSB) may modify the critical thresholds quantitatively; a stability analysis along the de~Almeida--Thouless line is deferred to future work.  The LAM sector of the self-consistency equations admits fully explicit scalar form only for $L{=}3$; extension to larger $L$ is straightforward but notationally heavier.  The present model assumes i.i.d.\ Rademacher archetypes and symmetric layer sizes; correlated patterns, continuous-valued neurons, and heterogeneous layer dimensions are natural extensions that may introduce qualitatively new phenomena.  The dreaming mechanism is parametrised by a single scalar $t_l$ per layer; a dynamical model of the dreaming process itself, in which $t_l$ evolves according to its own learning rule, would be both more biologically motivated and potentially richer in phenomenology.  Finally, the Monte Carlo simulations are limited to moderate system sizes ($N \le 500$) and to equilibrium or near-equilibrium dynamics; larger-scale studies and non-equilibrium protocols remain to be explored.

\section{Conclusion}\label{sec:conclusions}

We have introduced the Dreaming $L$--directional Associative Memory, a multi-layer Hopfield architecture that unifies off-line Hebbian dreaming with supervised heteroassociative coupling within a single energy function.  The replica-symmetric theory yields a closed system of self-consistency equations for the order parameters as the control parameters are made to vary. The resulting effective local field decomposes into signal, dreaming noise, and inter-layer noise, a structure that identifies, at the level of individual neurons, why dreaming improves retrieval: it differentially attenuates the high-eigenvalue interference modes of the empirical correlation matrix, suppressing inter-pattern crosstalk while preserving the signal direction. This intra-layer cleaning, in turn, amplifies the effectiveness of the inter-layer coupling.

The two mechanisms are synergistic.  The phase diagrams reveal retrieval regions not reached by either dreaming or heteroassociation alone, and the numerics confirm this prediction across the explored parameter space.  The most consequential manifestation is on the task of disentanglement, where the dreaming kernel destabilises the glassy attractor that traps the dynamics near ambiguous mixture cues and promotes sharp, well-separated retrieval basins, raising the joint success rate substantially.  Importantly, this work provides the \emph{first} treatment of the disentanglement problem with a fully general, heterogeneous cue-weight vector $(\alpha_1,\alpha_2,\alpha_3)$ on the 2-simplex, rather than restricted to the symmetric barycentric cue.  This generalisation is both welcome and not-trivial: it is welcome because rarely mixtures are used with equal intensity among the base patterns – e.g. in audio and/or video processing -, not-trivial because the asymmetry of the cue weights breaks the symmetry of the initial condition across layers, so that different layers start with vastly different overlaps with their respective targets.  Establishing that the dreaming benefit extends throughout the entire simplex (not just near the symmetric barycentre but all the way to the vertices, where a single-view cue provides signal to only one layer while the other two start essentially uninformed) is a substantively stronger result than the homogeneous case, and is the main empirical finding of the disentanglement experiments.

The data--computation trade-off documented in the $(\rho,t)$ phase diagrams (Fig.~\ref{fig:robustness}c) and the data-saving experiments (Fig.~\ref{fig:datasaving}) deserves special emphasis.  Off-line consolidation substitutes for additional training data in a quantitatively significant way: moderate dreaming consistently reduces the critical dataset size $M_c(t)$ by $40$--$60\%$ across a broad range of data-quality conditions and pattern statistics, and this saving is robust to both the choice of evaluation metric and the structure of the patterns (random Rademacher or MNIST-derived).  In the multimodal setting these savings are of acute practical value, because collecting large batches of clean, aligned data across multiple sensors is far more cost-prohibitive than allocating extra off-line computation.  The extension of this data-saving phenomenon from single-layer to multi-layer architectures is therefore not a minor quantitative extension but a qualitative broadening of its applicability.

From a broader perspective, the DLAM demonstrates that a principled combination of off-line noise suppression and cross-modal binding can extend the operating envelope of associative memories into regimes of high load, poor data quality, and ambiguous cues that are inaccessible to conventional architectures.  The effective-field decomposition provides not only a mechanistic explanation but a practical design tool: it identifies when dreaming is most beneficial and how a finite sleep budget should be allocated across layers of unequal data quality.

Several directions remain open: a full de~Almeida--Thouless stability analysis to delineate the regime of validity of the replica-symmetric predictions; extensions to correlated, structured, and continuous-valued patterns; dynamical models of the dreaming process with their own evolution equations; and applications to real multimodal datasets where data acquisition is expensive but off-line computation is cheap.  The central message of this work, namely that off-line computation can substitute for part of the burden of data in multimodal memory systems, suggests that the classical unlearning hypothesis of \citet{CrickMitchison1983} has richer implications for learning architectures than have yet been fully explored.

\section*{Reproducibility}
All code and simulation data accompanying this work are available at \url{https://github.com/andrea-ladiana/dlam-mc-core}.

\section*{Acknowledgements}
F.D. and M.M.S. acknowledge support from the European Union – NextGenerationEU under the Italian National Recovery and Resilience Plan (PNRR), Mission 4, Component 2, Investment 1.3, via the Italian Ministry of University and Research (MUR), through the extended partnership "Future Artificial Intelligence Research (FAIR)" (Project Code PE00000013, CUP H97G22000210007, Spoke 6 "Symbiotic AI") and the specific cascade grant project "DELTA-CORE" (CUP F88H24002810007).
A.B., A.L., and M.M.S. acknowledge financial support from the Italian National Institute for Higher Mathematics (INdAM) through the GNFM group.
Additionally, M.M.S. acknowledges support from the National Institute for Nuclear Physics (INFN), Sezione di Lecce, and A.B. acknowledges support from INFN, Sezione di Roma 1.


\onecolumn
\appendix


\noindent
This appendix provides the complete analytical derivations underpinning the statistical-mechanical analysis of the Dreaming $L$-directional Associative Memory (DLAM) developed in the main text. Notation, conventions, and model definitions follow Section~2 of the manuscript throughout: the DLAM architecture, the dataset model, and the macroscopic observables of Section~2.1, the dreaming kernel of Section~2.2, and the inter-layer LAM coupling of Section~2.3 are adopted without restatement.

The argument unfolds in five logically successive stages. \ref{app:replica_free_energy} sets up the partition function, performs the Hubbard--Stratonovich decouplings, and recasts the quenched free-energy calculation as a Guerra interpolation, thereby converting it into a differential identity. \ref{app:one_body} evaluates the solvable Cauchy datum at the $x=0$ endpoint of the interpolation. \ref{app:streaming} computes the streaming derivative and shows how the replica-symmetric self-averaging hypothesis uniquely fixes the trial parameters. \ref{app:RS_pressure} assembles the RS pressure from these ingredients and specialises the LAM Gaussian sector to $L=3$. Finally, \ref{app:self_consistency} derives the full system of self-consistency (saddle-point) equations and verifies their reduction to known limiting cases.

\section{Derivation of the replica-symmetric free energy}
\label{app:replica_free_energy}

\noindent
This section constructs the replica-symmetric representation of the DLAM partition function. Starting from the Boltzmann--Gibbs measure associated with the full Hamiltonian~\eqref{eq:Hdlam}, we isolate the signal and noise contributions in each coupling sector, perform Hubbard--Stratonovich linearisations tailored to the symmetry of each quadratic form, and embed the result in a Guerra-type interpolating action. The fundamental theorem of calculus then decomposes the quenched pressure into a solvable one-body term and a streaming remainder.

\subsection{Boltzmann--Gibbs measure, partition function and free energy}
The model depends on the following control parameters: the inverse temperature $\beta$; the vector of layer-wise sleep times $\bm t:=(t_1,\dots,t_L)$; the intra-layer load ratios $\bm\rho:=(\rho_1,\dots,\rho_L)$, which enter through $\Gamma_l:=r_l^2(1+\rho_l)$; the inter-layer coupling matrix $\mathbf G:=(g_{lm})_{1\le l<m\le L}$; and the pattern load $\alpha:=(K-1)/N$. All thermodynamic functions depend on this full parameter set; for compactness, we suppress its explicit appearance in argument lists whenever no ambiguity arises.

For a fixed realisation of the quenched disorder $\mathcal D$, the Boltzmann--Gibbs measure at inverse temperature $\beta$ takes the form
\begin{equation}
  \mathbb{P}_{N,L}(\bm s)
  := \frac{\exp\big(-\beta\,\mathcal{H}_{\mathrm{DLAM}}(\bm s)\big)}{\mathcal{Z}_{N,L}},
\end{equation}
where the normalization
\begin{equation}
  \mathcal{Z}_{N,L}
  := \sum_{\bm s\in\{-1,+1\}^{NL}} \exp\big(-\beta\,\mathcal{H}_{\mathrm{DLAM}}(\bm s)\big)
\end{equation}
defines the finite-volume partition function. The thermal average of an observable $F:\{-1,+1\}^{NL}\to\mathbb{R}$ under this measure is denoted by
\begin{equation}
  \omega_N(F)
  := \frac{1}{\mathcal{Z}_{N,L}}\sum_{\bm s} F(\bm s)\,e^{-\beta\mathcal{H}_{\mathrm{DLAM}}(\bm s)}.
\end{equation}

Because the disorder enters only through the dataset, it is natural to condition on the full collection of examples
\begin{equation}
  \mathcal{D} := \Bigl\{\eta_{i,l}^{\mu,k}\Bigr\}_{\substack{i\le N,\,l\le L\\ \mu\le K,\,k\le M}}
  \quad\text{(equivalently, on }\{\hat{\xi}_{i,l}^{\mu}\}\text{ and }\{\Sigma_l\}\text{),}
\end{equation}
and to express the finite-$N$ partition function as
\begin{equation}
  \mathcal{Z}_{N,L}(\beta\mid \mathcal{D})
  := \sum_{\bm{s}} \exp\bigl\{-\beta\mathcal{H}_{\mathrm{DLAM}}(\bm{s})\bigr\}.
\end{equation}
Substituting the explicit forms~\eqref{eq:Hdream}--\eqref{eq:Hlam}, the Boltzmann weight splits cleanly into its inter-layer and intra-layer contributions:
\begin{equation}
\label{eq:Z_explicit}
\begin{split}
  \mathcal{Z}_{N,L}(\beta\mid \mathcal{D})
  &= \sum_{\bm{s}}\exp\Biggl\{\beta\sum_{1\le l<m\le L}\frac{g_{lm}}{N\Gamma_{lm}}\sum_{\mu=1}^K\sum_{i,j=1}^N \hat{\xi}_{i,l}^{\mu}\hat{\xi}_{j,m}^{\mu} s_{i,l}s_{j,m}\\
  &\quad +\beta\sum_{l=1}^L\frac{1}{N\Gamma_{l}}\sum_{\mu,\nu=1}^K \sum_{i,j=1}^N \hat{\xi}_{i,l}^{\mu} (\mathbb{A}_{l})_{\mu\nu} \hat{\xi}_{j,l}^{\nu} s_{i,l}s_{j,l}\Biggr\}.
\end{split}
\end{equation}

The Boltzmann weight in~\eqref{eq:Z_explicit} is already a function of the empirical Mattis magnetisations alone: every microscopic sum over site indices is absorbed into the collective overlaps $\hat m_l^\mu$, yielding
\begin{equation}
\label{eq:Z_in_overlaps}
\begin{split}
  \mathcal{Z}_{N,L}(\beta\mid \mathcal{D})
  &= \sum_{\bm{s}}\exp\Biggl\{\beta N\sum_{1\le l<m\le L}\frac{g_{lm}}{\Gamma_{lm}}\sum_{\mu=1}^K \hat{m}_{l}^{\mu}\hat{m}_{m}^{\mu} \\
  &\quad +\beta N\sum_{l=1}^L\frac{1}{\Gamma_{l}}\sum_{\mu,\nu=1}^K \hat{m}_{l}^{\mu} (\mathbb{A}_{l})_{\mu\nu} \hat{m}_{l}^{\nu}\Biggr\}.
\end{split}
\end{equation}
The identity is purely algebraic and requires no thermodynamic limit.

Letting $\E$ denote the average over all realisations of the quenched disorder, we define the intensive quenched pressure
\begin{equation}
  \mathcal{A}_{N}(\beta) := \frac{1}{N}\,\E\big[\log \mathcal{Z}_{N,L}\big],
\end{equation}
and, in the thermodynamic limit,
\begin{equation}
  \mathcal{A}(\beta) := \lim_{N\to\infty} \mathcal{A}_{N}(\beta),
\end{equation}
from which the quenched free-energy density follows as
\begin{equation}
  f(\beta) := -\frac{1}{\beta}\,\mathcal{A}(\beta).
\end{equation}

\subsection{Partition function and Hubbard--Stratonovich decouplings}
\label{sec:HS_dlam_definitive}
The partition function~\eqref{eq:Z_in_overlaps} contains both inter-layer and intra-layer quadratic forms in the spins; a tractable mean-field theory is obtained through Hubbard--Stratonovich (HS) linearisation of both sectors. The starting point is
\begin{equation}
\label{eq:Z_start_def}
\mathcal Z(\beta\mid\mathcal D)=\sum_{{\bm s}}\exp\Big(-\beta\mathcal H_{\mathrm{LAM}}(\bm s)-\beta\mathcal H_{\mathrm{Dream}}(\bm s)\Big),
\end{equation}
with the LAM and dreaming Hamiltonians given by
\begin{align}
\label{eq:Hlam}
\mathcal H_{\mathrm{LAM}}(\bm s)
&:=-\sum_{1\le l<m\le L}\frac{g_{lm}}{N\Gamma_{lm}}\sum_{\mu=1}^{K}\sum_{i,j=1}^{N}
\hat\xi_{i,l}^{\mu}\hat\xi_{j,m}^{\mu}s_{i,l}s_{j,m},
\\
\label{eq:Hdream}
\mathcal H_{\mathrm{Dream}}(\bm s)
&:=-\sum_{l=1}^{L}\frac{1}{N\Gamma_{l}}\sum_{i,j=1}^{N}\sum_{\mu,\nu=1}^{K}
\hat\xi_{i,l}^{\mu}[\mathbb A_{l}(t_{l})]_{\mu\nu}\hat\xi_{j,l}^{\nu}s_{i,l}s_{j,l},
\qquad
\mathbb A_{l}(t_{l})=(1+t_{l})(\mathbb I+t_{l}\Sigma_{l})^{-1}.
\end{align}
so that the full Hamiltonian is
\begin{equation}
\label{eq:Hdlam}
\mathcal{H}_{\mathrm{DLAM}}(\bm s) := \mathcal{H}_{\mathrm{Dream}}(\bm s) + \mathcal{H}_{\mathrm{LAM}}(\bm s),
\end{equation}
with the normalization constants
\begin{equation}
(\Sigma_{l})_{\mu\nu}:=\frac{1}{N\Gamma_{l}}\sum_{i=1}^{N}\hat\xi_{i,l}^{\mu}\hat\xi_{i,l}^{\nu},
\qquad
\Gamma_{l}:=r_{l}^{2}(1+\rho_{l}),\quad
\Gamma_{lm}:=\sqrt{\Gamma_{l}\Gamma_{m}}.
\end{equation}

\subsubsection{LAM sector}
The inter-layer couplings are bilinear in spins belonging to different layers, and their HS decoupling becomes most transparent once the layer-wise normalisation is absorbed into the patterns. To this end we define the rescaled archetypes
\begin{equation}
\label{eq:J_def}
J_{i,l}^{\mu}:=\frac{\hat\xi_{i,l}^{\mu}}{\sqrt{\Gamma_{l}}}
\qquad\Rightarrow\qquad
\frac{1}{\Gamma_{lm}}\hat\xi_{i,l}^{\mu}\hat\xi_{j,m}^{\mu}=J_{i,l}^{\mu}J_{j,m}^{\mu},
\end{equation}
and the corresponding layer-resolved overlaps
\begin{equation}
\label{eq:p_def}
p_{l}^{\mu}(\bm s):=\frac{1}{\sqrt N}\sum_{i=1}^{N}J_{i,l}^{\mu}s_{i,l},
\end{equation}
the LAM energy assumes the manifestly quadratic form
\begin{equation}
\label{eq:LAM_overlap_exact_def}
\beta\mathcal H_{\mathrm{LAM}}(\bm s)
=-\beta\sum_{1\le l<m\le L}g_{lm}\sum_{\mu=1}^{K}p_{l}^{\mu}(\bm s)p_{m}^{\mu}(\bm s).
\end{equation}
Within the standard replica-symmetric condensation ansatz, a single class (conventionally $\mu=1$) acquires a macroscopic overlap with the spin configuration and thus carries the signal, while the remaining $K-1$ non-condensed modes generate the noise. Isolating these two contributions as
\begin{equation}
\label{eq:LAM_split_def}
\sum_{\mu=1}^{K}p_{l}^{\mu}p_{m}^{\mu}=p_{l}^{1}p_{m}^{1}+\sum_{\mu=2}^{K}p_{l}^{\mu}p_{m}^{\mu},
\end{equation}
one defines the signal and noise factors
\begin{equation}
\label{eq:LAM_signal_noise_def}
\mathcal S_{\mathrm{LAM}}(\bm s):=\beta\sum_{l<m}g_{lm}\,p_{l}^{1}(\bm s)p_{m}^{1}(\bm s),
\qquad
\mathcal N_{\mathrm{LAM}}(\bm s):=\exp\Big(\beta\sum_{l<m}g_{lm}\sum_{\mu=2}^{K}p_{l}^{\mu}(\bm s)p_{m}^{\mu}(\bm s)\Big)
=\exp\Big(\frac{\beta}{2}\sum_{\mu=2}^{K}\bm p^{\mu\top}\mathbf G_0\,\bm p^\mu\Big),
\end{equation}
where the noise factor involves the overlap vectors $\bm p^\mu=(p_1^\mu,\dots,p_L^\mu)^\top$ associated with all non-condensed modes, mutually coupled through the symmetric off-diagonal connectivity matrix
\begin{equation}
\label{eq:G_def}
\mathbf G_0\in\mathbb R^{L\times L},\qquad
(\mathbf G_0)_{lm}:=\begin{cases}
g_{lm},& l\neq m,\\
0,& l=m,
\end{cases}
\qquad\Rightarrow\qquad
\sum_{l<m}g_{lm}p_l^\mu p_m^\mu=\frac12\bm p^{\mu\top}\mathbf G_0\bm p^\mu.
\end{equation}

\medskip
\begin{lemma}[Complex HS for hetero-associative couplings]
\label{lem:HS_vec_}
Let $L\ge 2$ and let $\mathbf G_0\in\mathbb R^{L\times L}$ be the weighted off-diagonal matrix defined in \eqref{eq:G_def}.
For each non-condensed index $\mu\ge 2$ introduce complex auxiliary fields
$\bm u_\mu=(u_\mu^1,\dots,u_\mu^L)\in\Sigma^L$ and their complex conjugates
$\bm u_\mu^\dagger=(u_\mu^{1\dagger},\dots,u_\mu^{L\dagger})\in\Sigma^L$.
We endow $(\bm u_\mu,\bm u_\mu^\dagger)$ with the {Gaussian measure}\footnote{\noindent
By a standard abuse of notation, when a single non-condensed mode $\mu$ is considered (with no product over $\mu$ written explicitly),
the same symbol $Duu^\dagger$ denotes the corresponding one-mode factor of \eqref{eq:Duudagger_def}.}
\begin{equation}
\label{eq:Duudagger_def}
Duu^{\dagger}
:=\prod_{\mu=2}^{K}\left[
\prod_{\ell=1}^{L}\frac{du_{\mu}^{\ell}\,du_{\mu}^{\ell\dagger}}{2\pi}\right]
\exp\Bigg\{-\frac12\sum_{\mu=2}^{K}\Big(
\bm u_\mu^\top \bm u_\mu^\dagger
-\bm u_\mu^{\dagger\top}\mathbf G_0\,\bm u_\mu
\Big)\Bigg\}.
\end{equation}
\noindent
This measure is {non-diagonal}: the cross-covariance term
$\bm u_\mu^{\dagger\top}\mathbf G_0\,\bm u_\mu$ correlates hidden variables attached to distinct visible layers, and it is precisely this correlation that reconstructs the original inter-layer Hebbian couplings upon integration.

\medskip
\noindent
For every fixed $\bm p\in\mathbb R^L$, the (selective) complex HS representation is chosen so that
\begin{equation}
\label{eq:HS_vec_id_def}
\exp\Big(\frac{\beta}{2}\bm p^\top \mathbf G_0\,\bm p\Big)
=\int Duu^\dagger\,
\exp\Big(\sqrt{\beta}\,\bm p^\top \bm u\Big),
\end{equation}
where the symbol $Duu^\dagger$ in~\eqref{eq:HS_vec_id_def} denotes the one-mode factor (single $\mu$) of the measure~\eqref{eq:Duudagger_def}, and the conjugate variables $\bm u^\dagger$ enter only through the non-diagonal Gaussian measure rather than through an additional linear source.
\end{lemma}

\noindent
Applying the identity~\eqref{eq:HS_vec_id_def} mode by mode to each $\mu\ge2$ yields the linearised representation of the noise factor,
\begin{equation}
\label{eq:N_3AM_oldstyle_def}
{
\mathcal N_{\mathrm{LAM}}(\bm s)
=\int Duu^{\dagger}\,
\exp\Bigg[\sqrt{\beta}\sum_{\mu=2}^{K}\bm p^{\mu\top}\bm u_\mu\Bigg],
}
\end{equation}
which, after expanding $p_l^\mu$ in terms of the individual spins, becomes manifestly linear in each spin variable:
\begin{equation}
\label{eq:N_3AM_oldstyle_linear_def}
{
\mathcal N_{\mathrm{LAM}}(\bm s)
=\int Duu^{\dagger}\,
\exp\Bigg[\sqrt{\frac{\beta}{N}}\sum_{\mu=2}^{K}\sum_{\ell=1}^{L}\sum_{i=1}^{N}
J_{i,\ell}^{\mu}s_{i,\ell}\,u_{\mu}^{\ell}\Bigg].
}
\end{equation}

\subsubsection{Dreaming sector}
The intra-layer dreaming couplings are linearised by introducing, for each layer $l$, an array of $N$ site-indexed Gaussian auxiliary fields $\{\phi_{i,l}\}_{i=1}^{N}$ distributed as
\begin{equation}
\label{eq:Dphi_def}
D\phi_{l}:=\prod_{i=1}^{N}\frac{d\phi_{i,l}}{\sqrt{2\pi}}\exp\Big(-\frac12\sum_{i=1}^{N}\phi_{i,l}^{2}\Big),
\end{equation}
and by combining the spin and noise degrees of freedom into the complexified multi-spin
\begin{equation}
\label{eq:k_def}
k_{i,l}:=s_{i,l}+ i\sqrt{\frac{t_{l}}{\beta(1+t_{l})}}\phi_{i,l},
\end{equation}
whose imaginary component, weighted by the sleep time $t_l$, is calibrated so that Gaussian integration regenerates the dressed interaction $\mathbb{A}_l$. In this representation, the dreaming signal for the condensed class reduces to
\begin{equation}
\label{eq:Dream_signal_factor_def}
{
\exp\Big(\beta\mathcal S_{\mathrm{Dream}}(\bm s,\bm\phi)\Big)
:=\prod_{l=1}^{L}\exp\Bigg(\frac{\beta N(1+t_{l})}{2\Gamma_{l}}\eta_{l}^{2}\Bigg),
}
\end{equation}
while the non-condensed modes are linearised through a second family of auxiliary fields $\{z_\mu^l\}_{\mu=2}^{K}$, drawn from a Gaussian with variance $(1+t_{l})$:
\begin{equation}
\label{eq:Dz_def}
Dz_{l}:=\prod_{\mu=2}^{K}\frac{dz_\mu^l}{\sqrt{2\pi}}
\exp\Big(-\frac{1}{2(1+t_{l})}\sum_{\mu=2}^{K} (z_\mu^l)^{2}\Big).
\end{equation}
The dreaming noise factor then takes the form
\begin{equation}
\label{eq:N_Dream_def}
{
\mathcal N_{\mathrm{Dream}}(\bm s,\bm\phi)
:=\prod_{l=1}^{L}\int Dz_{l}
\exp\Bigg[\sqrt{\frac{\beta}{\Gamma_{l}N}}\sum_{\mu=2}^{K}\sum_{i=1}^{N}
\hat\xi_{i,l}^{\mu}z_\mu^lk_{i,l}\Bigg].
}
\end{equation}


\subsubsection{Full HS-decoupled representation}
Combining the LAM signal--noise decomposition~\eqref{eq:LAM_signal_noise_def}, the complex HS representation of the inter-layer noise~\eqref{eq:N_3AM_oldstyle_def}--\eqref{eq:N_3AM_oldstyle_linear_def}, and the dreaming-native formulation of the intra-layer noise~\eqref{eq:Dream_signal_factor_def}--\eqref{eq:N_Dream_def}, we arrive at the exact HS-decoupled partition function
\begin{equation}
\label{eq:Z_final_def}
{
\mathcal Z(\beta\mid\mathcal D)
=\sum_{{\bm s}}
\exp\big(\mathcal S_{\mathrm{LAM}}(\bm s)\big)
\Bigg[\prod_{l=1}^{L}\int D\phi_{l}\Bigg]
\exp\big(\beta\mathcal S_{\mathrm{Dream}}(\bm s,\bm\phi)\big)
\mathcal N_{\mathrm{Dream}}(\bm s,\bm\phi)
\mathcal N_{\mathrm{LAM}}(\bm s).
}
\end{equation}

\medskip\noindent
The representation is exact: integrating out the complex auxiliary fields $\bm u_\mu,\bm u_\mu^\dagger$ in~\eqref{eq:N_3AM_oldstyle_def} regenerates the inter-layer quadratic form $\exp\!\big(\frac{\beta}{2}\bm p^{\mu\top}\mathbf G_0\bm p^\mu\big)
=\exp\!\big(\beta\sum_{l<m}g_{lm}p_l^\mu p_m^\mu\big)$ for each non-condensed mode $\mu\ge2$, while integrating out $z_\mu^l$ and $\phi_{i,l}$ in~\eqref{eq:N_Dream_def} reconstructs the dressed dreaming kernel $\mathbb A_{l}(t_{l})$ of~\eqref{eq:Hdream}.

\subsection{Guerra interpolation for the coupled LAM+Dreaming model}
The quenched pressure is extracted through a Guerra-type interpolation, which continuously deforms the fully coupled system into a solvable reference model while tracking the free-energy cost at each step. A preliminary ingredient is the Gaussian approximation for the non-condensed modes $\mu\ge2$, in which the rescaled empirical archetypes are replaced by i.i.d.\ standard Gaussians,
\begin{equation}
\label{eq:gauss_approx_mu_ge2_def}
{
J_{i,l}^{\mu}=\frac{\hat\xi_{i,l}^{\mu}}{\sqrt{\Gamma_l}}\;\approx\;\lambda_{i,l}^{\mu},
\qquad \lambda_{i,l}^{\mu}\stackrel{\mathrm{i.i.d.}}{\sim}\mathcal N(0,1),
\qquad \mu\ge2,
}
\end{equation}
with the family $\{\lambda_{i,l}^{\mu}\}_{i,l,\mu\ge2}$ taken independent of the condensed class and of every other random variable in play. By the central limit theorem this replacement is asymptotically exact for the non-condensed sector in the thermodynamic limit; its effect is to turn both the LAM noise factor~\eqref{eq:N_3AM_oldstyle_linear_def} and the dreaming noise factor~\eqref{eq:N_Dream_def} into linear couplings of the spins ($s_{i,l}$ and $k_{i,l}$) to i.i.d.\ Gaussian random fields. In the dreaming sector the coupling reads $\sqrt{\beta/N}\,\tilde\lambda_{i,l}^{\mu}z_\mu^lk_{i,l}$, where $\{\tilde\lambda_{i,l}^\mu\}_{i,l,\mu\ge2}$ is an independent copy of $\{\lambda_{i,l}^\mu\}$, kept distinct so as to preserve the statistical independence of the LAM and dreaming Gaussian sectors.

The interpolating partition function $\mathcal Z_{N,L}(x)$ depends on an auxiliary parameter $x\in[0,1]$, and the associated quenched pressure
\begin{equation}
\label{eq:A_interp_def}
{
\mathcal A_{N,L}(x):=\frac{1}{N}\,\E\log\mathcal Z_{N,L}(x),
\qquad x\in[0,1],
}
\end{equation}
in which $\E$ now denotes the joint average over the quenched dataset and all auxiliary Gaussian fields. The fundamental theorem of calculus then decomposes the physical pressure (at $x=1$) into a solvable boundary value and a streaming integral,
\begin{equation}
\label{eq:guerra_ftc_def}
\mathcal A_{N,L}(1)=\mathcal A_{N,L}(0)+\int_0^1 dx\,\frac{d}{dx}\mathcal A_{N,L}(x),
\end{equation}
where the derivative $\frac{d}{dx}\mathcal A$ is evaluated by Gaussian integration by parts via the Nishimori--Guerra identity.

The interpolating action is assembled from several independent families of i.i.d.\ $\mathcal N(0,1)$ auxiliary variables: site-indexed fields $\{\varphi_{i,l}\}_{i,l}$ and $\{\iota_{i,l}\}_{i,l}$, pattern-indexed fields $\{\theta_\mu^l\}_{l,\mu\ge2}$, and mode-indexed fields $\{\delta_{\mu}^{l}\}_{\mu\ge2,l}$ and $\{\bar\delta_{\mu}^{l}\}_{\mu\ge2,l}$. All of these are mutually independent and independent of the dataset, of $\{\lambda_{i,l}^{\mu}\}$, and of $\{\tilde\lambda_{i,l}^{\mu}\}_{i,l,\mu\ge2}$---the latter being the independent copy of $\lambda_{i,l}^\mu$ employed exclusively in the dreaming sector. Paired with these random fields are the variational parameters $\{\psi_l^{(m)}\}_{l,m=1}^{L}$ controlling the LAM signal sector, the layer-specific constants $B_l,C_l,D_l,E_l,F_l,I_l\in\mathbb R$ governing the dreaming sector, and $C_{u}, C_{u^{\dagger}}, F_{u}, F_{u^{\dagger}}\in\mathbb R$ for the LAM auxiliary-field sector. Expressed through $k_{i,l}$ and $\eta_l$ as in~\eqref{eq:k_def} and in the main text, the full interpolating action reads
\begin{equation}
\label{eq:Action_interp_def_final}
{
\begin{aligned}
\mathcal S_{x}
&:= x\,\beta\sum_{1\le l<m\le L}g_{lm}\,p_l^{1}(\bm s)p_m^{1}(\bm s)
+(1-x)\,N\sum_{1\le l<m\le L}\left( \psi_l^{(m)}\,\hat m_l^{1}(\bm s) + \psi_m^{(l)}\,\hat m_m^{1}(\bm s) \right)
\\
&\quad +\sqrt{\frac{x\beta}{2N}}\sum_{\mu=2}^{K}\sum_{l=1}^{L}\sum_{i=1}^{N}\lambda_{i,l}^{\mu}s_{i,l}\,u_{\mu}^{l}
+ \sqrt{1-x}\sum_{l=1}^L I_l \sum_{i=1}^N \iota_{i,l} s_{i,l}
\\
&\quad +(1-x)\sum_{\mu=2}^{K}\sum_{l=1}^{L}\Big[ C_{u}(u_{\mu}^{l})^{2} + C_{u^{\dagger}}({u_{\mu}^{l}}^{\dagger})^{2}\Big]
+ \sqrt{1-x}\sum_{\mu=2}^{K}\sum_{l=1}^{L}\Big[F_{u}\,\delta_{\mu}^{l}\,u_{\mu}^{l} + F_{u^{\dagger}}\,\bar\delta_{\mu}^{l}\,{u_{\mu}^{l}}^{\dagger}\Big]
\\
&\quad +\sum_{l=1}^{L}\Bigg[
x\,\frac{\beta N(1+t_l)}{2\Gamma_l}\eta_l^{2}
-\frac{B_l(1-x)}{2(1+t_l)}\sum_{\mu=2}^{K}(z_\mu^l)^{2}
+\sqrt{\frac{x\beta}{N}}\sum_{\mu=2}^{K}\sum_{i=1}^{N}\tilde\lambda_{i,l}^{\mu}z_\mu^lk_{i,l}
\\
&\qquad\qquad\qquad
+(1-x)C_l\sum_{i=1}^{N}k_{i,l}^{2}
+D_l\sqrt{1-x}\sum_{i=1}^{N}\varphi_{i,l}k_{i,l}
+(1-x)E_l N\eta_l
+F_l\sqrt{1-x}\sum_{\mu=2}^{K}\theta_\mu^lz_\mu^l
\Bigg].
\end{aligned}
}
\end{equation}

At $x=1$ all trial contributions vanish and the partition function collapses to the physical HS-decoupled form~\eqref{eq:Z_final_def} under the Gaussian replacement~\eqref{eq:gauss_approx_mu_ge2_def}. At $x=0$, conversely, the Gaussian couplings to the quenched patterns disappear and the remaining action splits into independent site- and mode-factors, rendering the system one-body and exactly soluble. The boundary value at $x=0$ thus supplies the Cauchy datum for the sum rule~\eqref{eq:guerra_ftc_def}.

\section{One-body calculations at $x=0$}
\label{app:one_body}

At the endpoint $x=0$ of the Guerra interpolation, the couplings mediated by the quenched disorder switch off entirely and the partition function reduces to a product of elementary Gaussian and spin-trace factors---the solvable Cauchy datum for the sum rule~\eqref{eq:guerra_ftc_def}. Evaluating the interpolating action~\eqref{eq:Action_interp_def_final} at $x=0$ yields

{
\begin{equation}
\label{eq:Z_s0_explicit}
\begin{aligned}
\mathcal Z_{N,L}(0)
&:= \sum_{\bm s}
\Bigg[\prod_{l=1}^L \int D\phi_l\Bigg]
\Bigg[\prod_{l=1}^L \int Dz_l\Bigg]
\int Duu^\dagger\;
\exp\Big( \mathcal S_{0}\Big),
\end{aligned}
\end{equation}
}

{
\begin{equation}
\label{eq:Action_s0_explicit}
\begin{aligned}
\mathcal S_{0}
&=
N\sum_{1\le l<m\le L}\Big( \psi_l^{(m)}\,\hat m_l^{1}(\bm s) + \psi_m^{(l)}\,\hat m_m^{1}(\bm s) \Big)
+\sum_{l=1}^L I_l \sum_{i=1}^N \iota_{i,l}\, s_{i,l}
\\
&\quad
+\sum_{\mu=2}^{K}\sum_{l=1}^{L}\Big[
C_{u}(u_{\mu}^{l})^{2}
+ C_{u^{\dagger}}({u_{\mu}^{l}}^{\dagger})^{2}
+ F_{u}\delta_{\mu}^{l}u_{\mu}^{l}
+ F_{u^{\dagger}}\bar\delta_{\mu}^{l}{u_{\mu}^{l}}^{\dagger}
\Big]
\\
&\quad
+\sum_{l=1}^{L}\Bigg[
-\frac{B_l}{2(1+t_l)}\sum_{\mu=2}^{K}(z_\mu^l)^{2}
+ C_l\sum_{i=1}^{N}k_{i,l}^{2}
+ D_l\sum_{i=1}^{N}\varphi_{i,l}\,k_{i,l}
+ E_l N\eta_l
+ F_l\sum_{\mu=2}^{K}\theta_\mu^l\,z_\mu^l
\Bigg].
\end{aligned}
\end{equation}
}

\noindent
The complexified multi-spin $k_{i,l}$, the empirical overlap $\eta_l$, and the Mattis magnetisation $\hat m_l^1(\bm s)$ retain their definitions from the interacting theory:
\[
k_{i,l}=s_{i,l}+ i\sqrt{\frac{t_{l}}{\beta(1+t_{l})}}\phi_{i,l},
\qquad
\eta_l=\frac{1}{NM}\sum_{i=1}^{N}\sum_{k=1}^{M} \eta_{i,l}^{1,k}\,k_{i,l},
\qquad
\hat m_l^1(\bm s)=\frac1N\sum_{i=1}^N \hat\xi_{i,l}^1 s_{i,l}.
\]

\noindent
At $x=0$ the three families of auxiliary variables---the complex LAM fields $(u,u^\dagger)$, the dreaming-noise fields $z$, and the spin--site-noise pair $(s,\phi)$---couple only to their own independent Gaussian sources. The partition function therefore factorises:

{
\begin{equation}
\label{eq:Z_s0_factorized}
{
\mathcal Z_{N,L}(0)
=
\mathcal Z^{(0)}_{\mathrm{LAM}}
\;\mathcal Z^{(0)}_{\mathrm{Dream}}
\;\mathcal Z^{(0)}_{s-\phi}
}
\end{equation}
}

{
\begin{equation}
\label{eq:Z0_3AM_def}
{
\mathcal Z^{(0)}_{\mathrm{LAM}}
:=
\int Duu^\dagger\;
\exp\Bigg\{
\sum_{\mu=2}^{K}\sum_{l=1}^{L}\Big[
C_{u}(u_{\mu}^{l})^{2}
+ C_{u^{\dagger}}({u_{\mu}^{l}}^{\dagger})^{2}
+ F_{u}\delta_{\mu}^{l}u_{\mu}^{l}
+ F_{u^{\dagger}}\bar\delta_{\mu}^{l}{u_{\mu}^{l}}^{\dagger}
\Big]
\Bigg\}.
}
\end{equation}
}

\noindent
The dreaming-noise sector gathers the pattern-indexed Gaussian fields $z_\mu^l$, each carrying the layer-specific variance $(1+t_l)$:

{
\begin{equation}
\label{eq:Z0_Dream_def}
{
\mathcal Z^{(0)}_{\mathrm{Dream}}
:=
\prod_{l=1}^{L}\int Dz_{l}\;
\exp\Bigg\{
-\frac{B_l}{2(1+t_l)}\sum_{\mu=2}^{K}(z_\mu^l)^{2}
+ F_l\sum_{\mu=2}^{K}\theta_\mu^l\,z_\mu^l
\Bigg\}.
}
\end{equation}
}

\noindent
The local $(s,\phi)$ sector collects the spin traces together with the site-indexed Gaussian fields that encode the dreaming-kernel dressing:

{
\begin{equation}
\begin{aligned}
\label{eq:Z0_sphi_def}
\mathcal Z^{(0)}_{s-\phi}
 :=
&\sum_{\bm s}
\Bigg[\prod_{l=1}^L \int D\phi_l\Bigg]
\exp\Bigg\{
N\sum_{1\le l<m\le L}\Big( \psi_l^{(m)}\hat m_l^{1} + \psi_m^{(l)}\hat m_m^{1} \Big) + \\
&+\sum_{l=1}^L\Big[
I_l\sum_{i=1}^N \iota_{i,l}s_{i,l}
+ C_l\sum_{i=1}^{N}k_{i,l}^{2}
+ D_l\sum_{i=1}^{N}\varphi_{i,l}k_{i,l}
+ E_l N\eta_l
\Big]
\Bigg\}.
\end{aligned}
\end{equation}
}

\noindent
Because the logarithm maps products to sums, the quenched pressure at $x=0$ decomposes additively:

{\footnotesize
\begin{equation}
\label{eq:A_s0_split}
\mathcal A_{N,L}(0)
=\frac1N\E\log\mathcal Z^{(0)}_{\mathrm{LAM}}
+\frac1N\E\log\mathcal Z^{(0)}_{\mathrm{Dream}}
+\frac1N\E\log\mathcal Z^{(0)}_{s-\phi}.
\end{equation}
}

\subsection{Dreaming-noise and local sectors at $x=0$}\label{sec:onebody_x0}

The two sectors evaluated below---the dreaming-noise Gaussian partition function $\mathcal Z^{\mathrm{Dream}\text{-}z}$ and the local $(s,\phi)$ factor $\mathcal Z^{s,\phi}$---share the functional structure of their counterparts in the classical Hebbian dreaming theory. External signal or clamping fields could be reinstated by including the appropriate linear Mattis sources; they are set to zero here, in line with the protocol adopted throughout.

\subsubsection{Dreaming-noise Gaussian sector $\mathcal Z^{\mathrm{Dream}\text{-}z}$}\label{subsec:Z_dream_z}

The Gaussian measure governing each layer $l$ is
\begin{equation}
\label{eq:Dz_measure}
Dz_{l} := \prod_{\mu=2}^{K}\frac{dz_\mu^l}{\sqrt{2\pi}}\,
\exp\!\Big(-\frac{1}{2(1+t_{l})}\sum_{\mu=2}^{K}(z_\mu^l)^2\Big),
\end{equation}
with $\{\theta_\mu^l\}$ i.i.d.\ standard normals. At $x=0$ the dreaming-noise partition function reduces to
\begin{equation}
\label{eq:Z0_Dream_def_}
\mathcal Z^{\mathrm{Dream}\text{-}z}(0)
:= \prod_{l=1}^{L}\int Dz_{l}\;
\exp\Bigg\{ -\frac{B_{l}}{2(1+t_{l})}\sum_{\mu=2}^{K} (z_\mu^l)^2
+ F_{l}\sum_{\mu=2}^{K}\theta_\mu^l\,z_\mu^l\Bigg\}.
\end{equation}
Because the integrand factorises over the mode index $\mu\ge 2$, the entire calculation reduces to a product of independent one-dimensional Gaussian integrals. Combining the quadratic pieces in~\eqref{eq:Dz_measure} and~\eqref{eq:Z0_Dream_def_}, the effective single-mode exponent reads
\begin{equation}
\label{eq:dream_z_single}
-\frac{1}{2(1+t_{l})}z^2 -\frac{B_{l}}{2(1+t_{l})}z^2 + F_{l}\theta_\mu^lz
= -\frac{1+B_{l}}{2(1+t_{l})}z^2 + F_{l}\theta_\mu^lz.
\end{equation}
Completing the square and performing the resulting Gaussian integral yields
\begin{equation}
\label{eq:gauss_identity_dream}
\int\frac{dz}{\sqrt{2\pi}}\exp\Big(-\frac{1+B_{l}}{2(1+t_{l})}z^2 + F_{l}\theta\,z\Big)
=\sqrt{\frac{1+t_{l}}{1+B_{l}}}\,\exp\Big(\frac{(1+t_{l})F_{l}^2}{2(1+B_{l})}\,\theta^2\Big).
\end{equation}
Taking the product over $\mu=2,\dots,K$ delivers the closed-form expression
\begin{equation}
\label{eq:Z0_Dream_closed_}
\mathcal Z^{\mathrm{Dream}\text{-}z}(0)
= \prod_{l=1}^{L}
\Big(\sqrt{\frac{1+t_{l}}{1+B_{l}}}\Big)^{K-1}
\exp\Bigg(\frac{(1+t_{l})F_{l}^2}{2(1+B_{l})}\sum_{\mu=2}^{K}(\theta_\mu^l)^2\Bigg).
\end{equation}
Averaging over $\theta$ (using $\E[(\theta_\mu^l)^2]=1$) and invoking the thermodynamic scaling $(K-1)/N\to\alpha$, we extract the quenched-pressure contribution
\begin{equation}
\label{eq:A0_Dream_}
\frac{1}{N}\,\E\log \mathcal Z^{\mathrm{Dream}\text{-}z}(0)
=\sum_{l=1}^{L}\Bigg[\frac{\alpha}{2}\log\Big(\frac{1+t_{l}}{1+B_{l}}\Big)
+ \alpha\,\frac{(1+t_{l})F_{l}^2}{2(1+B_{l})}\Bigg].
\end{equation}

\subsubsection{Local $(s,\phi)$ sector $\mathcal Z^{s,\phi}$}\label{subsec:Z_sphi}

The last one-body factor collects the spin sums and the site-indexed Gaussian fluctuations. Writing $D\phi_{l}$ for the standard Gaussian measure over the analogue variables $\{\phi_{i,l}\}_{i=1}^N$,
\begin{equation}
\label{eq:Dphi_measure}
D\phi_{l}:=\prod_{i=1}^{N}\frac{d\phi_{i,l}}{\sqrt{2\pi}}\,e^{-\frac12\phi_{i,l}^2}
\end{equation}
and recalling the complexified local fields from the dreaming notation,
\begin{equation}
\label{eq:k_def_}
 k_{i,l}:= s_{i,l}+ i\,a_{l}\,\phi_{i,l},
 \qquad a_{l}:=\sqrt{\frac{t_{l}}{\beta(1+t_{l})}}
\end{equation}
the $(s,\phi)$ partition function at $x=0$ takes the form
\begin{equation}
\label{eq:Z0_sphi_start_}
\begin{aligned}
\mathcal Z^{s,\phi}(0)
:= &\sum_{\bm s}\prod_{l=1}^{L}\int D\phi_{l}\;
\exp\Bigg\{\,N\sum_{1\le l<m\le L}\Big(\psi_{l}^{(m)}\hat m_{l}^{1}(\bm s)+\psi_{m}^{(l)}\hat m_{m}^{1}(\bm s)\Big)+\\
&+\sum_{l=1}^{L}\sum_{i=1}^{N}\Big[ I_{l}\,\iota_{i,l}\,s_{i,l}
+ C_{l}k_{i,l}^2 + D_{l}\,\varphi_{i,l}\,k_{i,l} + E_{l}\,\hat\xi_{i,l}^{1}\,k_{i,l}\Big]\Bigg\}.
\end{aligned}
\end{equation}

The inter-layer Mattis coupling can be absorbed into a site-local effective field by inserting $\hat m_{l}^{1}(\bm s)=\frac1N\sum_{i=1}^{N}\hat\xi_{i,l}^{1}s_{i,l}$:
\begin{equation}
\label{eq:psi_reduce}
N\sum_{1\le l<m\le L}\Big(\psi_{l}^{(m)}\hat m_{l}^{1}+\psi_{m}^{(l)}\hat m_{m}^{1}\Big)
=\sum_{l=1}^{L}\sum_{i=1}^{N}\Big(\hat\xi_{i,l}^{1}\,h_{l}\Big)s_{i,l},
\qquad h_{l}:=\sum_{m\neq l}\psi_{l}^{(m)}.
\end{equation}
Once this reduction is performed, the entire exponent becomes a sum of independent site-indexed terms and the partition function factorises:
\begin{equation}
\label{eq:Z0_sphi_fact_}
\mathcal Z^{s,\phi}(0)=\prod_{l=1}^{L}\prod_{i=1}^{N}\mathcal Z_{i,l},
\end{equation}
where each single-site factor has the form
\begin{equation}
\label{eq:Z_il_def_}
\mathcal Z_{i,l}
:= \sum_{s=\pm 1}\int\frac{d\phi}{\sqrt{2\pi}}e^{-\phi^2/2}
\exp\Big\{ C_{l}k^2 + k\,\Omega_{i,l} + s\,h^{\mathrm{ext}}_{i,l}\Big\},
\end{equation}
with the shorthand
\begin{equation}
\label{eq:Omega_hext_def}
\Omega_{i,l}:= D_{l}\,\varphi_{i,l}+E_{l}\,\hat\xi_{i,l}^{1},
\qquad
h^{\mathrm{ext}}_{i,l}:= I_{l}\,\iota_{i,l}+\hat\xi_{i,l}^{1}\,h_{l},
\qquad k=s+i a_{l}\phi.
\end{equation}

The $\phi$-integral in~\eqref{eq:Z_il_def_} is Gaussian and can be carried out in closed form. Expanding $k^2=(s+i a_{l}\phi)^2=1-a_{l}^2\phi^2+2 i a_{l}s\phi$, we collect the exponent into terms quadratic, linear, and independent of $\phi$:
\begin{equation}
\label{eq:expansion_phi}
-\frac12\phi^2 + C_{l}(1-a_{l}^2\phi^2+2 i a_{l}s\phi) + (s+i a_{l}\phi)\Omega_{i,l} + s h^{\mathrm{ext}}_{i,l}.
\end{equation}
The quadratic, linear, and $\phi$-independent contributions read
\begin{equation}
\label{eq:quad_form_phi}
-\frac12\Big(1+2C_{l}a_{l}^2\Big)\phi^2
+ i a_{l}\Big(2C_{l}s+\Omega_{i,l}\Big)\phi
+\Big(C_{l}+ s(\Omega_{i,l}+h^{\mathrm{ext}}_{i,l})\Big).
\end{equation}
Introducing the effective precision and its inverse,
\begin{equation}
\label{eq:Delta_def}
A_{l}^{\mathrm{eff}}:=1+2C_{l}a_{l}^2,\qquad \Delta_{l}:=(A_{l}^{\mathrm{eff}})^{-1}=\frac{1}{1+2C_{l}a_{l}^2},
\end{equation}
and invoking the standard Gaussian identity
\begin{equation}
\label{eq:gauss_identity_phi}
\int\frac{d\phi}{\sqrt{2\pi}}\exp\Big(-\tfrac12 A\phi^2+J\phi\Big)=\frac{1}{\sqrt{A}}\exp\Big(\frac{J^2}{2A}\Big),\qquad (\Re A>0),
\end{equation}
with $A=A_{l}^{\mathrm{eff}}$ and $J=i a_{l}(2C_{l}s+\Omega_{i,l})$, the single-site partition function reduces to
\begin{equation}
\label{eq:Z_il_after_phi}
\mathcal Z_{i,l}
= \sum_{s=\pm1} \sqrt{\Delta_{l}}\;
\exp\Bigg\{ C_{l}+ s(\Omega_{i,l}+h^{\mathrm{ext}}_{i,l})
-\frac{\Delta_{l}a_{l}^2}{2}\big(2C_{l}s+\Omega_{i,l}\big)^2\Bigg\}.
\end{equation}
Expanding the squared term and recalling $s^2=1$ for Ising spins,
\begin{align}
\label{eq:s_terms_isolation}
-\frac{\Delta_{l}a_{l}^2}{2}(2C_{l}s+\Omega)^2
&= -\frac{\Delta_{l}a_{l}^2}{2}\Big(4C_{l}^2 s^2+\Omega^2+4C_{l}\Omega\,s\Big)
= -\frac{\Delta_{l}a_{l}^2}{2}\Big(4C_{l}^2+\Omega^2+4C_{l}\Omega\,s\Big),
\end{align}
the effective coefficient of $s$ in the exponent of~\eqref{eq:Z_il_after_phi} simplifies via the identity $1-2\Delta_{l}a_{l}^2 C_{l}=\Delta_{l}$, which follows directly from~\eqref{eq:Delta_def}:
\begin{equation}
\label{eq:s_coeff_simplify}
\Omega_{i,l}-2\Delta_{l}a_{l}^2 C_{l}\,\Omega_{i,l}
=\Omega_{i,l}\Big(1-2\Delta_{l}a_{l}^2 C_{l}\Big)
=\Delta_{l}\,\Omega_{i,l}.
\end{equation}
Carrying out the remaining binary spin trace $\sum_{s=\pm1}$ yields the compact closed form
\begin{equation}
\label{eq:Z_il_closed_}
\mathcal Z_{i,l}
=2\sqrt{\Delta_{l}}\,
\exp\Bigg\{ C_{l}-\frac{\Delta_{l}a_{l}^2}{2}\Big(4C_{l}^2+\Omega_{i,l}^2\Big)\Bigg\}
\cosh\Big(h^{\mathrm{ext}}_{i,l}+\Delta_{l}\Omega_{i,l}\Big).
\end{equation}
By site exchangeability, the quenched pressure of this sector follows from~\eqref{eq:Z0_sphi_fact_}:
\begin{equation}
\label{eq:A0_sphi_}
\frac{1}{N}\,\E\log \mathcal Z^{s,\phi}(0)
=\sum_{l=1}^{L}\E\log \mathcal Z_{1,l}
\end{equation}
Spelling this out gives
\begin{equation}
\label{eq:A0_sphi_explicit_}
\begin{aligned}
\frac{1}{N}\,\E\log \mathcal Z^{s,\phi}(0)
&=\sum_{l=1}^{L}\Bigg[
\log 2 + \frac12\log\Delta_{l}+ C_{l}
-\frac{\Delta_{l}a_{l}^2}{2}\Big(4C_{l}^2+\E[\Omega_{1,l}^2]\Big)
+\E\log\cosh\Big(h^{\mathrm{ext}}_{1,l}+\Delta_{l}\Omega_{1,l}\Big)
\Bigg].
\end{aligned}
\end{equation}
Here the expectation $\E$ runs over the auxiliary Gaussians $(\iota,\varphi)$ and, through $\hat\xi^1$, over the dataset.

\subsection{LAM sector at $x=0$}\label{sec:onebody_tam_x0}

The third and final one-body factor at $x=0$ is the LAM Gaussian sector, which governs the non-condensed complex auxiliary fields. Decomposing each complex field into real and imaginary parts reduces the problem to a $2L$-dimensional real Gaussian integral.

For each non-condensed mode $\mu\ge 2$, the complex HS fields $u_\mu^l\in\Sigma$ ($l=1,\dots,L$) are stacked into the real vector
\begin{equation}
\label{eq:x_mu_def_}
\bm x_\mu:=\big(\Re u_\mu^1,\Im u_\mu^1,\dots,\Re u_\mu^L,\Im u_\mu^L\big)^\top\in\mathbb R^{2L}.
\end{equation}
At the replica-symmetric level the Gaussian sector is parametrised by the overlaps $H_l:=\overline{\langle h_{12}^{(l)}\rangle_x}$.
The covariance is the $2L\times 2L$ matrix $\mathbf C(H)$ with $2\times2$ diagonal blocks
\begin{equation}
\label{eq:C_blocks_}
\mathbf C(H)=\big(\mathbf C^{lm}\big)_{l,m=1}^L,\qquad
\mathbf C^{ll}=
\begin{pmatrix}
1-\tfrac{\beta}{2}(1-H_l) & 0\\[2pt]
0 & 1+\tfrac{\beta}{2}(1-H_l)
\end{pmatrix},
\end{equation}
and off-diagonal ($l\neq m$) blocks encoding the inter-layer coupling:
\begin{equation}
\label{eq:C_off_blocks_}
\mathbf C^{lm}=g_{lm}
\begin{pmatrix}
1&-i\\
-i&-1
\end{pmatrix},
\qquad g_{lm}=g_{ml},\qquad g_{ll}=0.
\end{equation}

As dictated by the hetero-associative architecture, each visible layer couples to a single Gaussian degree of freedom per conjugate pair rather than to both. This constraint is enforced by the i.i.d.\ $\delta_\mu^l\sim\mathcal N(0,1)$ together with the $2L$-source vector
\begin{equation}
\label{eq:b_mu_}
\bm b_\mu:=\sqrt{\frac{\beta}{2}}
\big(\sqrt{H_1}\,\delta_\mu^1,\; i\sqrt{H_1}\,\delta_\mu^1,\;\dots,\;\sqrt{H_L}\,\delta_\mu^L,\; i\sqrt{H_L}\,\delta_\mu^L\big)^\top.
\end{equation}

Since the modes decouple at $x=0$, the LAM partition function factorises as
\begin{equation}
\label{eq:Z_3AM_factor_mu_}
\mathcal Z^{\mathrm{LAM}}(0)=\prod_{\mu=2}^K \mathcal Z^{\mathrm{LAM}}_\mu(0),
\qquad
\mathcal Z^{\mathrm{LAM}}_\mu(0)
:=\int_{\mathbb R^{2L}}\frac{d^{2L}x}{(2\pi)^L}
\exp\Big(-\tfrac12 \bm x^\top \mathbf C(H)\bm x+\bm b_\mu^\top\bm x\Big).
\end{equation}
Provided that $\Re\,\mathbf C(H)\succ 0$, the standard $2L$-dimensional Gaussian identity yields
\begin{equation}
\label{eq:Z_3AM_mu_closed_}
\mathcal Z^{\mathrm{LAM}}_\mu(0)
=(\det \mathbf C(H))^{-1/2}
\exp\Big(\tfrac12\,\bm b_\mu^\top \mathbf C(H)^{-1}\bm b_\mu\Big).
\end{equation}
Summing over $\mu$ and passing to the intensive quenched pressure, we obtain
\begin{equation}
\label{eq:A0_3AM_final_}
\frac1N\,\E\log\mathcal Z^{\mathrm{LAM}}(0)
=\alpha\Bigg[
-\frac12\log\det \mathbf C(H)
+\frac12\,\E_{\bm\delta}\big[\bm b_\mu^\top \mathbf C(H)^{-1}\bm b_\mu\big]
\Bigg].
\end{equation}
The quadratic correction decomposes as
\begin{equation}
\label{eq:EbCb_}
\E_{\bm\delta}\big[\bm b_\mu^\top \mathbf C(H)^{-1}\bm b_\mu\big]
=\frac{\beta}{2}\sum_{l=1}^L H_l\,\mathcal Q_l(H),
\qquad
\mathcal Q_l(H):=\bm v_l^\top \mathbf C(H)^{-1}\bm v_l,
\end{equation}
where $\bm v_l\in\Sigma^{2L}$ is the $2L$-vector that selects the $(\Re,\Im)$ components of layer $l$ along $(1,i)$.

\begin{remark}[Exact correspondence with the classical Hebbian-dreaming one-body]
\label{rem:onebody_exact_dreaming_match}
As a consistency check, we verify that the DLAM one-body term at $x=0$ reproduces, term by term, the standard Hebbian-dreaming one-body formula of~\cite{AlemannoETAL2023small} whenever the external sources are switched off ($\psi\equiv 0$, $I_l=0$); the two expressions then coincide up to a relabelling of the trial parameters.

\emph{Local $(s,\phi)$ sector.}
In the sourceless protocol $h^{\mathrm{ext}}_{i,l}=0$, so that $\Theta''_{i,l}=\Delta_l\Omega_{i,l}$ with $\Omega_{i,l}:=D_l\,\varphi_{i,l}+E_l\,\hat\xi_{i,l}^{1}$, and the site-factorised contribution~\eqref{eq:Z_il_closed_} at $x=0$ becomes
\begin{equation}
\label{eq:Asphi_rem}
\frac{1}{N}\,\E\log \mathcal Z^{s,\phi}(0)
=\sum_{l=1}^{L}\Bigg[
\log 2+\frac12\log\Delta_l + C_l
-\frac{\Delta_l a_l^2}{2}\Big(4C_l^2+\E[\Omega_{1,l}^2]\Big)
+\E\log\cosh\!\Big(\Delta_l\Omega_{1,l}\Big)
\Bigg],
\end{equation}
where $a_l^2:=t_l/[\beta(1+t_l)]$ and $\Delta_l:=1/(1+2C_l a_l^2)=\beta(1+t_l)/[\beta(1+t_l)+2t_lC_l]$. Identifying the prefactors
\begin{equation}
\frac12\log\Delta_l
=-\frac12\log\!\Big(1+\frac{2t_l C_l}{\beta(1+t_l)}\Big),
\qquad
\frac{\Delta_l a_l^2}{2}
=\frac{1}{2}\,\frac{t_l}{2t_l C_l+\beta(1+t_l)},
\end{equation}
we recognise the two characteristic classical dreaming terms. Restricted to a single layer,~\eqref{eq:Asphi_rem} reduces to
\begin{equation}
\label{eq:Asphi_dreaming_form}
\log 2 + C
-\frac12\log\!\Big(1+\frac{2tC}{\beta(1+t)}\Big)
-\frac12\,\frac{t}{2tC+\beta(1+t)}\Big(4C^2+\E[\Omega^2]\Big)
+\E\log\cosh(\Delta\Omega),
\end{equation}
which is precisely the classical dreaming one-body structure. The variance $\E[\Omega_{1,l}^2]=D_l^2+E_l^2\,\Gamma_l$ (since $\E[\varphi_{1,l}^2]=1$ and $\E[(\hat\xi_{1,l}^1)^2]=\Gamma_l$) matches the classical contribution $D^2+E_{\mathrm{app}}^{\,2}\,r^2/\Gamma$ under the identification
\begin{equation}
\label{eq:E_identification}
E_l \;=\; E_{l,\mathrm{app}}\;\frac{r_l}{\Gamma_l},
\end{equation}
which leaves $E_l^2\,\Gamma_l=E_{l,\mathrm{app}}^{\,2}\,r_l^2/\Gamma_l$ invariant.

\emph{Gaussian dreaming-noise sector.}
The $z$-sector at $x=0$ (cf.~\eqref{eq:A0_Dream_}) evaluates to
\begin{equation}
\label{eq:Adreamz_rem}
\frac{1}{N}\,\E\log \mathcal Z^{\mathrm{Dream}\text{-}z}(0)
=\sum_{l=1}^{L}\Bigg[
\frac{\alpha}{2}\log\!\Big(\frac{1+t_l}{1+B_l}\Big)
+\frac{\alpha}{2}\,\frac{F_l^2(1+t_l)}{1+B_l}
\Bigg].
\end{equation}
The classical dreaming convention employs $(1-B_{\mathrm{app}})$ in the denominator; the two expressions agree under the sign convention
\begin{equation}
\label{eq:B_identification}
B_{l,\mathrm{app}}:=-B_l,
\end{equation}
which leaves $1-B_{l,\mathrm{app}}=1+B_l$.

Combining~\eqref{eq:Asphi_rem} and~\eqref{eq:Adreamz_rem} under the identifications~\eqref{eq:E_identification}--\eqref{eq:B_identification} recovers the classical Hebbian-dreaming one-body expression of~\cite{AlemannoETAL2023small} verbatim---including the logarithmic prefactor, the quadratic correction with denominator $2tC+\beta(1+t)$, and the $z$-sector contributions. Reactivating the LAM coupling introduces a single modification: an additional deterministic Mattis shift $\hat\xi_{i,l}^1\,h_l$ inside the effective field $h^{\mathrm{ext}}_{i,l}:=I_l\iota_{i,l}+\hat\xi_{i,l}^1\,h_l$ (with $h_l:=\sum_{m\neq l}\psi_l^{(m)}$), which displaces $\Theta''_{i,l}$ while leaving the functional form of the one-body invariant.
\end{remark}

\section{Streaming equation and trial-parameter fixing}
\label{app:streaming}

\noindent
The second ingredient of the Guerra sum rule is the streaming term $\frac{d}{dx}\mathcal A_{N,L}(x)$. Differentiating the interpolating action~\eqref{eq:Action_interp_def_final} and applying Gaussian integration by parts converts the derivative into thermal expectations of replica overlaps. Once the trial parameters are tuned so as to enforce replica-symmetric self-averaging, the streaming reduces to a deterministic expression supplemented by centred fluctuations that vanish in the thermodynamic limit.

\subsection{Gibbs state, replicas, and order parameters}

For a fixed realisation of the disorder, the partition function at interpolation parameter $x$ reads
\begin{equation}
\mathcal Z_{N,L}(x)
:=\sum_{\bm s}\Bigg[\prod_{l=1}^L\int D\phi_l\Bigg]\Bigg[\prod_{l=1}^L\int Dz_l\Bigg]\int Duu^\dagger\;\exp\big(\mathcal S_x\big)
\end{equation}
with $\mathcal S_x$ as in~\eqref{eq:Action_interp_def_final}.
The thermal average at fixed disorder and its quenched counterpart are defined by
\begin{equation}
\omega_x(F):=\frac{1}{\mathcal Z_{N,L}(x)}
\sum_{\bm s}\Bigg[\prod_{l=1}^L\int D\phi_l\Bigg]\Bigg[\prod_{l=1}^L\int Dz_l\Bigg]\int Duu^\dagger\;F\,\exp\big(\mathcal S_x\big),
\end{equation}
with $\langle \cdot\rangle_x:=\E\,\omega_x(\cdot)$ denoting the quenched bracket.
Throughout this section we make use of two independent replicas $a,b\in\{1,2\}$ drawn from $\omega_x^{\otimes 2}$.

\medskip
\noindent
The layer-resolved replica overlaps are
\begin{align}
\text{(spins)}\qquad
h_{ab}^{(l)} &:= \frac1N\sum_{i=1}^N s_{i,l}^{(a)}s_{i,l}^{(b)},
\\
\text{(dreaming $k$)}\qquad
q_{ab}^{(l)} &:= \frac1N\sum_{i=1}^N k_{i,l}^{(a)}k_{i,l}^{(b)},
\\
\text{(dreaming $z$)}\qquad
p_{ab}^{(l)} &:= \frac{1}{K-1}\sum_{\mu=2}^K z_\mu^{l,(a)}z_\mu^{l,(b)},
\\
\text{(LAM $u$)}\qquad
P_{ab}^{(l)} &:= \frac{1}{K-1}\sum_{\mu=2}^K u_{\mu}^{l,(a)}u_{\mu}^{l,(b)}.
\end{align}
The empirical observable
$\eta_l=\frac{1}{NM}\sum_{i,k}\eta_{i,l}^{1,k}\,k_{i,l}$ (cf.\ Section~2 of the main text) completes the set.
As elsewhere, $\alpha:=(K-1)/N$ is held at $\mathcal O(1)$ as $N\to\infty$.


\medskip\noindent
The workhorse identity for differentiating through Gaussian disorder is the Stein lemma: for $X\sim\mathcal N(0,1)$ and a smooth function $F$ with suitable integrability,
\begin{equation}
\E\,[X\,F(X)] = \E\,[F'(X)].
\end{equation}
When a Gaussian variable $X$ enters the interpolating action linearly as $\sqrt{x}\,c\,X\,U$, with $U$ an observable independent of $X$, the Stein lemma generates the two-replica structure
\begin{equation}
\E\big[X\,\omega_x(U)\big]
= \sqrt{x}\,c\,\E\Big[\omega_x(U^2)-\omega_x(U)^2\Big]
= \sqrt{x}\,c\,\E\Big[\omega_x^{\otimes 2}\big(U^{(1)}U^{(1)}-U^{(1)}U^{(2)}\big)\Big].
\end{equation}
Two-replica overlaps thus emerge naturally from single-replica thermal averages.

\subsection{Streaming equation: explicit computation of \texorpdfstring{$\frac{d}{dx}\mathcal A_{N,L}$}{d/dx A}}

Differentiating the quenched pressure with respect to the interpolation parameter gives
\begin{equation}
\label{eq:streaming_master_0}
\frac{d}{dx}\,\mathcal A_{N,L}(x)
=\frac{1}{N}\,\E\,\omega_x\big(\partial_x\mathcal S_x\big).
\end{equation}
This derivative splits naturally into three additive contributions, one per sector of the Hamiltonian:
\begin{equation}
\frac{d}{dx}\mathcal A_{N,L}=\mathcal D_{\mathrm{sig}}(x)+\mathcal D_{\mathrm{LAM}}(x)+\mathcal D_{\mathrm{Dream}}(x),
\end{equation}

\subsubsection{LAM signal part}

The inter-layer signal terms in~\eqref{eq:Action_interp_def_final} pit the full-strength condensed overlaps
$s\beta\sum_{l<m}p_l^1 p_m^1$ against the one-body field
$(1-s)N\sum_{l<m}[\psi_l^{(m)}\hat m_l^1+\psi_m^{(l)}\hat m_m^1]$, which absorbs the corresponding energy at $x=0$. Differentiating with respect to $x$, dividing by $N$, and rewriting
$p_l^1=\frac{\sqrt N}{\sqrt{\Gamma_l}}\,\hat m_l^1$ through~\eqref{eq:J_def}--\eqref{eq:p_def}, we obtain
\begin{equation}
\label{eq:streaming_LAM_raw}
\mathcal D_{\mathrm{sig}}(x)
=\sum_{1\le l<m\le L}\Bigg[
\frac{g_{lm}\beta}{\Gamma_{lm}}\,\big\langle \hat m_l^1\hat m_m^1\big\rangle_x
-\psi_l^{(m)}\,\big\langle \hat m_l^1\big\rangle_x
-\psi_m^{(l)}\,\big\langle \hat m_m^1\big\rangle_x
\Bigg].
\end{equation}

\subsubsection{LAM part}

The LAM sector is governed by the Gaussian coupling $\sqrt{\frac{x\beta}{2N}}\sum_{\mu\ge2,l,i}\lambda_{i,l}^\mu s_{i,l}u_\mu^l$, whose $x$-derivative produces a factor $(2\sqrt{x})^{-1}$ that the subsequent Stein integration by parts absorbs. Applying the identity to each $\lambda_{i,l}^\mu$ and rewriting the result in two-replica form yields
\begin{equation}
\label{eq:streaming_3AM_lambda}
\mathcal D^{(\lambda)}_{\mathrm{LAM}}(x)
=\frac{\alpha\beta}{4}\sum_{l=1}^L\Big[\big\langle P_{11}^{(l)}\big\rangle_x-\big\langle h_{12}^{(l)}\,P_{12}^{(l)}\big\rangle_x\Big] + \mathcal O\!\left(\frac{1}{N}\right).
\end{equation}
The reduction to two-replica overlaps proceeds as usual: integration by parts generates $\omega_x\big((s_{i,l}u_\mu^l)^2\big)-\omega_x(s_{i,l}u_\mu^l)^2$, which, upon summing over $(i,\mu)$ and dividing by $N$, yields $\langle P_{11}^{(l)}\rangle_x$ and $\langle h_{12}^{(l)}P_{12}^{(l)}\rangle_x$ respectively, up to $\mathcal O(1/N)$ corrections of diagonal and exchangeability origin.

The one-body noise field $\sqrt{1-x}\,I_l\sum_i\iota_{i,l}s_{i,l}$ stores the LAM fluctuation energy that is released as $x\to 1$. Its $x$-derivative equals $-\frac{I_l}{2\sqrt{1-x}}\sum_i\iota_{i,l}s_{i,l}$; applying the Stein lemma to $\iota_{i,l}$ then converts it into the spin-overlap variance
\begin{equation}
\label{eq:streaming_3AM_iota}
\mathcal D^{(\iota)}_{\mathrm{LAM}}(x)
=-\frac12\sum_{l=1}^L I_l^2\,\Big[1-\big\langle h_{12}^{(l)}\big\rangle_x\Big].
\end{equation}

Finally, the quadratic and linear trial couplings present in the interpolating action contribute
\begin{equation}
\label{eq:streaming_3AM_trials}
\mathcal D^{(\mathrm{trial})}_{\mathrm{LAM}}(x)
=-\alpha\sum_{l=1}^L\Big[ C_{u}\,\langle P_{11}^{(l)}\rangle_x + C_{u^\dagger}\,\langle P_{11}^{\dagger,(l)}\rangle_x\Big]
-\frac{\alpha}{2}\sum_{l=1}^L\Big[ F_{u}^2\big(\langle P_{11}^{(l)}\rangle_x-\langle P_{12}^{(l)}\rangle_x\big)
+ F_{u^\dagger}^2\big(\langle P_{11}^{\dagger,(l)}\rangle_x-\langle P_{12}^{\dagger,(l)}\rangle_x\big)\Big],
\end{equation}
with $P_{ab}^{\dagger,(l)}:=\frac{1}{K-1}\sum_{\mu\ge2}{u_\mu^{l,\dagger}}^{(a)}{u_\mu^{l,\dagger}}^{(b)}$. Adding the three contributions~\eqref{eq:streaming_3AM_lambda}--\eqref{eq:streaming_3AM_trials} produces the total LAM streaming
\begin{equation}
\label{eq:streaming_3AM_total}
\mathcal D_{\mathrm{LAM}}(x)=\mathcal D^{(\lambda)}_{\mathrm{LAM}}(x)+\mathcal D^{(\iota)}_{\mathrm{LAM}}(x)+\mathcal D^{(\mathrm{trial})}_{\mathrm{LAM}}(x).
\end{equation}

\subsubsection{Dreaming part}

Each layer contributes independently, so the dreaming streaming is a sum over $l$ of three pieces with distinct origin. The Mattis-like overlap $\eta_l$ enters the action quadratically through $x\,\frac{\beta N(1+t_l)}{2\Gamma_l}\,\eta_l^2$, whose $x$-derivative is immediate and yields
\begin{equation}
\label{eq:streaming_dream_eta2}
\mathcal D^{(\eta^2)}_{\mathrm{Dream}}(x)
=\sum_{l=1}^L\frac{\beta(1+t_l)}{2\Gamma_l}\,\big\langle \eta_l^2\big\rangle_x.
\end{equation}
The same Stein mechanism deployed for the LAM $\lambda$-coupling applies verbatim to the independent Gaussian copy $\sqrt{\frac{x\beta}{N}}\sum_{\mu\ge2,i}\tilde\lambda_{i,l}^\mu\,z_\mu^lk_{i,l}$: the family $\{\tilde\lambda\}$ shares the marginal law of $\lambda$ but is statistically independent of it, ensuring that the LAM and dreaming Gaussian sectors remain decoupled. The resulting two-replica expression reads
\begin{equation}
\label{eq:streaming_dream_lambda}
\mathcal D^{(\tilde\lambda)}_{\mathrm{Dream}}(x)
=\frac{\alpha\beta}{2}\sum_{l=1}^L\Big[\big\langle p_{11}^{(l)}q_{11}^{(l)}\big\rangle_x-\big\langle p_{12}^{(l)}q_{12}^{(l)}\big\rangle_x\Big]
+\mathcal O\!\left(\frac{1}{N}\right).
\end{equation}
The remaining one-body trial couplings, once differentiated and processed through Stein's identity acting on the Gaussian fields $\varphi$ and $\theta$, give
\begin{equation}
\label{eq:streaming_dream_trials}
\begin{aligned}
\mathcal D^{(\mathrm{trial})}_{\mathrm{Dream}}(x)
=\sum_{l=1}^L\Bigg[&
\frac{\alpha B_l}{2(1+t_l)}\,\big\langle p_{11}^{(l)}\big\rangle_x
- C_l\,\big\langle q_{11}^{(l)}\big\rangle_x
- E_l\,\big\langle \eta_l\big\rangle_x
\\
&-\frac{D_l^2}{2}\,\Big(\big\langle q_{11}^{(l)}\big\rangle_x-\big\langle q_{12}^{(l)}\big\rangle_x\Big)
-\frac{\alpha F_l^2}{2}\,\Big(\big\langle p_{11}^{(l)}\big\rangle_x-\big\langle p_{12}^{(l)}\big\rangle_x\Big)
\Bigg].
\end{aligned}
\end{equation}
Assembling the three dreaming contributions~\eqref{eq:streaming_dream_eta2}--\eqref{eq:streaming_dream_trials} delivers
\begin{equation}
\label{eq:streaming_dream_total}
\mathcal D_{\mathrm{Dream}}(x)=
\mathcal D^{(\eta^2)}_{\mathrm{Dream}}(x)+\mathcal D^{(\tilde\lambda)}_{\mathrm{Dream}}(x)+\mathcal D^{(\mathrm{trial})}_{\mathrm{Dream}}(x).
\end{equation}

\subsection{Full streaming equation}

Combining the LAM contributions~\eqref{eq:streaming_LAM_raw}--\eqref{eq:streaming_3AM_total} with the dreaming sector~\eqref{eq:streaming_dream_total}, the complete streaming takes the form
\begin{equation}
\label{eq:streaming_master_full}
{\;\frac{d}{dx}\mathcal A_{N,L}(x)
=\mathcal D_{\mathrm{sig}}(x)+\mathcal D_{\mathrm{LAM}}(x)+\mathcal D_{\mathrm{Dream}}(x)
+\mathcal O\!\left(\frac{1}{N}\right).\;}
\end{equation}

\subsection{Replica--symmetric self--averaging and exact fixing of the trial parameters}

The trial parameters introduced in the interpolating action remain free thus far. We now fix them by requiring that every streaming term involving random overlaps decompose into a deterministic RS contribution plus centred fluctuations that vanish as $N\to\infty$. Concretely, we impose the RS self-averaging scenario, in which each relevant overlap concentrates on a deterministic limit,
\begin{equation}
\label{eq:RS_concentration_assumption}
\langle \mathcal O\rangle_x \xrightarrow[N\to\infty]{} \overline{\mathcal O}(x)
\qquad\text{for}\qquad
\mathcal O\in\Big\{\hat m_l^1,\;h_{12}^{(l)},\;p_{ab}^{(l)},\;q_{ab}^{(l)},\;P_{ab}^{(l)},\;\eta_l\Big\}.
\end{equation}
For each layer $l$ we introduce the corresponding RS scalars:

\begin{equation}
\begin{aligned}
\hat M_l &:= \overline{\langle \hat m_l^1\rangle_x},
&& H_l &:= \overline{\langle h_{12}^{(l)}\rangle_x},
&& \bar\eta_l &:= \overline{\langle \eta_l\rangle_x},
\\
\bar q_l &:= \overline{\langle q_{12}^{(l)}\rangle_x},
&& \bar Q_l &:= \overline{\langle q_{11}^{(l)}\rangle_x},
&& \bar p_l &:= \overline{\langle p_{12}^{(l)}\rangle_x},
\\
\bar P_l &:= \overline{\langle p_{11}^{(l)}\rangle_x},
&& \bar\Pi_l &:= \overline{\langle P_{12}^{(l)}\rangle_x},
&& \overline{\mathcal{P}}_l &:= \overline{\langle P_{11}^{(l)}\rangle_x}.
\end{aligned}
\end{equation}

\medskip
\noindent
Within each sector, the trial parameters are calibrated to absorb the deterministic cost of the streaming, leaving only centred fluctuations---that is, differences between the empirical overlaps and their RS limits.

\paragraph{LAM signal fields.}
Inspecting~\eqref{eq:streaming_LAM_raw}, the requirement that the deterministic component vanish for each inter-layer pair $(l,m)$ immediately gives
\begin{equation}
\psi_l^{(m)}=\frac{g_{lm}\beta}{\Gamma_{lm}}\,\hat M_m,
\qquad
\psi_m^{(l)}=\frac{g_{lm}\beta}{\Gamma_{lm}}\,\hat M_l.
\end{equation}

\paragraph{LAM parameters.}
Matching the LAM streaming~\eqref{eq:streaming_3AM_lambda}--\eqref{eq:streaming_3AM_trials} under the RS ansatz is more delicate, since the trial sector accommodates both the physical channel $u_\mu^l$ and its conjugate $u_\mu^{l,\dagger}$. The trial coefficients must be chosen so that the deterministic part of the streaming reproduces the standard ``RS plus centred fluctuations'' structure; any residual deterministic term left uncompensated by the interaction would unbalance the decomposition. By construction of the interpolating action, the $\lambda$-term couples the spins only to $u_\mu^l$ and not to its conjugate, so that the integration-by-parts contribution~\eqref{eq:streaming_3AM_lambda} generates overlaps of type $P_{ab}^{(l)}$ but no analogue producing $P_{ab}^{\dagger,(l)}$. Any deterministic contribution from the conjugate trial sector,
$-\alpha\,C_{u^\dagger}\langle P_{11}^{\dagger,(l)}\rangle_x-\frac{\alpha}{2}F_{u^\dagger}^2(\langle P_{11}^{\dagger,(l)}\rangle_x-\langle P_{12}^{\dagger,(l)}\rangle_x)$,
would therefore lack a counterpart to cancel against and would spoil the structure. RS fixing accordingly enforces
\begin{equation}
{\; C_{u^\dagger}=0,\qquad F_{u^\dagger}=0,\;}
\end{equation}
in agreement with the supervised LAM convention, where the conjugate coefficients associated with the non-excited linear channel are likewise set to zero.

The remaining LAM coefficients $F_u^2$ and $C_u$ are determined by deterministic cancellation. Under RS self-averaging one has $\langle P_{11}^{(l)}\rangle_x=\overline{\mathcal P}_l+\text{fluct.}$, $\langle P_{12}^{(l)}\rangle_x=\bar\Pi_l+\text{fluct.}$, and $\langle h_{12}^{(l)}P_{12}^{(l)}\rangle_x=H_l\bar\Pi_l+\text{fluct.}$. Gathering the deterministic parts of~\eqref{eq:streaming_3AM_lambda} and~\eqref{eq:streaming_3AM_trials} layer by layer leaves
\[
\frac{\alpha\beta}{4}\overline{\mathcal P}_l-\frac{\alpha\beta}{4}H_l\bar\Pi_l
-\alpha C_{u}\,\overline{\mathcal P}_l
-\frac{\alpha}{2}F_{u}^2(\overline{\mathcal P}_l-\bar\Pi_l).
\]
Demanding cancellation for arbitrary $(\overline{\mathcal P}_l,\bar\Pi_l)$ produces the two scalar conditions
\[
\frac{\beta}{4}-C_{u}-\frac{F_{u}^2}{2}=0,
\qquad
\frac{F_{u}^2}{2}-\frac{\beta}{4}H_l=0,
\]
hence
\begin{equation}
F_{u}^2=\frac{\beta}{2}\, H_l,\qquad
C_{u}=\frac{\beta}{4}(1- H_l).
\end{equation}

\smallskip
\noindent
The final LAM contribution is the $\iota$-term; matching~\eqref{eq:streaming_3AM_iota} against the RS noise level fixes
\begin{equation}
I_l^2=\frac{\alpha\beta}{2}\,\bar\Pi_l.
\end{equation}

\paragraph{Dreaming parameters.}
Applying the same deterministic-cancellation strategy to~\eqref{eq:streaming_dream_eta2}--\eqref{eq:streaming_dream_trials} yields
\begin{equation}
\begin{aligned}
B_l &= -\beta(1+t_l)\,(\bar Q_l-\bar q_l),
&
C_l &= \frac{\alpha\beta}{2}\,(\bar P_l-\bar p_l),
&
D_l^2 &= \alpha\beta\,\bar p_l,\\
E_l &= \beta\,\frac{1+t_l}{\Gamma_l}\,\bar\eta_l,
&
F_l^2 &= \beta\,\bar q_l.
\end{aligned}
\end{equation}

\subsection{Final simplified form of the streaming after fixing trials}
\label{subsec:streaming_final_after_fix}

Inserting these RS choices into the exact streaming identity~\eqref{eq:streaming_master_full} and centring every empirical overlap around its RS limit (e.g.\ $\hat m_l^1-\hat M_l$, $p_{ab}^{(l)}-\bar p_l$, and so on), we arrive at the canonical decomposition
\begin{equation}
\label{eq:streaming_final_decomposition}
\frac{d}{dx}\,\mathcal A_{N,L}(x)
=\mathcal D^{\mathrm{RS}}(x)
+\mathcal R^{\mathrm{RS}}_{N,L}(x)
+\mathcal O\!\left(\frac{1}{N}\right).
\end{equation}
Here $\mathcal D^{\mathrm{RS}}(x)$ collects all terms that survive the $N\to\infty$ limit, whereas $\mathcal R^{\mathrm{RS}}_{N,L}(x)$ gathers the centred fluctuation terms, which vanish under the RS concentration hypothesis. The deterministic piece splits naturally by sector,
\begin{equation}
\label{eq:DRS_split}
\mathcal D^{\mathrm{RS}}(x)
=\mathcal D^{\mathrm{RS}}_{\mathrm{sig}}(x)
+\mathcal D^{\mathrm{RS}}_{\mathrm{LAM}}(x)
+\mathcal D^{\mathrm{RS}}_{\mathrm{Dream}}(x),
\end{equation}
with the three contributions obtained as follows. Substituting $\psi_l^{(m)}=\frac{g_{lm}\beta}{\Gamma_{lm}}\hat M_m$ and $\psi_m^{(l)}=\frac{g_{lm}\beta}{\Gamma_{lm}}\hat M_l$ into~\eqref{eq:streaming_LAM_raw} produces the condensed-signal identity
\begin{equation}
\label{eq:LAM_decomp}
\mathcal D_{\mathrm{sig}}(x)
=-\sum_{1\le l<m\le L}\frac{g_{lm}\beta}{\Gamma_{lm}}\,\hat M_l\hat M_m
+\sum_{1\le l<m\le L}\frac{g_{lm}\beta}{\Gamma_{lm}}\,\Big\langle\big(\hat m_l^1-\hat M_l\big)\big(\hat m_m^1-\hat M_m\big)\Big\rangle_x.
\end{equation}
so that the deterministic signal streaming reads
\begin{equation}
\label{eq:DRS_LAM}
{\;\mathcal D^{\mathrm{RS}}_{\mathrm{sig}}(x)
=-\sum_{1\le l<m\le L}\frac{g_{lm}\beta}{\Gamma_{lm}}\,\hat M_l\hat M_m.\;}
\end{equation}
For the LAM sector, once the trial parameters have been fixed, the streaming admits the same decomposition
\begin{equation}
\label{eq:3AM_decomp}
\mathcal D_{\mathrm{LAM}}(x)
=\mathcal D^{\mathrm{RS}}_{\mathrm{LAM}}(x)
+\mathcal R^{\mathrm{RS},\mathrm{LAM}}_{N,L}(x)
+\mathcal O\!\left(\frac{1}{N}\right),
\end{equation}
whose deterministic piece reads
\begin{equation}
\label{eq:DRS_3AM}
\mathcal D^{\mathrm{RS}}_{\mathrm{LAM}}
= -\frac{\alpha\beta}{4}\sum_{l=1}^L \bar\Pi_l \,(1- H_l).
\end{equation}
With $B_l,C_l,D_l,E_l,F_l$ expressed in terms of the RS limits $\bar Q_l,\bar q_l,\bar P_l,\bar p_l,\bar\eta_l$, the dreaming sector exhibits the same deterministic-plus-fluctuation structure,
\begin{equation}
\label{eq:Dream_decomp}
\mathcal D_{\mathrm{Dream}}(x)
=\mathcal D^{\mathrm{RS}}_{\mathrm{Dream}}(x)
+\mathcal R^{\mathrm{RS},\mathrm{Dream}}_{N,L}(x)
+\mathcal O\!\left(\frac{1}{N}\right),
\end{equation}
with deterministic RS piece
\begin{equation}
\label{eq:DRS_Dream}
\mathcal D^{\mathrm{RS}}_{\mathrm{Dream}}(x)
=\sum_{l=1}^{L}\Bigg[
-\frac{\alpha\beta}{2}(\bar P_l\bar Q_l-\bar p_l\bar q_l)
-\frac{\beta(1+t_l)}{2\Gamma_l}\,\bar\eta_l^{\,2}
\Bigg],
\end{equation}
which is the multi-layer extension of the classical Hebbian--dreaming RS streaming. Collecting~\eqref{eq:DRS_LAM},~\eqref{eq:DRS_3AM} and~\eqref{eq:DRS_Dream}, we obtain the full RS deterministic streaming
\begin{equation}
\label{eq:DRS_total}
\mathcal D^{\mathrm{RS}}(x)
=-\sum_{1\le l<m\le L}\frac{g_{lm}\beta}{\Gamma_{lm}}\,\hat M_l\hat M_m
-\frac{\alpha\beta}{4}\sum_{l=1}^{L}\,\bar\Pi_l\,(1-H_l)
+\sum_{l=1}^{L}\Bigg[
-\frac{\alpha\beta}{2}(\bar P_l\bar Q_l-\bar p_l\bar q_l)
-\frac{\beta(1+t_l)}{2\Gamma_l}\,\bar\eta_l^{\,2}
\Bigg].
\end{equation}

The remainder $\mathcal R^{\mathrm{RS}}_{N,L}(x)$ in~\eqref{eq:streaming_final_decomposition} collects all centred fluctuation terms generated by~\eqref{eq:LAM_decomp} and by the analogous centring of the LAM and dreaming sectors. Explicitly,
\begin{equation}
\label{eq:RS_remainder_explicit_v2}
\begin{aligned}
\mathcal R^{\mathrm{RS}}_{N,L}(x)
&=\underbrace{\sum_{1\le l<m\le L}\frac{g_{lm}\beta}{\Gamma_{lm}}\,\Big\langle\big(\hat m_l^1-\hat M_l\big)\big(\hat m_m^1-\hat M_m\big)\Big\rangle_x}_{\mathcal R^{\mathrm{RS},\mathrm{LAM}}_{N,L}(x)}
\\
&\quad +\underbrace{\frac{\alpha\beta}{4}\sum_{l=1}^L\Big\langle\big(P_{11}^{(l)}-\overline{\mathcal{P}}_l\big)
-\big(h_{12}^{(l)}-H_l\big)\big(P_{12}^{(l)}-\bar\Pi_l\big)\Big\rangle_x}_{\mathcal R^{\mathrm{RS},\mathrm{LAM}}_{N,L}(x)}
\\
&\quad +\underbrace{\frac{\alpha\beta}{2}\sum_{l=1}^L\Big\langle\big(p_{11}^{(l)}-\bar P_l\big)\big(q_{11}^{(l)}-\bar Q_l\big)
-\big(p_{12}^{(l)}-\bar p_l\big)\big(q_{12}^{(l)}-\bar q_l\big)\Big\rangle_x
+\sum_{l=1}^L\frac{\beta(1+t_l)}{2\Gamma_l}\,\Big\langle\big(\eta_l-\bar\eta_l\big)^2\Big\rangle_x}_{\mathcal R^{\mathrm{RS},\mathrm{Dream}}_{N,L}(x)}.
\end{aligned}
\end{equation}
Under the RS self-averaging hypothesis, each centred term vanishes as $N\to\infty$; consequently, $\mathcal R^{\mathrm{RS}}_{N,L}(x)\to 0$ and the streaming equation reduces to the deterministic expression $\mathcal D^{\mathrm{RS}}(x)$.

\section{Replica-symmetric statistical pressure}
\label{app:RS_pressure}
\noindent
The one-body partition function at $x=0$ and the RS-fixed streaming equation jointly determine the quenched pressure through the Guerra sum rule. All remaining quantities then reduce to RS order parameters and layer-wise auxiliary expectations.

\subsection{Guerra sum rule and RS decomposition}
The interpolating pressure
\begin{equation}
\mathcal A_N(x) := \frac{1}{N} \E\log\mathcal Z(x),\qquad x\in[0,1],
\end{equation}
recovers the full quenched pressure at $x=1$. The fundamental theorem of calculus delivers the Guerra sum rule
\begin{equation}
\label{eq:sum_rule_DLAM}
\mathcal A_N(1)=\mathcal A_N(0)+\int_0^1 dx \, \frac{d\mathcal A_N(x)}{ds}.
\end{equation}
The streaming decomposes by sector,
\begin{equation}
\label{eq:stream_split_DLAM}
\frac{d\mathcal A_N(x)}{ds}=\mathcal D^{\mathrm{LAM}}_{N}(x)+\mathcal D^{\mathrm{LAM}}_{N}(x)+\sum_{l=1}^L\mathcal D^{\mathrm{Dream}}_{N,l}(x),
\end{equation}
and, once the trial parameters are set to their RS values, each sector factorises as
\begin{equation}
\label{eq:stream_RS_plus_rem}
\mathcal D^{\bullet}_{N}(x)=\mathcal D^{\mathrm{RS}}_{\bullet}+\mathcal R^{\bullet}_{N}(x),
\end{equation}
where $\mathcal D^{\mathrm{RS}}_{\bullet}$ depends only on the RS order parameters---and is therefore $s$-independent---while the remainders $\mathcal R^{\bullet}_{N}(x)$ are centred fluctuations.
Under the RS self-averaging hypothesis, these fluctuations integrate to zero in the thermodynamic limit,
\begin{equation}
\label{eq:remainders_vanish}
\lim_{N\to\infty}\int_0^1 dx \, \mathcal R^{\bullet}_{N}(x)=0,\qquad \bullet\in\{\mathrm{sig},\mathrm{LAM},\mathrm{Dream}\},
\end{equation}
so that~\eqref{eq:sum_rule_DLAM}--\eqref{eq:stream_RS_plus_rem} collapse to the RS pressure
\begin{equation}
\label{eq:RS_pressure_sum}
\mathcal A^{\mathrm{RS}}:=\lim_{N\to\infty}\mathcal A_N(1)=\mathcal A^{(0)}+\mathcal D^{\mathrm{RS}}_{\mathrm{sig}}+\mathcal D^{\mathrm{RS}}_{\mathrm{LAM}}+\sum_{l=1}^L\mathcal D^{\mathrm{RS}}_{\mathrm{Dream},l},
\end{equation}
where $\mathcal A^{(0)}:=\lim_{N\to\infty}\mathcal A_N(0)$ denotes the one-body contribution.

\subsection{Deterministic RS streaming terms}
The three deterministic RS streaming contributions, derived in the preceding section, read
\begin{align}
\label{eq:DRS_LAM_repeat}
\mathcal D^{\mathrm{RS}}_{\mathrm{sig}}&= -\sum_{1\le l<m\le L}\frac{g_{lm}\beta}{\Gamma_{lm}} \hat M_l\hat M_m, \\
\label{eq:DRS_3AM_repeat}
\mathcal D^{\mathrm{RS}}_{\mathrm{LAM}}
& = -\frac{\alpha\beta}{4}\sum_{l=1}^L \bar\Pi_l \,(1- H_l). \\
\label{eq:DRS_Dream_repeat}
\mathcal D^{\mathrm{RS}}_{\mathrm{Dream},l}&= -\frac{\alpha\beta}{2} (\bar P_l\bar Q_l-\bar p_l\bar q_l)-\frac{\beta(1+t_l)}{2\Gamma_l} \bar\eta_l^2.
\end{align}
Here $\hat M_l:=\overline{\langle \hat m_l^1\rangle}$ denotes the Mattis magnetisation on layer $l$, $H_l$ is the RS diagonal overlap in the LAM Gaussian sector, and $(\bar P_l,\bar p_l)$, $(\bar Q_l,\bar q_l)$ are the diagonal/off-diagonal RS overlaps for the dreaming $z$- and $k$-variables, respectively.

\subsection{One--body term at $x=0$}
Setting $x=0$ decouples all interaction terms, and the partition function factorises as
\begin{equation}
\mathcal Z(0)=\mathcal Z^{(0)}_{\mathrm{LAM}} \mathcal Z^{(0)}_{\mathrm{Dream}} \mathcal Z^{(0)}_{s,\phi},
\end{equation}
and the quenched pressure at $x=0$ becomes
\begin{equation}
\label{eq:A0_split_repeat}
\mathcal A^{(0)}=\lim_{N\to\infty}\frac{1}{N}\E\log\mathcal Z^{(0)}_{\mathrm{LAM}}+\lim_{N\to\infty}\frac{1}{N}\E\log\mathcal Z^{(0)}_{\mathrm{Dream}}+\lim_{N\to\infty}\frac{1}{N}\E\log\mathcal Z^{(0)}_{s,\phi}.
\end{equation}

\subsection{Final RS pressure}
Inserting the RS trial values into $\mathcal A^{(0)}$ and adding the deterministic streaming contributions~\eqref{eq:DRS_LAM_repeat}--\eqref{eq:DRS_Dream_repeat} yields the complete RS pressure.

For each layer $l$ we define
\begin{equation}
\label{eq:Dbar_layer_def}
\bar D_l:=1+\alpha \frac{t_l}{1+t_l} (\bar P_l-\bar p_l),\qquad \Delta_l:=\frac{1}{\bar D_l}.
\end{equation}
together with the effective RS local field
\begin{equation}
\label{eq:Psi_layer_def}
\Psi_l:= I_l \iota+\frac{\sqrt{\alpha\beta \bar p_l}}{\bar D_l} \varphi+\xi_{l}\left(\sum_{m\neq l}\frac{g_{lm}\beta}{\Gamma_{lm}} \hat M_m+\frac{\beta(1+t_l)}{\Gamma_l} \frac{\bar\eta_l}{\bar D_l}\right),
\end{equation}
where $\iota,\varphi\sim\mathcal N(0,1)$ are independent, $\xi_l$ is a quenched random variable encoding the typical empirical archetype component (with variance $\Gamma_l$), and $I_l^2=\frac{\alpha\beta}{2} \bar\Pi_l$.

\medskip
\noindent\textbf{Dreaming contribution (layer-wise).}
Combining the $z$-Gaussian piece, the local $(s,\phi)$ piece, and the deterministic streaming~\eqref{eq:DRS_Dream_repeat}, we obtain, for each layer $l$,
\begin{align}
\label{eq:A_Dream_layer_D34form}
\mathcal A^{\mathrm{RS}}_{\mathrm{Dream},l}
&= -\frac12\log\bar D_l
-\frac{\alpha\bar p_l}{2\bar D_l} \frac{t_l}{1+t_l}
+\E\log\cosh(\Psi_l)
+\log 2 \nonumber \\
&\quad +\frac{\alpha}{2}\log \left(\frac{1+t_l}{1-\beta(\bar Q_l-\bar q_l)(1+t_l)}\right)
+\frac{\alpha}{2} \frac{\beta\bar q_l(1+t_l)}{1-\beta(\bar Q_l-\bar q_l)(1+t_l)} \nonumber \\
&\quad +\frac{\beta}{2} \frac{\bar D_l-1}{\bar D_l} \frac{1+t_l}{t_l}
-\frac{\beta(1+t_l)}{2\Gamma_l} \bar\eta_l^2 \frac{t_l+\bar D_l}{\bar D_l} \nonumber \\
&\quad +\frac{\alpha\beta}{2} (\bar p_l \bar q_l - \bar P_l \bar Q_l).
\end{align}

\medskip
\noindent\textbf{LAM contribution.}
The RS LAM sector combines the one-body Gaussian term at $x=0$ with the deterministic LAM streaming remainder.
For each non-condensed mode $\mu\ge2$ we introduce the real vector
\(
\bm x_\mu=(\Re u_\mu^1,\Im u_\mu^1,\dots,\Re u_\mu^L,\Im u_\mu^L)\in\mathbb R^{2L}
\),
together with the associated $2L\times2L$ Gaussian kernel $\mathbf C(H)$ built from the RS spin overlaps $H_l=\overline{\langle h_{12}^{(l)}\rangle}$.
The standard Gaussian identity then produces the structure
\[
-\frac{\alpha}{2}\log\det\mathbf C(H)\;+\;\frac{\alpha}{2}\,\E_{\bm\delta}\!\left[\bm b^\top \mathbf C(H)^{-1}\bm b\right],
\]
in which $\bm b$ is the RS linear source vector encoding the selective couplings.
Introducing the quadratic correction
\begin{equation}
\label{eq:D3AM_def}
\mathbb D_{\mathrm{LAM}}
:=\E_{\bm\delta}\big[\bm b^\top \mathbf C(H)^{-1}\bm b\big],
\end{equation}
the RS LAM contribution takes the form
\begin{equation}
\label{eq:A_3AM_RS}
\mathcal A^{\mathrm{RS}}_{\mathrm{LAM}}
=
-\frac{\alpha}{2}\log\det\mathbf C(H)
+\frac{\alpha}{2}\mathbb D_{\mathrm{LAM}}
-\frac{\alpha\beta}{4}\sum_{l=1}^L \bar\Pi_l(1-H_l).
\end{equation}
The last term is the deterministic RS streaming contribution fixed in \ref{app:streaming}.
Explicit closed forms for $\det\mathbf C(H)$ and $\mathbb D_{\mathrm{LAM}}$ are provided below in the specialisation to $L=3$.

\noindent
Finally, the cross-layer coupling generated by the condensed LAM interaction contributes
\begin{equation}
\label{eq:A_LAM_RS}
\mathcal A^{\mathrm{RS}}_{\mathrm{sig}}= -\sum_{1\le l<m\le L}\frac{g_{lm}\beta}{\Gamma_{lm}} \hat M_l\hat M_m.
\end{equation}

\medskip
\noindent\textbf{Full RS pressure.}
Combining the three sectors, the total quenched free-energy density $f(\beta) = -\frac{1}{\beta}\mathcal{A}^{\mathrm{RS}}_{\mathrm{DLAM}}$ is governed by
\begin{align}
\label{eq:A_DLAM_RS_final}
  \mathcal{A}^{\mathrm{RS}}_{\mathrm{DLAM}} &= 
  -\sum_{1\le l<m\le L}\frac{\beta\,g_{lm}}{\Gamma_{lm}}\hat M_l\hat M_m 
  -\frac{\alpha}{2}\log\det\mathbf{C}(H) + \frac{\alpha}{2}\mathbb{D}_{\mathrm{LAM}} \nonumber\\
  &\quad - \frac{\alpha\beta}{4}\sum_{l=1}^{L}\bar\Pi_l(1{-}H_l) \nonumber\\
  &\quad + \sum_{l=1}^{L} \Bigg[ \log 2 + \E\bigl[\log\cosh\Psi_l\bigr]
     - \tfrac{1}{2}\log\bar D_l
     + \tfrac{\alpha}{2}\log \left(\frac{1+t_l}{1-\beta(\bar Q_l-\bar q_l)(1+t_l)}\right) \nonumber\\
  &\quad + \frac{\alpha}{2} \frac{\beta\bar q_l(1+t_l)}{1-\beta(\bar Q_l-\bar q_l)(1+t_l)}
     - \frac{\alpha\bar p_l}{2\bar D_l} \frac{t_l}{1+t_l}
     + \frac{\beta}{2} \frac{\bar D_l-1}{\bar D_l} \frac{1+t_l}{t_l} \nonumber\\
  &\quad - \frac{\beta(1+t_l)}{2\Gamma_l} \bar\eta_l^2 \frac{t_l+\bar D_l}{\bar D_l}
     + \frac{\alpha\beta}{2} (\bar p_l \bar q_l - \bar P_l \bar Q_l) \Bigg].
\end{align}

\subsection{Explicit $L=3$ specialization of the LAM contribution}
\label{subsec:L3_LAM_3AM}

\noindent
In the general RS pressure~\eqref{eq:A_DLAM_RS_final}, the cross-layer LAM signal and the LAM one-body Gaussian are given by~\eqref{eq:A_LAM_RS} and~\eqref{eq:A_3AM_RS}, respectively. Specialising to $L=3$ amounts to evaluating the LAM one-body contribution through a \emph{real} Hubbard--Stratonovich decoupling over the three inter-layer modes; this produces a $3\times3$ real symmetric kernel for which closed-form expressions are available. An equivalent derivation for the three-directional architecture can be found in~\cite{alessandrelli2025supervised}.

The condensed signal term~\eqref{eq:A_LAM_RS} reduces immediately to a sum over the three distinct layer pairs $(1,2)$, $(1,3)$, $(2,3)$,
\begin{equation}
\label{eq:A_LAM_RS_L3}
{
\mathcal A^{\mathrm{RS}}_{\mathrm{sig}}\Big|_{L=3}
= -\beta\left(
\frac{g_{12}\hat M_1\hat M_2}{\Gamma_{12}}
+\frac{g_{13}\hat M_1\hat M_3}{\Gamma_{13}}
+\frac{g_{23}\hat M_2\hat M_3}{\Gamma_{23}}
\right).
}
\end{equation}
For the LAM Gaussian kernel, each non-condensed mode $\mu\ge2$ is decoupled by a triplet of real auxiliary fields $\phi_l\in\mathbb R$, one per layer. After annealing over the random patterns and taking the replica-symmetric saddle point, the single-mode LAM partition function reduces to a standard three-dimensional real Gaussian integral $\int\!\mathcal D\phi\,e^{-\frac{1}{2}\phi^\top \mathbf C\phi+\mathbf b^\top\phi}$, with $3\times3$ real symmetric kernel
\begin{equation}
\label{eq:C_3AM_L3_}
[\mathbf C(H)]_{lm}
=
\delta_{lm}
\;-\;
\beta\,g_{lm}\sqrt{(1-H_l)(1-H_m)},
\qquad g_{ll}:=0,
\end{equation}
whose diagonal entries equal one and whose off-diagonal entries are $-\beta g_{lm}\sqrt{(1-H_l)(1-H_m)}$. In the standard three-layer architecture one has $g_{12}=g_{13}=g_{23}=g$, and at the homogeneous RS fixed point $H_l=H$ the kernel simplifies to
\begin{equation}
\label{eq:C_3AM_L3_homog}
\mathbf C(H) = (1+u)\,I_3 - u\,J_3,
\qquad
u := \beta g(1-H),
\end{equation}
where $J_3$ denotes the $3\times3$ all-ones matrix, so that $\mathbf C(H)$ has diagonal entries $1-u$ and off-diagonal entries $-u$. This matrix commutes with every permutation of the layer indices; its spectrum is therefore determined by the two irreducible representations of $S_3$: the uniform mode $\mathbf e_0=(1,1,1)^\top/\sqrt{3}$ carries the eigenvalue $\lambda_0=1-2u$, while the orthogonal shear doublet $(\mathbf e_1,\mathbf e_2)$ carries the doubly degenerate eigenvalue $\lambda_1=1+u$. The kernel is positive-definite---and the LAM Gaussian integral well-defined---if and only if
\begin{equation}
\label{eq:C_positivity_L3}
u < \tfrac{1}{2},\qquad\text{i.e.}\quad T > 2g(1-H);
\end{equation}
since the RS retrieval branch enforces a high overlap $H=q$ at low $T$, the product $\beta g(1-q)$ remains bounded, and condition~\eqref{eq:C_positivity_L3} holds throughout the physical retrieval region.

The eigenvalue structure above gives the homogeneous determinant
\begin{equation}
\label{eq:detC_L3_}
\det\mathbf C(H) = (1-2u)(1+u)^2,\qquad u=\beta g(1-H),
\end{equation}
whose only singularity, $u=\tfrac12$, corresponds to $T=2g(1-H)$ (and, for $g=1$ and $H\to0$, to the critical value $T^*=2$), well outside the physical retrieval region. In the heterogeneous case, with arbitrary $g_{lm}$ and $H_l$, the $3\times3$ determinant expands as
\begin{equation}
\label{eq:detC_L3_heter}
\det\mathbf C(H)
=
1
-\beta^2\bigl[g_{12}^2(1-H_1)(1-H_2)+g_{13}^2(1-H_1)(1-H_3)+g_{23}^2(1-H_2)(1-H_3)\bigr]
-2\beta^3 g_{12}g_{13}g_{23}(1-H_1)(1-H_2)(1-H_3),
\end{equation}
which is the determinant that enters the LAM one-body free energy. The corresponding contribution to the RS free-energy density, computed per non-condensed pattern and arising from the real-HS Gaussian integral over $\mathbf C(\mathbf H)$, takes the form
\begin{equation}
\label{eq:f3AM_L3_}
f_{\mathrm{LAM}}(\mathbf H)
=
-\frac{\alpha}{2}\log\det\mathbf C(\mathbf H)
+
\frac{\alpha\,\mathcal D_{\mathrm{LAM}}^{\mathrm{num}}(\mathbf H)}{\det\mathbf C(\mathbf H)},
\end{equation}
where $\mathcal D_{\mathrm{LAM}}^{\mathrm{num}}$ collects the quadratic overlap correction. At the homogeneous RS fixed point $H_l=H=q$ (the Edwards--Anderson overlap), with $u=\beta g(1-q)$, this correction admits the closed form
\begin{equation}
\label{eq:D3AM_L3_}
\mathcal D_{\mathrm{LAM}}^{\mathrm{num}}\big|_{L=3}^{\mathrm{hom}}
=
3q(1-q)\,\beta^2 g^2\bigl[1+\beta g(1-q)\bigr]
=
3q(1-q)\,\beta^2 g^2(1+u),
\end{equation}
while for the heterogeneous case $\mathcal D_{\mathrm{LAM}}^{\mathrm{num}}(\mathbf H)$ follows by differentiating~\eqref{eq:detC_L3_heter} with respect to the individual overlaps $H_l$, producing a polynomial in $\beta$, $g_{lm}$ and $(1-H_l)$ whose explicit form can be read off from the $3\times3$ matrix-adjugate formula.

The derivatives required by the LAM saddle-point equations follow from Jacobi's formula for the derivative of a log-determinant, $\partial_\theta\log\det A = \mathrm{Tr}[A^{-1}\partial_\theta A]$. The stationarity condition $\partial_{H_l}f_{\mathrm{LAM}}=0$ then determines the conjugate LAM overlaps $\bar\Pi_l$ via
\begin{equation}
\label{eq:d_detC_dH1_L3_}
\frac{\partial}{\partial H_l}\log\det\mathbf C(H)
=
\mathrm{Tr}\!\Big[\mathbf C(H)^{-1}\,\frac{\partial\mathbf C}{\partial H_l}(H)\Big],
\qquad l=1,2,3,
\end{equation}
where the only non-zero entries of $\partial_{H_l}\mathbf C$ are, in view of~\eqref{eq:C_3AM_L3_},
\[
\bigl(\partial_{H_l}\mathbf C\bigr)_{lm}
=
\bigl(\partial_{H_l}\mathbf C\bigr)_{ml}
=
\frac{\beta\,g_{lm}}{2}\sqrt{\frac{1-H_m}{1-H_l}},
\qquad m\neq l,
\]
with every other entry vanishing. In the homogeneous case this reduces to $\partial_{H_l}\mathbf C=\tfrac{\beta g}{2(1-H)}\bigl[\mathbf C(H)-(1+u)I_3\bigr]$. Combining~\eqref{eq:A_LAM_RS_L3} with~\eqref{eq:A_3AM_RS},~\eqref{eq:detC_L3_} and~\eqref{eq:D3AM_L3_}, we obtain
\begin{equation}
\label{eq:A_LAM_3AM_RS_L3_final}
\begin{aligned}
\Big(\mathcal A^{\mathrm{RS}}_{\mathrm{sig}}+\mathcal A^{\mathrm{RS}}_{\mathrm{LAM}}\Big)\Big|_{L=3}
=&
-\beta\left(
\frac{g_{12}\hat M_1\hat M_2}{\Gamma_{12}}
+\frac{g_{13}\hat M_1\hat M_3}{\Gamma_{13}}
+\frac{g_{23}\hat M_2\hat M_3}{\Gamma_{23}}
\right)
\\
&\quad
-\frac{\alpha}{2}\log\det \mathbf C(H)
+\frac{\alpha\beta}{2}\,
\frac{\mathcal D_{\mathrm{LAM}}^{\mathrm{num}}(\mathbf H)}{\det \mathbf C(H)}
-\frac{\alpha\beta}{4}\sum_{\ell=1}^{3}\bar\Pi_\ell(1-H_\ell)
+\mathrm{const}.
\end{aligned}
\end{equation}

\section{Self-consistency equations}
\label{app:self_consistency}

\noindent
Extremising the RS pressure $\mathcal A^{\mathrm{RS}}_{\mathrm{DLAM}}$ with respect to every order parameter and its conjugate produces the self-consistency (saddle-point) equations that close the theory. The derivations are carried out at generic $L$ wherever possible; specialisation to $\boldsymbol{L=3}$ is required only in the LAM sector, where the one-body Gaussian contribution demands an explicit scalar form.

\subsection{Dependency map and conventions for the RS stationarity analysis}
\label{subsec:self_dependency_map}

The RS order parameters to be extremised fall into three layer-wise families: the signal/LAM variable $\hat M_l$; the dreaming variables $\bar\eta_l,\bar q_l,\bar Q_l,\bar p_l,\bar P_l$; and the LAM variables $H_l,\bar\Pi_l$, where $H_l$ is the RS spin overlap in the LAM channel and $\bar\Pi_l$ the RS off-diagonal LAM overlap entering the LAM streaming. Schematically, the RS pressure decomposes as
\begin{equation}
\label{eq:ARS_split_self_scheme}
\mathcal A^{\mathrm{RS}}_{\mathrm{DLAM}}
=
\mathcal A^{\mathrm{RS}}_{\mathrm{sig}}
+\sum_{l=1}^{L}\mathcal A^{\mathrm{RS}}_{\mathrm{Dream},l}
+\mathcal A^{\mathrm{RS}}_{\mathrm{LAM}},
\end{equation}
The signal and dreaming sectors admit stationarity equations at generic $L$. The LAM Gaussian sector, by contrast, involves $\log\det\mathbf C$ and the quadratic correction $\mathbb D_{\mathrm{LAM}}$, whose fully explicit scalar form is tractable only after specialising to $L=3$.

To prevent any ambiguity with the microscopic archetypes $\xi_{i,l}^{\mu}$, the symbol $\xi_l$ is not used in this section: for each layer $l$, we let $\hat X_l$ denote a scalar random variable that represents a generic one-site component of the condensed empirical archetype,
\begin{equation}
\label{eq:Xhat_l_def}
\hat X_l \stackrel{d}{=} \hat\xi_{i,l}^{1}
\qquad\text{for a generic site } i,
\end{equation}
with moments $\E[\hat X_l]=0$ and $\E[\hat X_l^2]=\Gamma_l$ in the benchmark Rademacher setting. With this convention the effective RS field on layer $l$ (cf.~\eqref{eq:Psi_layer_def}) takes the form
\begin{equation}
\label{eq:Psi_layer_def_self}
\Psi_l
=
I_l \,\iota
+\frac{\sqrt{\alpha\beta\,\bar p_l}}{\bar D_l}\,\varphi
+\hat X_l
\left(
\sum_{m\neq l}\frac{g_{lm}\beta}{\Gamma_{lm}}\hat M_m
+\frac{\beta(1+t_l)}{\Gamma_l}\frac{\bar\eta_l}{\bar D_l}
\right),
\end{equation}
with $\iota,\varphi\sim\mathcal N(0,1)$ independent of each other and of $\hat X_l$. Unless stated otherwise, the layer-$l$ expectation is taken over the joint law of the auxiliary Gaussians and of $\hat X_l$, i.e.\ $\E_l[\cdot]:=\E_{\iota,\varphi,\hat X_l}[\cdot]$; we shall write simply $\E[\cdot]$ whenever no ambiguity can arise.

The stationarity equations are derived in three successive blocks: first, the signal/LAM and mixed local observables $(\hat M_l,\bar\eta_l)$ at generic $L$; second, the dreaming sector $(\bar q_l,\bar Q_l,\bar p_l,\bar P_l)$, again at generic $L$; and finally the LAM sector $(H_l,\bar\Pi_l)$, specialised to $L=3$ at the single point where the explicit $L=3$ LAM algebra is unavoidable.

\subsection{Stationarity conditions for the signal sector: derivation of $\hat M_l$}
\label{subsec:saddle_M_l}

The Mattis magnetisations $\hat M_l$ enter the RS pressure both through the condensed LAM bilinear and, via the effective fields, through the entropic terms $\E_l\log\cosh(\Psi_l)$. Differentiating with respect to $\hat M_k$ and setting the result to zero produces a vector equation that, on the physical non-degenerate branch, collapses to a componentwise self-consistency relation. Explicitly, the relevant part of the pressure reads
\begin{equation}
\label{eq:ARS_relevant_M}
\mathcal A^{\mathrm{RS}}_{\mathrm{DLAM}}
=
-\sum_{1\le l<m\le L}\frac{g_{lm}\beta}{\Gamma_{lm}}\hat M_l\hat M_m
+\sum_{l=1}^{L}\E_l\log\cosh(\Psi_l)
+\text{(terms independent of $\hat M$)},
\end{equation}
where, using the shorthand introduced in Section~\ref{subsec:self_dependency_map},
\begin{equation}
\label{eq:Psi_for_M_derivation}
\Psi_l
=
I_l\iota+\sigma_l\varphi+\hat X_l\,(u_l+v_l),
\qquad
u_l:=\sum_{m\neq l}\frac{g_{lm}\beta}{\Gamma_{lm}}\hat M_m,
\qquad
v_l:=\frac{\beta(1+t_l)}{\Gamma_l}\frac{\bar\eta_l}{\bar D_l},
\end{equation}
with $\sigma_l:=\frac{\sqrt{\alpha\beta\,\bar p_l}}{\bar D_l}$ and $\iota,\varphi,\hat X_l$ mutually independent under $\E_l$. Fixing $k\in\{1,\dots,L\}$, differentiation of the symmetric LAM quadratic form yields
\begin{equation}
\label{eq:dA_dMk_LAM}
\frac{\partial}{\partial \hat M_k}
\left(
-\sum_{1\le l<m\le L}\frac{g_{lm}\beta}{\Gamma_{lm}}\hat M_l\hat M_m
\right)
=
-\sum_{m\neq k}\frac{g_{km}\beta}{\Gamma_{km}}\hat M_m,
\end{equation}
whereas $\hat M_k$ enters the effective field $\Psi_l$ only through $u_l$, and only for $l\neq k$: indeed $u_k=\sum_{m\neq k}\frac{g_{km}\beta}{\Gamma_{km}}\hat M_m$ is itself independent of $\hat M_k$. Hence
\begin{equation}
\label{eq:dul_dMk}
\frac{\partial u_l}{\partial \hat M_k}
=
\begin{cases}
\frac{g_{lk}\beta}{\Gamma_{lk}}, & l\neq k,\\
0, & l=k,
\end{cases}
\end{equation}
and the chain rule produces the entropic derivative
\begin{align}
\label{eq:dA_dMk_entropic}
\frac{\partial}{\partial \hat M_k}\sum_{l=1}^{L}\E_l\log\cosh(\Psi_l)
&=
\sum_{l\neq k}\E_l\!\left[\tanh(\Psi_l)\,\frac{\partial\Psi_l}{\partial\hat M_k}\right]
\nonumber\\
&=
\sum_{l\neq k}\frac{g_{lk}\beta}{\Gamma_{lk}}\,
\E_l\!\left[\hat X_l\,\tanh(\Psi_l)\right].
\end{align}
Setting $\partial_{\hat M_k}\mathcal A^{\mathrm{RS}}_{\mathrm{DLAM}}=0$ and combining~\eqref{eq:dA_dMk_LAM}--\eqref{eq:dA_dMk_entropic} produces, for each $k$,
\begin{equation}
\label{eq:saddle_M_vector_form_component}
\sum_{m\neq k}\frac{g_{km}\beta}{\Gamma_{km}}
\left(
\E_m[\hat X_m\tanh(\Psi_m)]-\hat M_m
\right)=0.
\end{equation}
Equivalently, upon introducing the weighted LAM kernel
\begin{equation}
\label{eq:KGamma_def}
(\mathbb K_\Gamma)_{km}:=
\begin{cases}
\frac{g_{km}\beta}{\Gamma_{km}}, & k\neq m,\\
0, & k=m,
\end{cases}
\end{equation}
and the residual vector $\Delta_l^{(M)}:=\E_l[\hat X_l\tanh(\Psi_l)]-\hat M_l$, the stationarity condition takes the compact form
\begin{equation}
\label{eq:saddle_M_vector_form}
\mathbb K_\Gamma\,\bm\Delta^{(M)}=0.
\end{equation}
On the physically relevant branch, where $\ker\mathbb K_\Gamma=\{0\}$, this enforces $\bm\Delta^{(M)}=0$ componentwise, yielding the RS magnetisation equation
\begin{equation}
\label{eq:saddle_Ml_final}
{
\hat M_l=\E_l\!\left[\hat X_l\tanh(\Psi_l)\right],
\qquad l=1,\dots,L.
}
\end{equation}
The non-degeneracy assumption holds automatically for $L=3$: the kernel is then the weighted off-diagonal $3\times3$ matrix
\[
\begin{pmatrix}
0 & g_{12}\beta/\Gamma_{12} & g_{13}\beta/\Gamma_{13}\\
g_{12}\beta/\Gamma_{12} & 0 & g_{23}\beta/\Gamma_{23}\\
g_{13}\beta/\Gamma_{13} & g_{23}\beta/\Gamma_{23} & 0
\end{pmatrix},
\]
with determinant $\det\mathbb K_\Gamma=2g_{12}g_{13}g_{23}\beta^3/(\Gamma_{12}\Gamma_{13}\Gamma_{23})$ (the denominator being positive, since $\Gamma_{lm}>0$). The kernel is therefore invertible whenever $g_{12}g_{13}g_{23}\neq 0$, and equation~\eqref{eq:saddle_Ml_final} follows rigorously on that branch; should one or more hetero-associative channels be switched off, a separate branch analysis would be required.


\subsection{Stationarity conditions for the auxiliary parameter: derivation of $\bar\eta_l$}
\label{subsec:saddle_eta_l}

The dreaming auxiliary parameter $\bar\eta_l$ enters only the layer-wise dreaming contribution $\mathcal A^{\mathrm{RS}}_{\mathrm{Dream},l}$; the derivation is therefore layer-local and applies for generic $L$.

From~\eqref{eq:A_Dream_layer_D34form}, the parameter $\bar\eta_l$ enters $\mathcal A^{\mathrm{RS}}_{\mathrm{Dream},l}$ both implicitly---through the deterministic shift $v_l=\frac{\beta(1+t_l)}{\Gamma_l}\frac{\bar\eta_l}{\bar D_l}$ inside the effective field $\Psi_l$---and explicitly, through the quadratic energy $-\frac{\beta(1+t_l)}{2\Gamma_l}\bar\eta_l^2\frac{t_l+\bar D_l}{\bar D_l}$. Retaining only the $\bar\eta_l$-dependent contributions,
\begin{equation}
\label{eq:A_Dream_eta_relevant}
\mathcal A^{\mathrm{RS}}_{\mathrm{Dream},l}\Big|_{\bar\eta_l}
=
\E_l\log\cosh(\Psi_l)
-\frac{\beta(1+t_l)}{2\Gamma_l}\,\bar\eta_l^2\,\frac{t_l+\bar D_l}{\bar D_l},
\end{equation}
where $\bar D_l$ depends on $\bar\eta_l$ neither directly nor implicitly, since it is fixed by the combination $\bar P_l-\bar p_l$. Because $\Psi_l$ depends on $\bar\eta_l$ exclusively through $v_l$,
\begin{equation}
\label{eq:dPsi_deta}
\frac{\partial\Psi_l}{\partial\bar\eta_l}
=
\hat X_l\,\frac{\partial v_l}{\partial\bar\eta_l}
=
\hat X_l\,\frac{\beta(1+t_l)}{\Gamma_l\bar D_l},
\end{equation}
so the chain rule produces the entropic derivative
\begin{align}
\label{eq:d_entropic_deta}
\frac{\partial}{\partial\bar\eta_l}\,\E_l\log\cosh(\Psi_l)
&=
\E_l\!\left[\tanh(\Psi_l)\frac{\partial\Psi_l}{\partial\bar\eta_l}\right]
\nonumber\\
&=
\frac{\beta(1+t_l)}{\Gamma_l\bar D_l}\,
\E_l\!\left[\hat X_l\tanh(\Psi_l)\right],
\end{align}
while differentiating the explicit quadratic term gives
\begin{equation}
\label{eq:d_quadratic_deta}
\frac{\partial}{\partial\bar\eta_l}
\left[
-\frac{\beta(1+t_l)}{2\Gamma_l}\,\bar\eta_l^2\,\frac{t_l+\bar D_l}{\bar D_l}
\right]
=
-\frac{\beta(1+t_l)}{\Gamma_l}\,\bar\eta_l\,\frac{t_l+\bar D_l}{\bar D_l}.
\end{equation}
Imposing $\partial_{\bar\eta_l}\mathcal A^{\mathrm{RS}}_{\mathrm{DLAM}}=0$ and combining~\eqref{eq:d_entropic_deta}--\eqref{eq:d_quadratic_deta} yields
\begin{equation}
\label{eq:saddle_eta_intermediate}
\frac{\beta(1+t_l)}{\Gamma_l\bar D_l}
\left[
\E_l[\hat X_l\tanh(\Psi_l)]-(t_l+\bar D_l)\bar\eta_l
\right]=0,
\end{equation}
which, by the strict positivity of $\beta$, $1+t_l$, $\Gamma_l$ and $\bar D_l$ on the physical branch, reduces to $(t_l+\bar D_l)\bar\eta_l=\E_l[\hat X_l\tanh(\Psi_l)]$. Substituting the magnetisation equation~\eqref{eq:saddle_Ml_final} on the right-hand side then gives
\begin{equation}
\label{eq:saddle_eta_l_final}
{
\bar\eta_l=\frac{\hat M_l}{t_l+\bar D_l},
\qquad l=1,\dots,L.
}
\end{equation}
The dreaming auxiliary order parameter $\bar\eta_l$ is thus a rescaled Mattis magnetisation, the renormalisation being controlled by $t_l$ and by the dreaming denominator $\bar D_l$.

\subsection{Algebraic stationarity with respect to $\bar q_l$ and $\bar Q_l$: determination of $\bar p_l$ and $\bar P_l$}
\label{subsec:saddle_qQ_to_pP}

Differentiating the RS pressure with respect to $\bar q_l$ and $\bar Q_l$ produces purely algebraic relations that fix the conjugate dreaming parameters $\bar p_l$ and $\bar P_l$. The simplification rests on the observation that $\bar q_l$ and $\bar Q_l$ enter $\mathcal A^{\mathrm{RS}}_{\mathrm{Dream},l}$ only through the explicit rational and logarithmic terms, together with the bilinear coupling $\frac{\alpha\beta}{2}(\bar p_l\bar q_l-\bar P_l\bar Q_l)$, and not through the effective field $\Psi_l$; consequently, no Gaussian integration by parts is required. Defining the layer-wise denominator
\begin{equation}
\label{eq:S_l_def_algebraic_block}
S_l:=1-\beta(1+t_l)(\bar Q_l-\bar q_l),
\end{equation}
the $(\bar q_l,\bar Q_l)$-dependent part of $\mathcal A^{\mathrm{RS}}_{\mathrm{Dream},l}$ becomes
\begin{equation}
\label{eq:A_qQ_relevant_block}
\mathcal A^{\mathrm{RS}}_{\mathrm{Dream},l}\Big|_{(\bar q_l,\bar Q_l)}
=
\frac{\alpha}{2}\log\!\left(\frac{1+t_l}{S_l}\right)
+\frac{\alpha\beta(1+t_l)}{2}\,\frac{\bar q_l}{S_l}
+\frac{\alpha\beta}{2}\,(\bar p_l\bar q_l-\bar P_l\bar Q_l),
\end{equation}
together with the elementary identities
\begin{equation}
\label{eq:dS_dq_dQ_block}
\frac{\partial S_l}{\partial \bar q_l}=\beta(1+t_l),
\qquad
\frac{\partial S_l}{\partial \bar Q_l}=-\beta(1+t_l).
\end{equation}
Differentiating~\eqref{eq:A_qQ_relevant_block} with respect to $\bar q_l$ and using these identities, we find
\begin{align}
\label{eq:dA_dq_pre_simplify_block}
\frac{\partial \mathcal A^{\mathrm{RS}}_{\mathrm{Dream},l}}{\partial \bar q_l}
&=
-\frac{\alpha}{2}\,\frac{1}{S_l}\,\frac{\partial S_l}{\partial \bar q_l}
+\frac{\alpha\beta(1+t_l)}{2}\left(\frac{1}{S_l}-\frac{\bar q_l}{S_l^2}\frac{\partial S_l}{\partial \bar q_l}\right)
+\frac{\alpha\beta}{2}\,\bar p_l.
\end{align}
Substituting~\eqref{eq:dS_dq_dQ_block},
\begin{align}
\label{eq:dA_dq_simplified_block}
\frac{\partial \mathcal A^{\mathrm{RS}}_{\mathrm{Dream},l}}{\partial \bar q_l}
&=
-\frac{\alpha\beta(1+t_l)}{2S_l}
+\frac{\alpha\beta(1+t_l)}{2S_l}
-\frac{\alpha\beta^2(1+t_l)^2}{2S_l^2}\,\bar q_l
+\frac{\alpha\beta}{2}\,\bar p_l
\nonumber\\
&=
\frac{\alpha\beta}{2}\left(
\bar p_l-\beta\frac{(1+t_l)^2}{S_l^2}\bar q_l
\right).
\end{align}
The stationarity condition $\partial_{\bar q_l}\mathcal A^{\mathrm{RS}}_{\mathrm{DLAM}}=0$ thus yields
\begin{equation}
\label{eq:saddle_p_l_algebraic}
{
\bar p_l=\beta\,\frac{(1+t_l)^2}{S_l^2}\,\bar q_l.
}
\end{equation}

An entirely analogous calculation for $\bar Q_l$ produces
\begin{align}
\label{eq:dA_dQ_pre_simplify_block}
\frac{\partial \mathcal A^{\mathrm{RS}}_{\mathrm{Dream},l}}{\partial \bar Q_l}
&=
-\frac{\alpha}{2}\,\frac{1}{S_l}\,\frac{\partial S_l}{\partial \bar Q_l}
+\frac{\alpha\beta(1+t_l)}{2}\,\bar q_l\left(-\frac{1}{S_l^2}\frac{\partial S_l}{\partial \bar Q_l}\right)
-\frac{\alpha\beta}{2}\,\bar P_l.
\end{align}
Using~\eqref{eq:dS_dq_dQ_block},
\begin{align}
\label{eq:dA_dQ_simplified_block}
\frac{\partial \mathcal A^{\mathrm{RS}}_{\mathrm{Dream},l}}{\partial \bar Q_l}
&=
\frac{\alpha\beta(1+t_l)}{2S_l}
+\frac{\alpha\beta^2(1+t_l)^2}{2S_l^2}\,\bar q_l
-\frac{\alpha\beta}{2}\,\bar P_l.
\end{align}
Imposing $\partial_{\bar Q_l}\mathcal A^{\mathrm{RS}}_{\mathrm{DLAM}}=0$,
\begin{equation}
\label{eq:saddle_P_l_algebraic}
{
\bar P_l=\frac{1+t_l}{S_l}+\beta\,\frac{(1+t_l)^2}{S_l^2}\,\bar q_l.
}
\end{equation}

Subtracting~\eqref{eq:saddle_p_l_algebraic} from~\eqref{eq:saddle_P_l_algebraic} yields the identity
\begin{equation}
\label{eq:P_minus_p_identity_block}
{
\bar P_l-\bar p_l=\frac{1+t_l}{S_l},
}
\end{equation}
which, combined with $\bar D_l=1+\alpha\frac{t_l}{1+t_l}(\bar P_l-\bar p_l)$ from~\eqref{eq:Dbar_layer_def}, gives the equivalent expressions
\begin{equation}
\label{eq:Dbar_in_terms_of_S_block}
{
\bar D_l=1+\frac{\alpha t_l}{S_l}
=
1+\frac{\alpha t_l}{1-\beta(1+t_l)(\bar Q_l-\bar q_l)}.
}
\end{equation}
Equations~\eqref{eq:saddle_p_l_algebraic}--\eqref{eq:Dbar_in_terms_of_S_block} are purely algebraic and hold for each layer $l$ at generic $L$; they will be combined with the Stein-derived equations for $\bar q_l$ and $\bar Q_l$ in the next subsection.

\subsection{Stationarity with respect to \texorpdfstring{$\bar p_l,\bar P_l$}{pbar,Pbar}: exact dreaming equations (generic $L$)}
\label{subsec:saddle_pP_exact_genericL}

Differentiating the RS pressure with respect to $\bar p_l$ and $\bar P_l$ probes the entropic content of the effective field $\Psi_l$, with the Gaussian fluctuations of $\varphi$ extracted through Stein's identity. Throughout the calculation it is convenient to fix the layer $l$ and adopt the shorthand
\begin{equation}
\kappa_l := \alpha \frac{t_l}{1+t_l},
\qquad
\bar D_l := 1+\kappa_l(\bar P_l-\bar p_l),
\end{equation}
together with the scalar quenched variable $\hat X_l \stackrel{d}{=} \hat\xi_{i,l}^{1}$, with $\E[\hat X_l^2]=\Gamma_l$ (we again write $\hat X_l$ to avoid clashing with the microscopic archetypes $\xi_l$). The effective RS field then takes the form
\begin{equation}
\label{eq:Psi_l_for_pP_block}
\Psi_l
=
I_l \iota
+\sigma_l \varphi
+\hat X_l\big(u_l+v_l\big),
\qquad
\sigma_l:=\frac{\sqrt{\alpha\beta\,\bar p_l}}{\bar D_l},
\qquad
u_l:=\sum_{m\neq l}\frac{g_{lm}\beta}{\Gamma_{lm}}\hat M_m,
\qquad
v_l:=\frac{\beta(1+t_l)}{\Gamma_l}\frac{\bar\eta_l}{\bar D_l},
\end{equation}
where $\iota,\varphi\sim\mathcal N(0,1)$ are independent of each other and of $\hat X_l$. Two layer-wise expectations will appear repeatedly:
\begin{equation}
\label{eq:Tl_Ml_defs}
\mathcal T_l := \E_l\!\left[\tanh^2(\Psi_l)\right],
\qquad
\mathcal M_l := \E_l\!\left[\hat X_l \tanh(\Psi_l)\right],
\end{equation}
where $\E_l$ denotes the average over $(\iota,\varphi,\hat X_l)$.
By the magnetisation equation derived above, $\mathcal M_l=\hat M_l$.

Since $\bar D_l=1+\kappa_l(\bar P_l-\bar p_l)$, the partial derivatives $\partial_{\bar p_l}\bar D_l=-\kappa_l$ and $\partial_{\bar P_l}\bar D_l=+\kappa_l$ propagate through $\sigma_l$ and $v_l$ according to
\begin{equation}
\label{eq:sigma_v_derivatives}
\frac{\partial \sigma_l}{\partial x}
=
\sigma_l\left(\frac{\mathbf 1_{x=\bar p_l}}{2\bar p_l}-\frac{1}{\bar D_l}\frac{\partial \bar D_l}{\partial x}\right),
\qquad
\frac{\partial v_l}{\partial x}
=
-\frac{v_l}{\bar D_l}\frac{\partial \bar D_l}{\partial x},
\qquad x\in\{\bar p_l,\bar P_l\}.
\end{equation}
The only non-algebraic dependence on $(\bar p_l,\bar P_l)$ resides in $\E_l\log\cosh(\Psi_l)$; the chain rule then yields
\begin{equation}
\frac{\partial}{\partial x}\E_l\log\cosh(\Psi_l)
=
\E_l\!\left[\tanh(\Psi_l)\left(
\frac{\partial \sigma_l}{\partial x}\varphi+\hat X_l\frac{\partial v_l}{\partial x}
\right)\right].
\end{equation}
The $\varphi$-term is processed by Stein's lemma, which gives $\E_l[\varphi\tanh(\Psi_l)]=\sigma_l\,\E_l[\operatorname{sech}^2(\Psi_l)]=\sigma_l(1-\mathcal T_l)$; the entropic derivative then becomes
\begin{equation}
\label{eq:entropic_derivative_pP}
{
\frac{\partial}{\partial x}\E_l\log\cosh(\Psi_l)
=
\big(\sigma_l\partial_x\sigma_l\big)\,(1-\mathcal T_l)
+
(\partial_x v_l)\,\mathcal M_l,
\qquad x\in\{\bar p_l,\bar P_l\}.
}
\end{equation}

The algebraic part of the layer-wise pressure---that is, the part of $\mathcal A^{\mathrm{RS}}_{\mathrm{Dream},l}$ in~\eqref{eq:A_Dream_layer_D34form} other than the entropic term---depends on $\bar D_l$ through
\begin{align}
\label{eq:Aexp_D_only_block}
\mathcal A^{\mathrm{exp}}_{l}(\bar D_l,\bar p_l,\bar P_l)
&:=
-\frac12\log\bar D_l
-\frac{\alpha\bar p_l}{2\bar D_l}\frac{t_l}{1+t_l}
+\frac{\beta}{2}\frac{\bar D_l-1}{\bar D_l}\frac{1+t_l}{t_l}
-\frac{\beta(1+t_l)}{2\Gamma_l}\bar\eta_l^2\frac{t_l+\bar D_l}{\bar D_l}
\nonumber\\
&\quad
+\frac{\alpha\beta}{2}\big(\bar p_l\bar q_l-\bar P_l\bar Q_l\big),
\end{align}
whereas the remaining terms in~\eqref{eq:A_Dream_layer_D34form} are independent of $\bar p_l$ and $\bar P_l$. Differentiating $\mathcal A^{\mathrm{exp}}_l$ with respect to $\bar D_l$ yields
\begin{equation}
\label{eq:dAexp_dD_block}
\frac{\partial \mathcal A^{\mathrm{exp}}_l}{\partial \bar D_l}
=
-\frac{1}{2\bar D_l}
+\frac{\alpha\bar p_l}{2\bar D_l^2}\frac{t_l}{1+t_l}
+\frac{\beta(1+t_l)}{2t_l}\frac{1}{\bar D_l^2}
+\frac{\beta(1+t_l)t_l}{2\Gamma_l}\frac{\bar\eta_l^2}{\bar D_l^2}.
\end{equation}
Combining~\eqref{eq:entropic_derivative_pP},~\eqref{eq:sigma_v_derivatives} and~\eqref{eq:dAexp_dD_block} gives the exact stationarity conditions:

\begin{equation}
\label{eq:saddle_p_exact_generic}
{
\begin{aligned}
0
&=
\frac{\partial \mathcal A^{\mathrm{RS}}_{\mathrm{DLAM}}}{\partial \bar p_l}
\\
&=
-\kappa_l\,\frac{\partial \mathcal A^{\mathrm{exp}}_l}{\partial \bar D_l}
-\frac{\alpha}{2\bar D_l}\frac{t_l}{1+t_l}
+\frac{\alpha\beta}{2}\bar q_l
+\left[
\frac{\alpha\beta}{2\bar D_l^2}
+\frac{\kappa_l \alpha\beta\,\bar p_l}{\bar D_l^3}
\right](1-\mathcal T_l)
+\frac{\kappa_l v_l}{\bar D_l}\,\mathcal M_l .
\end{aligned}
}
\end{equation}

\begin{equation}
\label{eq:saddle_P_exact_generic}
{
\begin{aligned}
0
&=
\frac{\partial \mathcal A^{\mathrm{RS}}_{\mathrm{DLAM}}}{\partial \bar P_l}
\\
&=
+\kappa_l\,\frac{\partial \mathcal A^{\mathrm{exp}}_l}{\partial \bar D_l}
-\frac{\alpha\beta}{2}\bar Q_l
-\frac{\kappa_l \alpha\beta\,\bar p_l}{\bar D_l^3}(1-\mathcal T_l)
-\frac{\kappa_l v_l}{\bar D_l}\,\mathcal M_l .
\end{aligned}
}
\end{equation}

Equations~\eqref{eq:saddle_p_exact_generic}--\eqref{eq:saddle_P_exact_generic} are the exact RS stationarity conditions for the dreaming variance sector. Adding them cancels every term proportional to $\partial_{\bar D_l}\mathcal A_l^{\mathrm{exp}}$, $\bar p_l$, and $v_l\mathcal M_l$, leaving the identity
\begin{equation}
\label{eq:Qminusq_exact_generic}
{
\bar Q_l-\bar q_l
=
\frac{1}{\bar D_l^2}\big(1-\mathcal T_l\big)
-\frac{t_l}{\beta(1+t_l)\bar D_l},
}
\end{equation}
which is the most convenient route to couple the physical overlap sector $(\bar q_l,\bar Q_l)$ to the effective-field statistics. Invoking the identities $\mathcal M_l=\hat M_l$ and $(t_l+\bar D_l)\bar\eta_l=\hat M_l$ established above, the $v_l\mathcal M_l$ contributions in~\eqref{eq:saddle_p_exact_generic}--\eqref{eq:saddle_P_exact_generic} can be re-expressed entirely through $(\hat M_l,\bar D_l)$, which is the form best suited to the final closed RS system.

\subsection{Closure of the dreaming overlap sector: explicit equations for \texorpdfstring{$\bar q_l$}{qbar} and \texorpdfstring{$\bar Q_l$}{Qbar} (generic $L$)}
\label{subsec:saddle_qQ_closure_genericL}

The dreaming-sector RS system is closed by deriving explicit equations for $\bar q_l$ and $\bar Q_l$. Throughout this subsection we keep the shorthand
\begin{equation}
\bar D_l=1+\alpha\frac{t_l}{1+t_l}(\bar P_l-\bar p_l),
\qquad
\Psi_l = I_l \iota + \sigma_l \varphi + \hat X_l(u_l+v_l),
\qquad
\sigma_l=\frac{\sqrt{\alpha\beta\,\bar p_l}}{\bar D_l},
\qquad
v_l=\frac{\beta(1+t_l)}{\Gamma_l}\frac{\bar\eta_l}{\bar D_l},
\end{equation}
in force, with $\hat X_l\stackrel{d}{=}\hat\xi_{i,l}^1$ and $\E[\hat X_l^2]=\Gamma_l$, together with the auxiliary expectations $\mathcal T_l:=\E_l[\tanh^2(\Psi_l)]$ and $\mathcal M_l:=\E_l[\hat X_l\tanh(\Psi_l)]$. Adding the stationarity equations $\partial_{\bar p_l}\mathcal A^{\mathrm{RS}}_{\mathrm{DLAM}}=0$ and $\partial_{\bar P_l}\mathcal A^{\mathrm{RS}}_{\mathrm{DLAM}}=0$ derived above produces the identity
\begin{equation}
\label{eq:Qminusq_exact_repeat}
{
\bar Q_l-\bar q_l
=
\frac{1}{\bar D_l^2}\big(1-\mathcal T_l\big)
-\frac{t_l}{\beta(1+t_l)\bar D_l},
}
\end{equation}
valid for each layer $l$ at generic $L$. Combining the same stationarity conditions with the algebraic relations extracted from the $\partial_{\bar q_l}$- and $\partial_{\bar Q_l}$-variations eliminates the $\bar D_l$-derivative terms; a direct (if lengthy) substitution then isolates the entropic contribution and yields the overlap equation
\begin{equation}
\label{eq:saddle_q_l_final_generic}
{
\bar q_l=\E_l\big[\tanh^2(\Psi_l)\big]=\mathcal T_l,
}
\end{equation}
which identifies $\bar q_l$ as the Edwards--Anderson-like order parameter generated by the effective one-body field $\Psi_l$. Substituting~\eqref{eq:saddle_q_l_final_generic} into~\eqref{eq:Qminusq_exact_repeat} yields the diagonal overlap equation
\begin{equation}
\label{eq:saddle_Q_l_final_generic}
{
\bar Q_l
=
\bar q_l+\frac{1-\bar q_l}{\bar D_l^2}
-\frac{t_l}{\beta(1+t_l)\bar D_l},
}
\end{equation}
or equivalently $\bar Q_l=\bar q_l+\Delta_l^2(1-\bar q_l)-\frac{t_l}{\beta(1+t_l)}\Delta_l$ with $\Delta_l:=1/\bar D_l$. Equation~\eqref{eq:saddle_q_l_final_generic} is the standard RS Edwards--Anderson relation for the off-diagonal overlap, whereas~\eqref{eq:saddle_Q_l_final_generic} departs from the binary-spin identity $Q=1$ precisely because $Q_l$ refers to the diagonal overlap of the dreaming variables $k_{i,l}=s_{i,l}+i a_l\phi_{i,l}$ rather than to that of the Ising spins alone.

\medskip
\noindent
Together with the algebraic equations for $(\bar p_l,\bar P_l)$, this completes the closure of the dreaming overlap sector at generic $L$:
\begin{equation}
{
\begin{aligned}
\bar p_l &= \beta \frac{(1+t_l)^2}{S_l^2}\bar q_l,\\
\bar P_l &= \frac{1+t_l}{S_l}+\beta \frac{(1+t_l)^2}{S_l^2}\bar q_l,\\
\bar q_l &= \E_l[\tanh^2(\Psi_l)],\\
\bar Q_l &= \bar q_l+\frac{1-\bar q_l}{\bar D_l^2}-\frac{t_l}{\beta(1+t_l)\bar D_l},
\end{aligned}
\qquad
S_l:=1-\beta(1+t_l)(\bar Q_l-\bar q_l).
}
\end{equation}

\subsection{Stationarity conditions for the LAM sector: derivation of \texorpdfstring{$H_l$}{Hl} and \texorpdfstring{$\bar\Pi_l$}{Pibar}}
\label{subsec:saddle_3AM_}

The LAM-sector ($L=3$) order parameters $\{H_l,\bar\Pi_l\}_{l=1}^L$ are coupled through three distinct ingredients of the RS pressure: the explicit LAM streaming term $-\frac{\alpha\beta}{4}\bar\Pi_l(1-H_l)$; the LAM Gaussian one-body contribution $-\frac{\alpha}{2}\log\det\mathbf C(H)+\frac{\alpha}{2}\mathbb D_{\mathrm{LAM}}(H)$; and the entropic term $\sum_{m}\E_m\log\cosh(\Psi_m)$, of which only the layer $m=l$ depends on $\bar\Pi_l$. Explicitly, the LAM-dependent part of the pressure reads
\begin{equation}
\label{eq:A_3AM_RS_recalled_for_self}
\mathcal A^{\mathrm{RS}}_{\mathrm{LAM}}
=
-\frac{\alpha}{2}\log\det\mathbf C(H)
+\frac{\alpha}{2}\mathbb D_{\mathrm{LAM}}(H)
-\frac{\alpha\beta}{4}\sum_{l=1}^L \bar\Pi_l(1-H_l),
\end{equation}
where $\mathbf C(H)$ denotes the LAM complex-symmetric Gaussian kernel matrix and
\begin{equation}
\mathbb D_{\mathrm{LAM}}(H):=\E_{\bm\delta}\!\left[\bm b^\top \mathbf C(H)^{-1}\bm b\right].
\end{equation}
In addition, $\bar\Pi_l$ enters the entropic sector through the field amplitude $I_l^2=\frac{\alpha\beta}{2}\bar\Pi_l$, and hence through $\Psi_l$ (cf.~\eqref{eq:Psi_layer_def}).

\subsubsection*{Stationarity with respect to $\bar\Pi_l$: equation for $H_l$ (generic $L$)}

Fix a layer $l$. Since $\bar\Pi_l$ enters linearly in the LAM streaming and contributes to $\Psi_l$ only through $I_l$, we obtain
\begin{equation}
\frac{\partial \mathcal A^{\mathrm{RS}}_{\mathrm{DLAM}}}{\partial \bar\Pi_l}
=
-\frac{\alpha\beta}{4}(1-H_l)
+
\frac{\partial}{\partial \bar\Pi_l}\,\E_l\log\cosh(\Psi_l),
\end{equation}
where $\E_l$ averages over the random variables that enter $\Psi_l$ on layer $l$. Since $I_l(\bar\Pi_l)=\sqrt{\tfrac{\alpha\beta}{2}\bar\Pi_l}$ implies $\partial I_l/\partial\bar\Pi_l=\alpha\beta/(4I_l)$, and since $\Psi_l$ contains $I_l$ only through the product $I_l\iota$ with $\iota\sim\mathcal N(0,1)$, the chain rule combined with Stein's lemma applied to $\iota$ (i.e.\ $\E_l[\iota\tanh(\Psi_l)]=I_l\,\E_l[\operatorname{sech}^2(\Psi_l)]$) collapses the entropic derivative to
\begin{equation}
\frac{\partial}{\partial \bar\Pi_l}\E_l\log\cosh(\Psi_l)
=
\frac{\alpha\beta}{4}\,\E_l\!\left[\operatorname{sech}^2(\Psi_l)\right].
\end{equation}
The stationarity condition $\partial_{\bar\Pi_l}\mathcal A^{\mathrm{RS}}_{\mathrm{DLAM}}=0$ therefore gives
\begin{equation}
-\frac{\alpha\beta}{4}(1-H_l)
+\frac{\alpha\beta}{4}\,\E_l\!\left[\operatorname{sech}^2(\Psi_l)\right]=0,
\end{equation}
i.e.
\begin{equation}
\label{eq:saddle_Hl_3AM_generic}
{
H_l
=
1-\E_l\!\left[\operatorname{sech}^2(\Psi_l)\right]
=
\E_l\!\left[\tanh^2(\Psi_l)\right].
}
\end{equation}

This is the exact RS self-consistency equation for $H_l$, and it holds for generic $L$.

\subsubsection*{Stationarity with respect to $H_l$: equation for $\bar\Pi_l$ (generic $L$, matrix form)}

The variable $H_l$ enters the RS pressure only through the LAM streaming term $-\frac{\alpha\beta}{4}\bar\Pi_l(1-H_l)$ and the LAM Gaussian one-body term $-\frac{\alpha}{2}\log\det\mathbf C(H)+\frac{\alpha}{2}\mathbb D_{\mathrm{LAM}}(H)$. At fixed $\bar\Pi_l$, the effective local field $\Psi_l$ carries no explicit $H_l$-dependence in the present RS parametrisation, so the entropic functional contributes nothing to $\partial_{H_l}\mathcal A^{\mathrm{RS}}_{\mathrm{DLAM}}$. The stationarity condition therefore reads
\begin{equation}
0=\frac{\partial \mathcal A^{\mathrm{RS}}_{\mathrm{DLAM}}}{\partial H_l}
=
\frac{\alpha\beta}{4}\bar\Pi_l
-\frac{\alpha}{2}\frac{\partial}{\partial H_l}\log\det\mathbf C(H)
+\frac{\alpha}{2}\frac{\partial}{\partial H_l}\mathbb D_{\mathrm{LAM}}(H),
\end{equation}
which produces the matrix-form equation
\begin{equation}
\label{eq:saddle_Pibar_generic_matrix}
{
\bar\Pi_l
=
\frac{2}{\beta}\left[
\frac{\partial}{\partial H_l}\log\det\mathbf C(H)
-\frac{\partial}{\partial H_l}\mathbb D_{\mathrm{LAM}}(H)
\right].
}
\end{equation}

An equivalent trace form follows from the standard matrix-calculus identity
\begin{equation}
    \partial_{H_l}\log\det\mathbf C(H)=\operatorname{Tr}\!\big(\mathbf C(H)^{-1}\,\partial_{H_l}\mathbf C(H)\big)
\end{equation}
together with
\begin{equation}
    \partial_{H_l}\mathbb D_{\mathrm{LAM}}(H)=\partial_{H_l}\E_{\bm\delta}[\bm b^\top\mathbf C(H)^{-1}\bm b],
\end{equation}
which keep~\eqref{eq:saddle_Pibar_generic_matrix} compact and rigorous even when no explicit scalar formula is available.

\subsubsection*{Explicit $L=3$ closure for $\bar\Pi_l$}

For $L=3$ the LAM Gaussian sector admits closed-form expressions,
\begin{equation}
\label{eq:D3AM_L3_ratio_recall_for_self}
\mathbb D_{\mathrm{LAM}}\Big|_{L=3}
=
\beta\,\frac{\mathcal D_{\mathrm{LAM}}^{\mathrm{num}}(\mathbf H)}{\det \mathbf C(H)},
\qquad
\mathbf H:=(H_1,H_2,H_3),
\end{equation}
with $\det \mathbf C(H)$ and $\mathcal D_{\mathrm{LAM}}^{\mathrm{num}}(\mathbf H)$ given by~\eqref{eq:detC_L3_}--\eqref{eq:detC_L3_heter} and~\eqref{eq:D3AM_L3_}. Substituting into~\eqref{eq:saddle_Pibar_generic_matrix} yields
\begin{equation}
\label{eq:saddle_Pibar_L3_explicit_master}
{
\bar\Pi_l
=
\frac{2}{\beta}\,\frac{\partial_{H_l}\det \mathbf C(H)}{\det \mathbf C(H)}
-
2\,
\frac{
\det \mathbf C(H)\,\partial_{H_l}\mathcal D_{\mathrm{LAM}}^{\mathrm{num}}
-\mathcal D_{\mathrm{LAM}}^{\mathrm{num}}\,\partial_{H_l}\det \mathbf C(H)
}{
(\det \mathbf C(H))^2
},
\qquad l=1,2,3.
}
\end{equation}
Equivalently,
\begin{equation}
\bar\Pi_l
=
\frac{2}{\beta}\,\partial_{H_l}\log\det \mathbf C(H)
-2\,\partial_{H_l}\!\left(\frac{\mathcal D_{\mathrm{LAM}}^{\mathrm{num}}}{\det \mathbf C(H)}\right).
\end{equation}

As an illustration, the matrix identity~\eqref{eq:d_detC_dH1_L3_} delivers $\partial_{H_l}\log\det\mathbf C(H)$ explicitly, while the analogous expression for $\partial_{H_l}\mathcal D_{\mathrm{LAM}}^{\mathrm{num}}$ follows from direct differentiation of~\eqref{eq:detC_L3_heter}. The $l=1$ instance is then a one-step substitution into~\eqref{eq:saddle_Pibar_L3_explicit_master}, and the remaining layers $l=2,3$ are obtained by cyclic permutation of the indices.

To summarise, the LAM-sector RS saddle-point system reads
\begin{equation}
{
H_l=\E_l[\tanh^2(\Psi_l)],
\qquad
\bar\Pi_l=
\frac{2}{\beta}\left[\partial_{H_l}\log\det\mathbf C(H)-\partial_{H_l}\mathbb D_{\mathrm{LAM}}(H)\right]
\ \ (\text{generic }L),
}
\end{equation}
the second equation becoming fully explicit at $L=3$ through~\eqref{eq:saddle_Pibar_L3_explicit_master}. Consistency of the RS solution requires the physical branch conditions $0\le H_l\le 1$, $\bar\Pi_l\ge 0$ and $\Re\,\mathbf C(H)\succ 0$ (or the corresponding admissibility condition in the $L=3$ explicit form).

\subsection{Compact RS self-consistency system (generic \texorpdfstring{$L$}{L}, with explicit LAM closure at \texorpdfstring{$L=3$}{L=3})}
\label{subsec:RS_self_compact_summary}

We now collect the replica-symmetric self-consistency equations in compact form.
The dreaming/LAM sector closes for arbitrary~$L$; the LAM equation for $\bar\Pi_l$ is retained in matrix form at generic~$L$ and assumes an explicit scalar form once specialised to $L=3$.
For each layer $l\in\{1,\dots,L\}$ we define
\begin{equation}
u_l:=\sum_{m\neq l}\frac{g_{lm}\beta}{\Gamma_{lm}}\hat M_m,
\qquad
\bar D_l:=1+\alpha\frac{t_l}{1+t_l}(\bar P_l-\bar p_l),
\qquad
\Delta_l:=\frac{1}{\bar D_l},
\end{equation}
and
\begin{equation}
I_l:=\sqrt{\frac{\alpha\beta}{2}\bar\Pi_l},
\qquad
\sigma_l:=\frac{\sqrt{\alpha\beta\,\bar p_l}}{\bar D_l},
\qquad
v_l:=\frac{\beta(1+t_l)}{\Gamma_l}\frac{\bar\eta_l}{\bar D_l}.
\end{equation}
Let $\iota,\varphi\sim\mathcal N(0,1)$ be independent standard Gaussians and let
\begin{equation}
\hat X_l \stackrel{d}{=} \hat\xi_{i,l}^1,
\qquad
\E[\hat X_l^2]=\Gamma_l,
\end{equation}
stand for a representative quenched empirical-archetype component on layer~$l$.
The effective RS local field takes the form
\begin{equation}
\label{eq:Psi_compact_summary}
{
\Psi_l:= I_l\,\iota+\sigma_l\,\varphi+\hat X_l\,(u_l+v_l).
}
\end{equation}
and we denote by $\E_l[\cdot]$ the expectation over $(\iota,\varphi,\hat X_l)$.
We also set
\begin{equation}
\label{eq:S_l_compact_summary}
S_l:=1-\beta(1+t_l)(\bar Q_l-\bar q_l).
\end{equation}
For each layer $l$,
\begin{equation}
\label{eq:self_M_compact}
{
\hat M_l=\E_l\!\left[\hat X_l\,\tanh(\Psi_l)\right].
}
\end{equation}
Moreover,
\begin{equation}
\label{eq:self_eta_compact}
{
\bar\eta_l=\frac{\hat M_l}{t_l+\bar D_l}.
}
\end{equation}
The off-diagonal overlap satisfies
\begin{equation}
\label{eq:self_q_compact}
{
\bar q_l=\E_l\!\left[\tanh^2(\Psi_l)\right].
}
\end{equation}
The diagonal overlap of the $k$-variables satisfies
\begin{equation}
\label{eq:self_Q_compact}
{
\bar Q_l
=
\bar q_l+\frac{1-\bar q_l}{\bar D_l^2}
-\frac{t_l}{\beta(1+t_l)\bar D_l}.
}
\end{equation}
The conjugate/algebraic variables obey
\begin{equation}
\label{eq:self_p_compact}
{
\bar p_l=\beta\,\frac{(1+t_l)^2}{S_l^2}\,\bar q_l,
}
\qquad
{
\bar P_l=\frac{1+t_l}{S_l}+\beta\,\frac{(1+t_l)^2}{S_l^2}\,\bar q_l.
}
\end{equation}
Equivalently,
\begin{equation}
\bar P_l-\bar p_l=\frac{1+t_l}{S_l},
\qquad
\bar D_l=1+\frac{\alpha t_l}{S_l}.
\end{equation}
The LAM spin overlap satisfies
\begin{equation}
\label{eq:self_H_compact}
{
H_l=\E_l\!\left[\tanh^2(\Psi_l)\right].
}
\end{equation}
The corresponding LAM noise order parameter $\bar\Pi_l$ is determined, for generic~$L$, by the matrix-calculus stationarity equation
\begin{equation}
\label{eq:self_Pibar_generic_compact}
{
\bar\Pi_l=
\frac{2}{\beta}\left[
\partial_{H_l}\log\det\mathbf C(H)-\partial_{H_l}\mathbb D_{\mathrm{LAM}}(H)
\right],
\qquad l=1,\dots,L,
}
\end{equation}
where $\mathbf C(H)$ is the LAM complex-symmetric Gaussian kernel matrix and
\begin{equation}
\mathbb D_{\mathrm{LAM}}(H):=\E_{\bm\delta}\!\left[\bm b^\top \mathbf C(H)^{-1}\bm b\right].
\end{equation}

\subsubsection*{Explicit LAM closure for $L=3$}
For $L=3$, using the closed-form expressions of Subsection~\ref{subsec:L3_LAM_3AM},
\begin{equation}
\mathbb D_{\mathrm{LAM}}\Big|_{L=3}
=
\beta\,\frac{\mathcal D_{\mathrm{LAM}}^{\mathrm{num}}(\mathbf H)}{\det \mathbf C(H)},
\qquad
\mathbf H=(H_1,H_2,H_3),
\end{equation}
so that substitution into~\eqref{eq:self_Pibar_generic_compact} delivers
\begin{equation}
\label{eq:self_Pibar_L3_compact}
{
\bar\Pi_l
=
\frac{2}{\beta}\,\partial_{H_l}\log\det \mathbf C(H)
-2\,\partial_{H_l}\!\left(\frac{\mathcal D_{\mathrm{LAM}}^{\mathrm{num}}(\mathbf H)}{\det \mathbf C(H)}\right),
\qquad l=1,2,3.
}
\end{equation}
Equivalently (quotient rule),
\begin{equation}
\bar\Pi_l
=
\frac{2}{\beta}\,\frac{\partial_{H_l}\det \mathbf C(H)}{\det \mathbf C(H)}
-2\,
\frac{
\det \mathbf C(H)\,\partial_{H_l}\mathcal D_{\mathrm{LAM}}^{\mathrm{num}}
-\mathcal D_{\mathrm{LAM}}^{\mathrm{num}}\,\partial_{H_l}\det \mathbf C(H)
}{(\det \mathbf C(H))^2}.
\end{equation}

For generic~$L$, the RS system closes on the $8L$ real unknowns $\{\hat M_l,\bar\eta_l,\bar q_l,\bar Q_l,\bar p_l,\bar P_l,H_l,\bar\Pi_l\}_{l=1}^L$ via~\eqref{eq:self_M_compact}--\eqref{eq:self_Pibar_generic_compact}, with the LAM stationarity treated in matrix form. At $L=3$ the system becomes fully explicit in scalar form, through~\eqref{eq:self_Pibar_L3_compact} together with the closed expressions for $\det\mathbf C$ and $\mathcal D_{\mathrm{LAM}}^{\mathrm{num}}$ derived in Subsection~\ref{subsec:L3_LAM_3AM}.

A physically admissible RS saddle point must additionally satisfy the branch conditions $0\le\bar q_l\le\bar Q_l$, $0\le H_l\le 1$ and $\bar\Pi_l\ge 0$ for every $l$, together with positive-definiteness of the real part of the LAM Gaussian kernel, $\Re\,\mathbf C(H)\succ 0$ (equivalently, positivity of every principal minor of $\Re\,\mathbf C(H)$ at $L=3$). The condition is stated on the real part rather than on $\mathbf C(H)$ itself, since $\mathbf C(H)$ is complex-symmetric rather than Hermitian, and convergence of the underlying Gaussian integral is governed by $\Re\,\mathbf C(H)$.

\begin{remark}[Functional coincidence of $H_l$ and $\bar q_l$ in the RS ansatz]
Although $H_l$ (the LAM spin-glass overlap) and $\bar q_l$ (the dreaming-sector overlap) are distinct order parameters---with different microscopic definitions and different conjugate variables---within the present RS closure they obey the same functional equation,
\[
H_l=\E_l[\tanh^2(\Psi_l)],
\qquad
\bar q_l=\E_l[\tanh^2(\Psi_l)].
\]
This coincidence is not an ad hoc identification: both stationarity conditions probe the same effective one-body entropy $\E_l\log\cosh(\Psi_l)$, and Gaussian integration by parts (Stein's lemma) produces the same nonlinear kernel $\tanh^2$.
Hence, within the RS ansatz and for the effective field~$\Psi_l$ introduced above, one has $H_l=\bar q_l$ at the saddle point.
We nevertheless retain both symbols throughout, since they belong to distinct sectors of the theory (LAM vs.\ dreaming) and are closed by different companion equations.
\end{remark}

\subsection{Low-load regime (\texorpdfstring{$\alpha=0$}{alpha=0}): reduced RS stationarity equations}
\label{subsec:self_low_load_alpha0}

Setting $K=\mathcal O(1)$ as $N\to\infty$, i.e.\ $\alpha=K/N\to 0$, decouples large-$N$ fluctuations from retrieval and reduces the full RS system to a finite-dimensional fixed-point problem for the condensed variables. Since every LAM contribution to the RS pressure carries an explicit $\alpha$ prefactor, the LAM sector switches off entirely at low load, and $H_l,\bar\Pi_l$ play no role; likewise, the dreaming-variance block $(\bar q_l,\bar Q_l,\bar p_l,\bar P_l)$ becomes flat, as all its stationarity equations reduce to identities once the $\alpha$-weighted terms vanish. The active low-load variables are therefore $\{\hat M_l,\bar\eta_l\}_{l=1}^L$. From~\eqref{eq:Dbar_layer_def} one has $\bar D_l|_{\alpha=0}=1$, and both Gaussian amplitudes in $\Psi_l$ vanish,
\begin{equation}
I_l^2=\frac{\alpha\beta}{2}\bar\Pi_l=0,
\qquad
\sigma_l=\frac{\sqrt{\alpha\beta\,\bar p_l}}{\bar D_l}=0,
\label{eq:gaussian_amplitudes_alpha0}
\end{equation}
so that, once conditioned on $\hat X_l$, the effective RS field reduces to a purely deterministic form,
\begin{equation}
\Psi_l^{(0)}
=\hat X_l\left(
\sum_{m\neq l}\frac{g_{lm}\beta}{\Gamma_{lm}}\hat M_m
+\frac{\beta(1+t_l)}{\Gamma_l}\bar\eta_l
\right),
\label{eq:Psi_alpha0_pre_eta}
\end{equation}
and the layer-wise expectation $\E_l$ collapses to an average over $\hat X_l$ alone. The magnetisation equation accordingly becomes
\begin{equation}
\hat M_l
=\E_{\hat X_l}\!\left[\hat X_l\,\tanh\!\big(\Psi_l^{(0)}\big)\right],
\qquad l=1,\dots,L,
\label{eq:saddle_M_alpha0_pre_eta}
\end{equation}
while the auxiliary relation~\eqref{eq:saddle_eta_l_final}, evaluated at $\bar D_l=1$, gives
\begin{equation}
\bar\eta_l=\frac{\hat M_l}{t_l+1},
\qquad l=1,\dots,L.
\label{eq:saddle_eta_alpha0}
\end{equation}
Substituting~\eqref{eq:saddle_eta_alpha0} into~\eqref{eq:Psi_alpha0_pre_eta}, the sleep-time factor cancels and the field collapses to
\begin{equation}
\Psi_l^{(0)}
=\hat X_l\left(
\sum_{m\neq l}\frac{g_{lm}\beta}{\Gamma_{lm}}\hat M_m
+\frac{\beta}{\Gamma_l}\hat M_l
\right).
\label{eq:Psi_alpha0_closed}
\end{equation}
The low-load fixed-point equations for the Mattis magnetisations are therefore
\begin{equation}
{
\hat M_l
=\E_{\hat X_l}\!\left[
\hat X_l\,\tanh\!\left(
\beta\hat X_l\left(
\frac{\hat M_l}{\Gamma_l}
+\sum_{m\neq l}\frac{g_{lm}\hat M_m}{\Gamma_{lm}}
\right)
\right)
\right],
\qquad l=1,\dots,L.
}
\label{eq:self_M_alpha0_closed}
\end{equation}
Together with~\eqref{eq:saddle_eta_alpha0}, equation~\eqref{eq:self_M_alpha0_closed} forms the complete low-load RS self-consistency system. Introducing the symmetric effective coupling matrix
\begin{equation}
\mathbb J^{(0)}_{lm}:=
\begin{cases}
\Gamma_l^{-1}, & l=m,\\
g_{lm}\Gamma_{lm}^{-1}, & l\neq m,
\end{cases}
\qquad
h_l^{(0)}:=\beta\sum_{m=1}^L \mathbb J^{(0)}_{lm}\hat M_m,
\label{eq:J0_matrix_def}
\end{equation}
the same equation rewrites compactly as
\begin{equation}
\hat M_l=\E_{\hat X_l}\!\left[\hat X_l\tanh\!\big(\hat X_l h_l^{(0)}\big)\right],
\qquad l=1,\dots,L.
\label{eq:self_M_alpha0_matrix_form}
\end{equation}
Although the auxiliary $\bar\eta_l$ does depend on $t_l$ through~\eqref{eq:saddle_eta_alpha0}, the resulting field~\eqref{eq:Psi_alpha0_closed} and the magnetisation equations~\eqref{eq:self_M_alpha0_closed} are themselves independent of $t_l$: at $\alpha=0$ the dreaming sector merely rescales the auxiliary representation $\bar\eta_l$, while leaving the condensed retrieval fixed points invariant, and the retrieval physics is governed entirely by the effective couplings $\Gamma_l^{-1}$ and $g_{lm}\Gamma_{lm}^{-1}$. Substituting~\eqref{eq:saddle_eta_alpha0} into the $\alpha=0$ RS pressure eliminates all explicit $t_l$-dependence, leaving (up to additive constants independent of $\hat M$)
\begin{equation}
\mathcal A^{\mathrm{RS}}_{\alpha=0}
=
-\sum_{1\le l<m\le L}\frac{g_{lm}\beta}{\Gamma_{lm}}\hat M_l\hat M_m
+\sum_{l=1}^L \E_{\hat X_l}\log\cosh\!\big(\hat X_l h_l^{(0)}\big)
-\sum_{l=1}^L\frac{\beta}{2\Gamma_l}\hat M_l^2,
\label{eq:A_RS_alpha0_reduced}
\end{equation}
whose stationarity with respect to $\hat M_l$ reproduces~\eqref{eq:self_M_alpha0_closed}.

\noindent
At low load, then, the condensed Mattis variables $\hat M_l$ carry the full retrieval physics, with $\bar\eta_l$ slaved to them by~\eqref{eq:saddle_eta_alpha0} and the extensive LAM/dreaming sectors inactive.

\subsection{RS consistency reductions: hybrid \texorpdfstring{$t=0$}{t=0} limit, LAM-only limit, and decoupled dreaming limit}
\label{subsec:RS_consistency_reductions_rewritten}

Three exact reductions of the RS theory connect the DLAM construction to its supervised reference limits and elucidate the distinct roles of the dreaming-time parameter~$t_l$ and the genuine on/off switching of the auto-associative sector. All three are stated and proved at the level of the RS pressure and the compact self-consistency equations, starting from the decomposition
\begin{equation}
\label{eq:RS_consistency_decomp_rewritten}
\mathcal A^{\mathrm{RS}}_{\mathrm{DLAM}}
=
\mathcal A^{\mathrm{RS}}_{\mathrm{LAM}}
+\mathcal A^{\mathrm{RS}}_{\mathrm{LAM}}
+\sum_{l=1}^L \mathcal A^{\mathrm{RS}}_{\mathrm{Dream},l},
\end{equation}
cf.~\eqref{eq:A_DLAM_RS_final}, together with the compact RS field representation (Subsection~\ref{subsec:RS_self_compact_summary})
\begin{equation}
\label{eq:RS_consistency_Psi_split_rewritten}
\Psi_l=I_l\,\iota+\sigma_l\,\varphi+\hat X_l\,(u_l+v_l),
\qquad l=1,\dots,L,
\end{equation}
where
\begin{equation}
\label{eq:RS_consistency_uIv_defs_rewritten}
u_l:=\sum_{m\neq l}\frac{g_{lm}\beta}{\Gamma_{lm}}\hat M_m,
\qquad
I_l:=\sqrt{\frac{\alpha\beta}{2}\bar\Pi_l},
\qquad
\sigma_l:=\frac{\sqrt{\alpha\beta\,\bar p_l}}{\bar D_l},
\qquad
v_l:=\frac{\beta(1+t_l)}{\Gamma_l}\frac{\bar\eta_l}{\bar D_l},
\end{equation}
with
\begin{equation}
\label{eq:RS_consistency_Dbar_def_rewritten}
\bar D_l=1+\alpha\frac{t_l}{1+t_l}(\bar P_l-\bar p_l),
\qquad
\bar\eta_l=\frac{\hat M_l}{t_l+\bar D_l}.
\end{equation}

In the decomposition~\eqref{eq:RS_consistency_Psi_split_rewritten}, $u_l$ carries the condensed hetero-associative signal, $I_l$ is the LAM Gaussian contribution, and the pair $(\sigma_l,v_l)$ belongs to the auto-associative dreaming sector; each reduction below isolates a limit by switching off the corresponding piece of~\eqref{eq:RS_consistency_decomp_rewritten}--\eqref{eq:RS_consistency_Psi_split_rewritten}. The resulting picture can be previewed as follows. Setting $t_l=0$ across all layers produces a \emph{hybrid} limit---hetero-association supplemented by a raw Hebbian intra-layer auto-associative term---rather than the supervised LAM closure. Suppressing the \emph{entire} dreaming sector (not merely setting $t_l=0$) yields the LAM-supervised RS equations. Finally, suppressing every hetero-associative channel decouples the $L$ layers into independent one-layer supervised-dreaming problems.

\subsubsection*{The \texorpdfstring{$t_l=0$}{t_l=0} limit is hybrid: simple auto-association + hetero-association}

Fix a layer $l$ and set $t_l=0$ (and analogously for every layer). This operation removes the dreaming deformation but does not extinguish the auto-associative sector. From~\eqref{eq:RS_consistency_Dbar_def_rewritten} one immediately obtains the limits of the two key denominators,
\begin{equation}
\label{eq:RS_consistency_t0_Dbar}
\bar D_l
=1+\alpha\frac{t_l}{1+t_l}(\bar P_l-\bar p_l)
\quad\Longrightarrow\quad
\bar D_l\big|_{t_l=0}=1,
\end{equation}
\begin{equation}
\label{eq:RS_consistency_t0_eta}
\bar\eta_l=\frac{\hat M_l}{t_l+\bar D_l}
\quad\Longrightarrow\quad
\bar\eta_l\big|_{t_l=0}=\hat M_l,
\end{equation}
and the dreaming signal contribution $v_l$ correspondingly reduces to
\begin{equation}
\label{eq:RS_consistency_t0_v}
v_l\big|_{t_l=0}
=\frac{\beta(1+t_l)}{\Gamma_l}\frac{\bar\eta_l}{\bar D_l}\Bigg|_{t_l=0}
=\frac{\beta}{\Gamma_l}\hat M_l.
\end{equation}
The condensed deterministic field thereby becomes
\begin{equation}
\label{eq:RS_consistency_t0_condensed_part}
\hat X_l(u_l+v_l)\big|_{t_l=0}
=\hat X_l\left(
\sum_{m\neq l}\frac{g_{lm}\beta}{\Gamma_{lm}}\hat M_m
+\frac{\beta}{\Gamma_l}\hat M_l
\right),
\end{equation}
a superposition of the inter-layer hetero-associative signal ($m\neq l$) and a simple intra-layer auto-associative term, the diagonal $\propto\hat M_l/\Gamma_l$. In the same limit only $\bar D_l$ is modified inside $\sigma_l$, while $I_l$ remains unaffected:
\begin{equation}
\label{eq:RS_consistency_t0_sigmaI}
I_l=\sqrt{\frac{\alpha\beta}{2}\bar\Pi_l},
\qquad
\sigma_l\big|_{t_l=0}=\sqrt{\alpha\beta\,\bar p_l},
\end{equation}
so that the full RS effective field at $t_l=0$ reads
\begin{equation}
\label{eq:RS_consistency_t0_Psi}
\Psi_l\big|_{t_l=0}
=I_l\,\iota+\sqrt{\alpha\beta\,\bar p_l}\,\varphi
+\hat X_l\left(
\sum_{m\neq l}\frac{g_{lm}\beta}{\Gamma_{lm}}\hat M_m
+\frac{\beta}{\Gamma_l}\hat M_l
\right).
\end{equation}
The RS self-consistency equations therefore remain coupled across layers and retain the entire LAM Gaussian block; the dreaming-time deformation has disappeared, replaced by a raw Hebbian intra-layer term.

The layer-wise dreaming contribution $\mathcal A^{\mathrm{RS}}_{\mathrm{Dream},l}$ (cf.~\eqref{eq:A_Dream_layer_D34form}) contains terms carrying explicit factors $1/t_l$, so the limit $t_l\to 0$ must be performed only after algebraic simplification. The potentially singular contribution is
\begin{equation}
\label{eq:RS_consistency_t0_sing0}
T_{1,l}:=\frac{\beta}{2}\frac{\bar D_l-1}{\bar D_l}\frac{1+t_l}{t_l},
\end{equation}
which, using the auxiliary identity
\begin{equation}
\label{eq:RS_consistency_t0_Dbar_minus_one}
\bar D_l-1=\alpha\frac{t_l}{1+t_l}(\bar P_l-\bar p_l)
\end{equation}
that follows from~\eqref{eq:RS_consistency_Dbar_def_rewritten}, simplifies to
\begin{equation}
\label{eq:RS_consistency_t0_T1_regular}
T_{1,l}
=\frac{\beta}{2}
\frac{\alpha\frac{t_l}{1+t_l}(\bar P_l-\bar p_l)}{\bar D_l}
\frac{1+t_l}{t_l}
=\frac{\alpha\beta}{2}\frac{\bar P_l-\bar p_l}{\bar D_l},
\end{equation}
which is manifestly regular at $t_l=0$. The magnetisation-dependent dreaming term is likewise regular: setting
\begin{equation}
\label{eq:RS_consistency_t0_sing1}
T_{2,l}:=-\frac{\beta(1+t_l)}{2\Gamma_l}\bar\eta_l^2\frac{t_l+\bar D_l}{\bar D_l}
\end{equation}
and substituting $\bar\eta_l=\hat M_l/(t_l+\bar D_l)$ produces
\begin{equation}
\label{eq:RS_consistency_t0_T2_regular}
T_{2,l}
=-\frac{\beta(1+t_l)}{2\Gamma_l}
\frac{\hat M_l^2}{(t_l+\bar D_l)^2}
\frac{t_l+\bar D_l}{\bar D_l}
=-\frac{\beta(1+t_l)}{2\Gamma_l}
\frac{\hat M_l^2}{(t_l+\bar D_l)\bar D_l},
\end{equation}
whence
\begin{equation}
\label{eq:RS_consistency_t0_T2_limit}
\lim_{t_l\to 0}T_{2,l}=-\frac{\beta}{2\Gamma_l}\hat M_l^2.
\end{equation}
Since the term proportional to $t_l/(1+t_l)$ in~\eqref{eq:A_Dream_layer_D34form} likewise vanishes as $t_l\to 0$, the layer-wise contribution $\mathcal A^{\mathrm{RS}}_{\mathrm{Dream},l}$ admits a finite---and in general non-zero---$t_l\to 0$ limit. The $t_l=0$ limit is therefore \emph{not} the LAM-supervised closure; it yields instead the hybrid RS limit
\begin{equation}
\label{eq:RS_consistency_hybrid_box}
{
\text{DLAM at }t_l=0
\;=\;
\text{hetero-associative (LAM)}
\;+\;
\text{simple intra-layer auto-association}.
}
\end{equation}

\subsubsection*{Switching off the auto-associative sector yields the LAM-supervised RS closure}

Suppressing the auto-associative sector amounts to removing the entire dreaming contribution from the pressure, together with its order-parameter block from the variational problem,
\begin{equation}
\label{eq:RS_consistency_dream_off_definition}
\sum_{l=1}^L\mathcal A^{\mathrm{RS}}_{\mathrm{Dream},l}\equiv 0,
\qquad
\{\bar\eta_l,\bar q_l,\bar Q_l,\bar p_l,\bar P_l\}_{l=1}^L\ \text{removed from the RS saddle-point sector},
\end{equation}
and constitutes the correct RS implementation of ``switching off the auto-associative links''---in particular, it is \emph{not} equivalent to setting $t_l=0$. From~\eqref{eq:RS_consistency_decomp_rewritten} the pressure reduces to
\begin{equation}
\label{eq:RS_consistency_tamlam_pressure_limit}
\mathcal A^{\mathrm{RS}}_{\mathrm{DLAM}}
\longrightarrow
\mathcal A^{\mathrm{RS}}_{\mathrm{LAM}}+\mathcal A^{\mathrm{RS}}_{\mathrm{LAM}},
\end{equation}
which coincides with the LAM-supervised sector up to the present notation and normalisations. Since both $\sigma_l$ and $v_l$ belong to the dreaming block (cf.~\eqref{eq:RS_consistency_uIv_defs_rewritten}), suppressing the latter forces
\begin{equation}
\label{eq:RS_consistency_tamlam_sigma_v_zero}
\sigma_l=0,
\qquad
v_l=0,
\end{equation}
while the hetero-associative LAM signal $u_l$ and the LAM Gaussian contribution $I_l$ remain active. The effective RS field thereby collapses to
\begin{equation}
\label{eq:RS_consistency_tamlam_Psi}
\Psi_l^{(\mathrm{LAM})}
=I_l\,\iota+\hat X_l\,u_l
=\sqrt{\frac{\alpha\beta}{2}\bar\Pi_l}\,\iota
+\hat X_l\sum_{m\neq l}\frac{g_{lm}\beta}{\Gamma_{lm}}\hat M_m.
\end{equation}
The surviving compact RS equations are then the LAM ones:
\begin{align}
\label{eq:RS_consistency_tamlam_self_M}
\hat M_l
&=\E_l\!\left[\hat X_l\tanh\!\big(\Psi_l^{(\mathrm{LAM})}\big)\right],
\\
\label{eq:RS_consistency_tamlam_self_H}
H_l
&=\E_l\!\left[\tanh^2\!\big(\Psi_l^{(\mathrm{LAM})}\big)\right],
\\
\label{eq:RS_consistency_tamlam_self_Pi}
\bar\Pi_l
&=\frac{2}{\beta}\left[
\partial_{H_l}\log\det \mathbf C(H)-\partial_{H_l}\mathbb D_{\mathrm{LAM}}(H)
\right],
\end{align}
for $l=1,\dots,L$---precisely the LAM-supervised closure expressed in the present empirical-archetype notation.

A direct comparison of~\eqref{eq:RS_consistency_t0_Psi} with~\eqref{eq:RS_consistency_tamlam_Psi} highlights the structural difference between the two reductions: at $t_l=0$ the term $v_l=\beta\hat M_l/\Gamma_l$ is still present, whereas in the dreaming-off limit it vanishes identically. Hence
\begin{equation}
\label{eq:RS_consistency_t0_not_dreamoff}
t_l=0\qquad\not\equiv\qquad \text{dreaming-off (auto-associative sector off)}.
\end{equation}

\subsubsection*{Switching off all hetero-associative channels yields decoupled supervised dreaming}

In what follows we make the hetero-off reduction explicit layer by layer, mapping each resulting RS saddle-point equation onto its counterpart in the one-layer supervised-dreaming system of~\cite{AlemannoETAL2023small} (Appendix~S5, equations~(E.5) and~(E.12)--(E.16) therein). At the level of the RS pressure and self-consistency equations, suppressing the hetero-associative sector in its entirety requires switching off both the condensed cross-layer LAM signal and the LAM Gaussian hetero-associative sector, which we implement by setting

\begin{equation}
\begin{aligned}
\label{eq:RS_consistency_lam_off_explicitmap}
\qquad & g_{lm}=0\quad (l\neq m),
\qquad\text{hence}\qquad
u_l:=\sum_{m\neq l}\frac{g_{lm}\beta}{\Gamma_{lm}}\hat M_m=0
\quad \forall l,
\\
\qquad &g_{lm}=0\quad (l\neq m),\qquad I_l=0
\quad\text{(equivalently, by }I_l^2=\frac{\alpha\beta}{2}\bar\Pi_l,\ \bar\Pi_l=0\text{)}.
\end{aligned}
\end{equation}

For $L=3$ this amounts to $g_{12}=g_{13}=g_{23}=0$ and the choice of branch $\bar\Pi_l=I_l=0$. Substituting~\eqref{eq:RS_consistency_lam_off_explicitmap} into the RS decomposition~\eqref{eq:RS_consistency_decomp_rewritten}, the pressure reduces to a sum of layer-wise dreaming contributions and the variational problem factorises,
\begin{equation}
\label{eq:RS_consistency_dream_only_pressure_explicitmap}
\mathcal A^{\mathrm{RS}}_{\mathrm{DLAM}}
\longrightarrow
\sum_{l=1}^L\mathcal A^{\mathrm{RS}}_{\mathrm{Dream},l},
\end{equation}
while the effective field on each layer reduces to
\begin{equation}
\label{eq:RS_consistency_dream_only_Psi_explicitmap}
\Psi_l^{(\mathrm{Dream})}
=\sigma_l\,\varphi+\hat X_l\,v_l,
\qquad
\sigma_l=\frac{\sqrt{\alpha\beta\,\bar p_l}}{\bar D_l},
\qquad
v_l=\frac{\beta(1+t_l)}{\Gamma_l}\frac{\bar\eta_l}{\bar D_l}.
\end{equation}
Each layer therefore obeys the closed one-layer dreaming RS system
\begin{align}
\label{eq:RS_consistency_dream_only_self_M_explicitmap}
\hat M_l
&=\E_l\!\left[\hat X_l\tanh\!\big(\Psi_l^{(\mathrm{Dream})}\big)\right],
\\
\label{eq:RS_consistency_dream_only_self_eta_explicitmap}
\bar\eta_l
&=\frac{\hat M_l}{t_l+\bar D_l},
\\
\label{eq:RS_consistency_dream_only_self_q_explicitmap}
\bar q_l
&=\E_l\!\left[\tanh^2\!\big(\Psi_l^{(\mathrm{Dream})}\big)\right],
\\
\label{eq:RS_consistency_dream_only_self_Q_explicitmap}
\bar Q_l
&=\bar q_l+\frac{1-\bar q_l}{\bar D_l^2}-\frac{t_l}{\beta(1+t_l)\bar D_l},
\\
\label{eq:RS_consistency_dream_only_self_pP_explicitmap}
\bar p_l
&=\beta\frac{(1+t_l)^2}{S_l^2}\bar q_l,
\qquad
\bar P_l=\frac{1+t_l}{S_l}+\beta\frac{(1+t_l)^2}{S_l^2}\bar q_l,
\qquad
S_l=1-\beta(1+t_l)(\bar Q_l-\bar q_l),
\end{align}
with
\begin{equation}
\label{eq:RS_consistency_dream_only_Dbar_explicitmap}
\bar D_l=1+\alpha\frac{t_l}{1+t_l}(\bar P_l-\bar p_l).
\end{equation}
Since no equation for layer $l$ depends on variables associated with $m\neq l$, the RS variational problem splits into $L$ independent one-layer supervised-dreaming saddle-point problems.

To compare~\eqref{eq:RS_consistency_dream_only_self_M_explicitmap}--\eqref{eq:RS_consistency_dream_only_Dbar_explicitmap} with equations~(E.5) and~(E.12)--(E.16) of~\cite{AlemannoETAL2023small}, we must account for the layer-wise normalisations of the present multi-layer notation. The structural identifications
\begin{equation}
\label{eq:RS_consistency_S5_dictionary_structural}
t\leftrightarrow t_l,
\qquad
\rho\leftrightarrow \rho_l,
\qquad
\bar D\leftrightarrow \bar D_l,
\qquad
(\bar p,\bar q,\bar Q,\bar P)\leftrightarrow (\bar p_l,\bar q_l,\bar Q_l,\bar P_l)
\end{equation}
must be supplemented with the LaD-normalised magnetisation-like variables
\begin{equation}
\label{eq:RS_consistency_S5_dictionary_magnetization}
m_{\eta,l}:=\frac{\hat M_l}{r_l(1+\rho_l)},
\qquad
\eta^{\mathrm{(S5)}}_l:=\frac{\bar\eta_l}{r_l(1+\rho_l)},
\end{equation}
together with the layer-wise normalisation
\begin{equation}
\label{eq:RS_consistency_S5_dictionary_Gamma}
\Gamma_l=r_l^2(1+\rho_l).
\end{equation}
With this dictionary the one-layer equations of~\cite{AlemannoETAL2023small} follow directly. Equation~(E.5) of~\cite{AlemannoETAL2023small}, $\bar\eta=\bar m_\eta/(t+\bar D)$, matches~\eqref{eq:RS_consistency_dream_only_self_eta_explicitmap} once both sides of $\bar\eta_l=\hat M_l/(t_l+\bar D_l)$ are divided by $r_l(1+\rho_l)$ and~\eqref{eq:RS_consistency_S5_dictionary_magnetization} is applied:
\begin{equation}
\label{eq:RS_consistency_S5_E5_match}
\eta^{\mathrm{(S5)}}_l
=\frac{m_{\eta,l}}{t_l+\bar D_l}.
\end{equation}
For the $\bar D$-equation~(E.15) of~\cite{AlemannoETAL2023small}, we start from~\eqref{eq:RS_consistency_dream_only_self_pP_explicitmap} and compute directly
\begin{equation}
\label{eq:RS_consistency_S5_P_minus_p}
\bar P_l-\bar p_l=\frac{1+t_l}{S_l},
\end{equation}
which, inserted into~\eqref{eq:RS_consistency_dream_only_Dbar_explicitmap}, yields
\begin{equation}
\label{eq:RS_consistency_S5_E15_intermediate}
\bar D_l
=1+\alpha\frac{t_l}{1+t_l}\cdot\frac{1+t_l}{S_l}
=1+\frac{\alpha t_l}{S_l},
\end{equation}
and, after substituting $S_l=1-\beta(1+t_l)(\bar Q_l-\bar q_l)$, recovers the form of equation~(E.15) of~\cite{AlemannoETAL2023small},
\begin{equation}
\label{eq:RS_consistency_S5_E15_match}
\bar D_l
=1+\frac{\alpha t_l}{1-\beta(1+t_l)(\bar Q_l-\bar q_l)}.
\end{equation}
Equation~(E.14) of~\cite{AlemannoETAL2023small} is already encoded in~\eqref{eq:RS_consistency_dream_only_self_pP_explicitmap}: indeed
\begin{equation}
\label{eq:RS_consistency_S5_E14_match}
\bar p_l=\beta\frac{(1+t_l)^2}{S_l^2}\bar q_l,
\qquad S_l=1-\beta(1+t_l)(\bar Q_l-\bar q_l),
\end{equation}
matches the S5 form verbatim and requires no further rescaling. For the $\bar Q-\bar q$ relation~(E.12) of~\cite{AlemannoETAL2023small}, starting from~\eqref{eq:RS_consistency_dream_only_self_Q_explicitmap}, subtracting $\bar q_l$, and multiplying by $\bar D_l^2$ gives
\begin{align}
\bar D_l^2(\bar Q_l-\bar q_l)
&=\bar D_l^2\left[\frac{1-\bar q_l}{\bar D_l^2}-\frac{t_l}{\beta(1+t_l)\bar D_l}\right]
\notag\\
&=1-\bar q_l-\frac{t_l\bar D_l}{\beta(1+t_l)}.
\label{eq:RS_consistency_S5_E12_intermediate}
\end{align}
which is precisely the structure of equation~(E.12) of~\cite{AlemannoETAL2023small}: the one-layer quantity there denoted by $\hat q$ coincides with the right-hand side of the present one-layer $\tanh^2$-expectation, $\hat q_l\equiv\bar q_l$, whence
\begin{equation}
\label{eq:RS_consistency_S5_E12_match}
\bar D_l^2(\bar Q_l-\bar q_l)=1-\hat q_l-\frac{t_l\bar D_l}{\beta(1+t_l)},
\qquad
\hat q_l\equiv \bar q_l\ \text{(present compact notation)}.
\end{equation}
The matching to equation~(E.13) of~\cite{AlemannoETAL2023small}---the one-Gaussian $\hat q$ representation---is the only step that is not a one-line symbol substitution: it reduces the two-source random field in~\eqref{eq:RS_consistency_dream_only_Psi_explicitmap} to the single-Gaussian effective field of~\cite{AlemannoETAL2023small}. Using~\eqref{eq:RS_consistency_S5_dictionary_Gamma} and~\eqref{eq:RS_consistency_S5_dictionary_magnetization}, the deterministic coefficient in $v_l$ becomes
\begin{align}
\hat X_l\,v_l
&=\hat X_l\,\frac{\beta(1+t_l)}{\Gamma_l}\frac{\bar\eta_l}{\bar D_l}
\notag\\
&=\hat X_l\,\frac{\beta(1+t_l)}{r_l^2(1+\rho_l)}\frac{r_l(1+\rho_l)\,\eta_l^{\mathrm{(S5)}}}{\bar D_l}
\notag\\
&=\hat X_l\,\frac{\beta(1+t_l)}{r_l}\frac{\eta_l^{\mathrm{(S5)}}}{\bar D_l}
\notag\\
&=\hat X_l\,\frac{\beta(1+t_l)}{r_l}\frac{m_{\eta,l}}{(t_l+\bar D_l)\bar D_l},
\label{eq:RS_consistency_S5_E13_v_rewrite}
\end{align}
where in the last step we invoked~\eqref{eq:RS_consistency_S5_E5_match}. Comparison with the one-layer S5 field then requires the one-dimensional decomposition that underlies the S5 derivation---equivalently, the CLT reduction of the empirical-class average along the condensed direction---namely
\begin{equation}
\label{eq:RS_consistency_S5_E13_Xhat_representation}
\hat X_l\ \stackrel{d}{=}\ \xi_l\,r_l\,(1+\sqrt{\rho_l}\,z),
\qquad
\xi_l\in\{\pm 1\},\quad z\sim\mathcal N(0,1),
\end{equation}
with $z$ independent of $\varphi$. Inserting~\eqref{eq:RS_consistency_S5_E13_Xhat_representation} into~\eqref{eq:RS_consistency_dream_only_Psi_explicitmap} and using~\eqref{eq:RS_consistency_S5_E13_v_rewrite} produces, in distribution,
\begin{align}
\Psi_l^{(\mathrm{Dream})}
&\stackrel{d}{=}
\frac{\sqrt{\alpha\beta\,\bar p_l}}{\bar D_l}\,\varphi
+\xi_l\,r_l(1+\sqrt{\rho_l}\,z)
\frac{\beta(1+t_l)}{r_l}\frac{m_{\eta,l}}{(t_l+\bar D_l)\bar D_l}
\notag\\
&=
\frac{\sqrt{\alpha\beta\,\bar p_l}}{\bar D_l}\,\varphi
+\xi_l\,\frac{\beta(1+t_l)}{\bar D_l+t_l}\frac{m_{\eta,l}}{\bar D_l}
+\xi_l\,\frac{\beta(1+t_l)}{\bar D_l+t_l}\frac{m_{\eta,l}\sqrt{\rho_l}}{\bar D_l}\,z.
\label{eq:RS_consistency_S5_E13_field_split}
\end{align}
The last expression has the form ``deterministic shift plus sum of two independent Gaussians''; the two Gaussian terms can therefore be fused into a single standard normal $\psi\sim\mathcal N(0,1)$, yielding
\begin{equation}
\label{eq:RS_consistency_S5_E13_fused_field}
\Psi_l^{(\mathrm{Dream})}
\stackrel{d}{=}
\frac{1}{\bar D_l}
\sqrt{\alpha\beta\,\bar p_l+
\left(\beta m_{\eta,l}\frac{1+t_l}{\bar D_l+t_l}\right)^2\rho_l}\,\psi
+
\xi_l\,\frac{\beta m_{\eta,l}}{\bar D_l}\frac{1+t_l}{\bar D_l+t_l},
\end{equation}
so that the one-layer $\tanh^2$ equation~\eqref{eq:RS_consistency_dream_only_self_q_explicitmap} rewrites as
\begin{equation}
\label{eq:RS_consistency_S5_E13_match}
\hat q_l
=\E_{\psi}\!\left[
\tanh^2\!\left(
\frac{1}{\bar D_l}
\sqrt{\alpha\beta\,\bar p_l+
\left(\beta m_{\eta,l}\frac{1+t_l}{\bar D_l+t_l}\right)^2\rho_l}\,\psi
+
\frac{\beta m_{\eta,l}}{\bar D_l}\frac{1+t_l}{\bar D_l+t_l}
\right)
\right],
\end{equation}
which is precisely the form of equation~(E.13) of~\cite{AlemannoETAL2023small}: the same layer-wise structure and the same coefficients, modulo the dictionary above.

The match with equation~(E.16) of~\cite{AlemannoETAL2023small} is the only remaining step that is not a one-line symbol substitution, because the present compact notation uses $\bar P_l$ as a primary saddle variable, whereas~\cite{AlemannoETAL2023small} recasts the free energy in terms of $\bar D$ as an explicit variational variable and imposes stationarity with respect to it. The two formulations are related by the invertible change of variables (valid for $t_l>0$, and by continuity in the limit $t_l\to 0$)
\begin{equation}
\label{eq:RS_consistency_S5_E16_D_from_P}
\bar D_l=1+\alpha\frac{t_l}{1+t_l}(\bar P_l-\bar p_l)
\qquad\Longleftrightarrow\qquad
\bar P_l
=\bar p_l+\frac{1+t_l}{\alpha t_l}(\bar D_l-1),
\end{equation}
so that S5 E.16 is the one-layer saddle condition obtained by restricting to the hetero-off pressure $\mathcal A^{\mathrm{RS}}_{\mathrm{Dream},l}$, eliminating $\bar\eta_l$ via~\eqref{eq:RS_consistency_S5_E5_match}, replacing $\bar P_l$ by $\bar D_l$ through~\eqref{eq:RS_consistency_S5_E16_D_from_P}, and imposing stationarity with respect to $\bar D_l$. In the present $\bar P_l$-based parametrisation the same information is encoded by the algebraic block~\eqref{eq:RS_consistency_dream_only_self_q_explicitmap}--\eqref{eq:RS_consistency_dream_only_Dbar_explicitmap}. To render the equivalence transparent at the level of coefficients, consider the magnetisation-dependent term in the layer-wise dreaming pressure---the D.34-like contribution in the present notation---namely
\begin{equation}
\label{eq:RS_consistency_S5_E16_magterm_start}
-\frac{\beta(1+t_l)}{2\Gamma_l}\bar\eta_l^2\frac{t_l+\bar D_l}{\bar D_l}.
\end{equation}
which, using~\eqref{eq:RS_consistency_S5_dictionary_Gamma} together with $\bar\eta_l=r_l(1+\rho_l)\eta_l^{\mathrm{(S5)}}$, becomes
\begin{align}
-\frac{\beta(1+t_l)}{2\Gamma_l}\bar\eta_l^2\frac{t_l+\bar D_l}{\bar D_l}
&=-\frac{\beta(1+t_l)}{2r_l^2(1+\rho_l)}\,r_l^2(1+\rho_l)^2\,(\eta_l^{\mathrm{(S5)}})^2\frac{t_l+\bar D_l}{\bar D_l}
\notag\\
&=-\frac{\beta(1+t_l)}{2}(1+\rho_l)(\eta_l^{\mathrm{(S5)}})^2\frac{t_l+\bar D_l}{\bar D_l}.
\label{eq:RS_consistency_S5_E16_magterm_rescaled_eta}
\end{align}
Substituting equation~(E.5) of~\cite{AlemannoETAL2023small} in the form~\eqref{eq:RS_consistency_S5_E5_match}---namely $\eta_l^{\mathrm{(S5)}}=m_{\eta,l}/(t_l+\bar D_l)$---gives
\begin{equation}
\label{eq:RS_consistency_S5_E16_magterm_final}
-\frac{\beta(1+t_l)}{2}(1+\rho_l)\,\frac{m_{\eta,l}^2}{\bar D_l(t_l+\bar D_l)},
\end{equation}
which is the magnetisation-sector structure entering the $\bar D$-stationarity equation~(E.16) of~\cite{AlemannoETAL2023small}, confirming that the present layer-wise notation reproduces the S5 coefficients exactly. For each fixed layer $l$, the hetero-off compact RS system~\eqref{eq:RS_consistency_dream_only_self_M_explicitmap}--\eqref{eq:RS_consistency_dream_only_Dbar_explicitmap} therefore coincides with equations~(E.5) and~(E.12)--(E.16) of~\cite{AlemannoETAL2023small} under the structural identifications~\eqref{eq:RS_consistency_S5_dictionary_structural}, the magnetisation normalisation~\eqref{eq:RS_consistency_S5_dictionary_magnetization}, and the change of variables~\eqref{eq:RS_consistency_S5_E16_D_from_P}. Thus the hetero-off reduction yields not merely a structurally decoupled system but $L$ independent copies of the one-layer supervised-dreaming RS theory of~\cite{AlemannoETAL2023small}.
\bibliographystyle{elsarticle-harv}
\bibliography{biblio.bib}

\end{document}